\documentclass{article}
\usepackage{amsfonts}
\usepackage{amsmath}
\usepackage{geometry}

\newtheorem{theorem}{Theorem}
\newtheorem{acknowledgement}[theorem]{Acknowledgement}

\input{tcilatex}
\begin{document}

\title{Applications of Two-Body Dirac Equations to the Meson Spectrum with
Three Versus Two Covariant Interactions, SU(3) Mixing, and Comparison to a
Quasipotential Approach}
\author{Horace W. Crater\thanks{%
hcrater@utsi.edu}~ and James Schiermeyer \\
The University of Tennessee Space Institute}
\maketitle

\begin{abstract}
In a previous paper Crater and Van Alstine applied the Two Body Dirac
equations of constraint dynamics to quark-antiquark bound states using a
relativistic extention of the Adler-Piran potential and compared their
spectral results to those from other approaches which also considered meson
spectroscopy as a whole and not in\ parts. In this paper we explore in more
detail the differences and similarities in an important subset of those
approaches, the quasipotential approach. In the earlier paper, the
transformation properties of the quark-antiquark potentials were limited to
a scalar and an electromagnetic-like four vector, with the former accounting
for the confining aspects of the overall potential, and the latter the short
range portion. The static Adler-Piran potential was first given an invariant
form and then apportioned between those two different types of potentials. \
Here we make a change in this apportionment that leads to a substantial
improvement in the resultant spectroscopy by including a time-like confining
vector potential over and above the scalar confining one and the
electromagnetic-like vector potential. \ Our fit includes 19 more mesons
than the earlier results and we modify the scalar portion of the potential
in such \ a way that allows this formalism to account for the isoscalar
mesons $\eta $ and$~\eta ^{\prime }$ not included in the previous work. \
Continuing the comparisons of formalisms and spectral results made in the
previous paper with other approaches to meson spectroscopy we examine in
this paper the quasipotential approach of \ Ebert, Faustov, and Galkin.
\end{abstract}

\bigskip

\section{\protect\bigskip Introduction}

There are a number of strategies in computational treatments of quantum
chromodynamics that emerge in the study of meson spectroscopy.\ \ One is to
set up a discrete lattice analog of the full quantum field theory. \ A
second is to first make analytic approximations which replace the quantum
field theoretic problem by a classical variational problem involving an
effective Lagrange function and action. \ The latter approach has been
exploited by Adler and Piran \cite{adl} \ and in a previous paper Crater and
Van Alstine gave a detailed account of applications of the Two Body Dirac
equations (TBDE) of constraint dynamics to the meson quark-antiquark bound
states \cite{crater2} using a relativistic extension of the Adler-Piran
potential. \ That paper also included a comparison of this approach to
others \cite{wisc}-\cite{isgr} who, like ours, considered the whole spectrum
instead of just selected parts.

Here we update the results presented in \cite{crater2} in four ways. First,
we include 19 more mesons not included in the previous work. \ Second, we
obtain a substantial improvement in our fit to most all of the mesons by
allowing the confining interaction, pure scalar in \cite{crater2}, to take
on a timelike vector portion. We still include the electromagnetic-like
vector potential used previously. Third, we extend the relativistic Schr\"{o}%
dinger-like form of the TBDE to include isoscalar mixing, thus incorporating
the isoscalar mesons $\eta $ and $~\eta ^{\prime }$.\ And finally we
critically examine, by comparison with the TBDE, aspects of quasipotential
approaches including a recent one presented in \cite{rusger} as well as in 
\cite{isgr}.

In Sec. 2 we give a short review of the relativistic two-body constraint
formalism, distinguishing between our new approach used for confining given
in this paper and the one presented in \cite{crater2} and including a
discussion of the closely related quasipotential approach. \ In Sec. 3 we
review the static Adler-Piran potential and how we apportion it between the
three invariant potential functions $A(r),\ S(r),$ and $V(r)$ used in our
TBDE. \ \ In Sec. 4 we present our main new results on meson spectroscopy. \
In Sec. 5 we include our treatment of SU(3) mixing and in Sec. 6 we discuss
the meson spectral results of \cite{rusger} including the advantages and
shortcomings of their quasipotentials bound state formalism. \ 

\section{Review of Relativistic Two-Body Formalisms}

\subsection{Two-Body Constraint Approach}

When the interaction and the masses are known, a common starting point in
describing the relativistic two-body bound state problem is the
Bethe-Salpeter equation \cite{bse}. The Bethe-Salpeter equation is, however,
usually not considered in its full four-dimensional form due to the
difficulty of treating the relative time coordinate \cite{nak}. Numerous
truncations of the Bethe-Salpeter equation have been proposed for the
relativistic two-body problem \cite{quasi,yaes}. Some of these types of
approximate methods have previously been applied with considerable success
to the $q\bar{q}$ meson spectrum \cite{prl84}-\cite{lu},~\cite{crater2}. The
ladder approximation and the instantaneous approximation of the Salpeter
equation have been widely used. It should be noted, however, that the simple
ladder approximation and the Salpeter equation do not lead to the correct
one-body limit \cite{oneb}, and do not respect gauge invariance \cite{gauge}%
. Crossed-ladder diagrams must be included to insure gauge-invariant
scattering amplitudes.

The Two Body Dirac equations of constraint dynamics provide a covariant
three-dimensional truncation of the Bethe-Salpeter equation. \ Sazdjian \cite%
{saz} has shown that the Bethe-Salpeter equation can be algebraically
transformed into two independent equations. The first yields a covariant
three-dimensional eigenvalue equation which for spinless particles takes the
form 
\begin{equation}
\biggl(\mathcal{H}_{10}+\mathcal{H}_{20}+2\Phi _{w}\biggr )\Psi
(x_{1},x_{2})=0,  \label{eq:sum}
\end{equation}%
where $\mathcal{H}_{i0}=p_{i}^{2}+m_{i}^{2}$ . The quasipotential $\Phi _{w}$
is a modified geometric series in the Bethe-Salpeter kernel $K$ such that in
lowest order in $K$ 
\begin{equation}
\Phi _{w}=\pi iw\delta (P\cdot p)K,  \label{bsen}
\end{equation}%
where $P=p_{1}+p_{2}$ is the total momentum, $p=\eta _{2}p_{1}-\eta _{1}p_{2}
$ is the relative momentum, $w$ is the invariant total center of momentum
(c.m.) energy with $P^{2}=-w^{2}$. $\ $The $\eta _{i}$ must be chosen so
that the relative coordinate $x=x_{1}-x_{2}$ and $p$ are canonically
conjugate, i.e. $\eta _{1}+\eta _{2}=1$. The second equation overcomes the
difficulty of treating the relative time in the center of momentum system by
setting an invariant condition on the relative momentum $p$, 
\begin{equation}
(\mathcal{H}_{10}-\mathcal{H}_{20})\Psi (x_{1},x_{2})=0=2P\cdot p\Psi
(x_{1},x_{2}).  \label{eq:dif}
\end{equation}%
Note that this implies $p^{\mu }\Psi =p_{\perp }^{\mu }\Psi \equiv (\eta
^{\mu \nu }+\hat{P}^{\mu }\hat{P}^{\nu })p_{\nu }\Psi $ in which $\hat{P}%
^{\mu }=P^{\mu }/w$ is a time like unit vector $(\hat{P}^{2}=-1)$ in the
direction of the total momentum.

One can further combine the sum and the difference of Eqs. (\ref{eq:sum})
and (\ref{eq:dif}) to obtain a set of two relativistic equations one for
each particle with each equation specifying two generalized mass-shell
constraints 
\begin{equation}
\mathcal{H}_{i}\Psi (x_{1},x_{2})=(p_{i}^{2}+m_{i}^{2}+\Phi _{w})\Psi
(x_{1},x_{2})=0,~i=1,2,  \label{dir}
\end{equation}%
including the interaction with the other particle. These constraint
equations are just those of Dirac's Hamiltonian constraint dynamics\footnote{%
These equations were originally proposed in the form of classical
generalized mass shell first class constraints $\mathcal{H}%
_{i}=(p_{i}^{2}+m_{i}^{2}+\Phi _{i})\approx 0$, and their quantization $%
\mathcal{H}_{i}\Psi =0$ without reference to a quantum field theory. For the
classical $\mathcal{H}_{i}$ to be compatible, their Poisson bracket with one
\ another must either vanish strongly or depend on the constraints
themselves, $\{\mathcal{H}_{1},\mathcal{H}_{2}\}\approx 0$. \ The simplest
solution of this equation is $\Phi _{1}=\Phi _{2}$, a kind of relativistic
third law condition, together with their common transverse coordinate
dependence $\Phi _{w}(x_{\perp }),$ just as with its quantum version.} \cite%
{dirac,cnstr}. In order for the two simultaneous wave equations of (\ref{dir}%
) to have solutions other than zero, Dirac's constraint dynamics stipulate
that these two constraints must be compatible among themselves, $[\mathcal{H}%
_{1},\mathcal{H}_{2}]\Psi =0$, that is, they must be first class. \ With no
external potentials, the coordinate dependence of the quasipotential $\Phi
_{w}$ $\ $would be through $x$ and the compatibility condition becomes $%
[p_{1}^{2}-p_{2}^{2},\Phi _{w}]\Psi =P^{\mu }\partial \Phi _{w}/\partial
x^{\mu }=0$. In order for this to be true in general, $\Phi _{w}$ must
depend on the relative coordinate $x$ only through its component, $x_{\perp
},$ perpendicular to $P,$%
\begin{equation}
x_{\perp }^{\mu }=(\eta ^{\mu \nu }+\hat{P}^{\mu }\hat{P}^{\nu
})(x_{1}-x_{2})_{\nu }.  \label{ti}
\end{equation}%
Since the total momentum is conserved, the single component wave function $%
\Psi ~$in coordinate space is a product of a plane wave eigenstate of $P$
~and an internal part $\psi $ \cite{cra87}, depending on this $x_{\perp }.$%
\footnote{%
We use the same symbol $P$ for the eigenvalue so that the $w$ dependence in
Eq. (\ref{em}) is regarded as an eigenvalue dependence. \ The wave function $%
\Psi $ can be viewed either as a relativistic 2-body wave function (similar
in interpretation to the Dirac wave function) or, if a close connection to
field theory is required, related directly to the Bethe Salpeter wave
function $\chi \mathbf{~}$by \cite{saz} $\Psi =-\pi i\delta (P\cdot p)%
\mathcal{H}_{10}\chi =-\pi i\delta (P\cdot p)\mathcal{H}_{20}\chi $.}

We find a plausible structure for the quasipotential $\Phi _{w}$ by
observing that the one-body Klein-Gordon equation $(p^{2}+m^{2})\psi =(%
\mathbf{p}^{2}-\varepsilon ^{2}+m^{2})\psi =0$ takes the form $(\mathbf{p}%
^{2}-\varepsilon ^{2}+m^{2}+2mS+S^{2}+2\varepsilon A-A^{2})\psi =0~$when one
introduces a scalar interaction and timelike vector interaction via $%
m\rightarrow m+S~$and $\varepsilon \rightarrow \varepsilon -A$. In the
two-body case, separate classical \cite{fw} and quantum field theory \cite%
{saz97} arguments show that when one includes world scalar\ and vector
interactions then $\Phi _{w}$ depends on two underlying invariant functions $%
S(r)$ and $A(r)$ through the two-body Klein-Gordon-like potential form with
the same general structure, that is%
\begin{equation}
\Phi _{w}=2m_{w}S+S^{2}+2\varepsilon _{w}A-A^{2}.  \label{em}
\end{equation}%
Those field theories further yield the c.m. energy dependent forms 
\begin{equation}
m_{w}=m_{1}m_{2}/w,  \label{mw}
\end{equation}%
and%
\begin{equation}
\varepsilon _{w}=(w^{2}-m_{1}^{2}-m_{2}^{2})/2w,  \label{ew}
\end{equation}%
ones that Tododov \cite{cnstr} introduced as the relativistic reduced mass
and effective particle energy for the two-body meson system.\ Similar to
what happens in the nonrelativistic two-body problem, in the relativistic
case\ we have the motion of this effective particle taking place as if it
were in an external field \ (here generated by $S$ and $A$). \ The two
kinematical variables (\ref{mw}) and (\ref{ew}) are related to one another
by the Einstein condition 
\begin{equation}
\varepsilon _{w}^{2}-m_{w}^{2}=b^{2}(w),
\end{equation}%
where the invariant 
\begin{equation}
b^{2}(w)\equiv
(w^{4}-2w^{2}(m_{1}^{2}+m_{2}^{2})+(m_{1}^{2}-m_{2}^{2})^{2})/4w^{2},
\label{bb}
\end{equation}%
is the c.m. value of the square of the relative momentum expressed as a
function of $w$. \ One also has%
\begin{equation}
b^{2}(w)=\varepsilon _{1}^{2}-m_{1}^{2}=\varepsilon _{2}^{2}-m_{2}^{2},
\end{equation}%
in which $\varepsilon _{1}$ and $\varepsilon _{2}$ are the invariant c.m.
energies of the individual particles satisfying%
\begin{equation}
\ \varepsilon _{1}+\varepsilon _{2}=w,\ \varepsilon _{1}-\varepsilon
_{2}=(m_{1}^{2}-m_{2}^{2})/w.  \label{es}
\end{equation}%
In terms of these invariants, the relative momentum appearing in Eq. (\ref%
{bsen}) and (\ref{eq:dif}) is given by%
\begin{equation}
p^{\mu }=(\varepsilon _{2}p_{1}^{\mu }-\varepsilon _{1}p_{2}^{\mu })/w%
\mathrm{,}  \label{relm}
\end{equation}%
so that $\eta _{1}+\eta _{2}=(\varepsilon _{1}+\varepsilon _{2})/w=1$. \ In 
\cite{tod} the forms for these two-body and effective particle variables are
given sound justifications based solely on relativistic kinematics,
supplementing the dynamical arguments of \cite{fw} and \cite{saz97}.

Originally, the Two Body Dirac equations of constraint dynamics arose from a
supersymmetric treatment of two pseudoclassical constraints (with Grassmann
variables in place of gamma matrices) which were then quantized \cite{cra82}-%
\cite{cww}. Sazdjian later derived \cite{saz} different forms of these same
equations, just as with their spinless counterparts above, as covariant and
three-dimensional truncation of the Bethe-Salpeter equation. \ The forms of
the equations are varied (see Appendix A), but the one that is the most
familiar is the "external potential" form similar in structure to the
ordinary Dirac equation. \ For two particles interacting through world
scalar and vector interactions they are 
\begin{align}
\mathcal{S}_{1}\psi & \equiv \gamma _{51}(\gamma _{1}\cdot (p_{1}-\tilde{A}%
_{1})+m_{1}+\tilde{S}_{1})\Psi =0,  \notag \\
\mathcal{S}_{2}\psi & \equiv \gamma _{52}(\gamma _{2}\cdot (p_{2}-\tilde{A}%
_{2})+m_{2}+\tilde{S}_{2})\Psi =0.  \label{tbde}
\end{align}%
Here $\Psi $ is a 16 component wave function consisting of an external plane
wave part that is an eigenstate of $P$ and an internal part $\psi =\psi
(x_{\perp })$. The vector potential$\ \tilde{A}_{i}^{\mu }$ was taken to be
an electromagnetic-like four-vector potential with the time and spacelike
portions both arising from a single invariant function $A.~$\footnote{%
In a perturbative context, i.e. for weak potentials, that would mean that
this aspects of $\tilde{A}_{i}^{\mu }$ is regarded as arising from a Feynman
gauge vertex coupling of a form proportional to $\gamma _{1}^{\mu }\gamma
_{2\mu }A$ (see Appendix A).} The tilde on these four-vector potentials as
well as on the scalar ones $\tilde{S}_{i}$ indicate that they are not only
position dependent but also spin-dependent by way of the gamma matrices. \
In this paper we allow for the presence of a timelike portion arising from
an independent invariant function $V.~$\footnote{%
In a perturbative or weak potential context, that would mean that this
aspect of $\tilde{A}_{i}^{\mu }$ is regarded as arising from an additional
vertex coupling proportional to $-\gamma _{1}\cdot \hat{P}~\gamma _{2}\cdot 
\hat{P}V.$ Similarly in a perturbative or weak potential context $\tilde{S}%
_{i}$ is regarded as arising from a vertex coupling proportional to $%
1_{1}1_{2}S.$ (See Appendix A).}\ \ In either case, the operators $\mathcal{S%
}_{1}$ and $\mathcal{S}_{2}$ must commute or at the very least $[\mathcal{S}%
_{1},\mathcal{S}_{2}]\psi =0$ since they operate on the same wave function. 
\footnote{%
The $\gamma _{5}$ matrices for each of the two particles are designated by $%
\gamma _{5i}$ $i=1,2$. \ The reason for putting these matrices out front of
the whole expression is that including them facilitates the proof of the
compatibility condition, see \cite{cra82}, \cite{cra87}.} This compatibility
condition gives restrictions on the spin dependence\ which the vector and
scalar potentials%
\begin{equation}
\tilde{A}_{i}^{\mu }=\tilde{A}_{i}^{\mu }(A(r),V(r),p_{\perp },\hat{P}%
,w,\gamma _{1},\gamma _{2}),~\ \tilde{S}_{i}=\tilde{S}_{i}(S(r),A(r),p_{%
\perp },\hat{P},w,\gamma _{1},\gamma _{2}).  \label{paul1}
\end{equation}%
are allowed to have \footnote{%
The dependence of the scalar potentials $\tilde{S}_{i}$ on the invariant $%
A(r)$ responsible for the electromagnetic-like potential is seen in \cite%
{cra87} and \cite{saz97} to result from the way the scalar and vector fields
combine. That combination without the presence of the independent time-like
portion leads to a two-body Klein-Gordon-like potential portion of $\Phi _{w}
$ to be of the form given in Eq. (\ref{em}).} in addition to requiring that
they depend on the invariant separation $r\equiv \sqrt{x_{\perp }^{2}}$
through the invariants $A(r),V(r),$ and $S(r)$ . The covariant constraint (%
\ref{eq:dif}) can also be shown to follow from Eq. (\ref{tbde}). \ We give
the explicit connections between $\tilde{A}_{i}^{\mu },\tilde{S}_{i}$ and
the invariants $A(r),V(r),$ and $S(r)$ in Appendix A. The Pauli reduction of
these coupled Dirac equations lead to a covariant Schr\"{o}dinger-like
equation for the relative motion with an explicit spin-dependent potential $%
\Phi _{w},$\footnote{%
In the presence of an additional and independent time-like vector
interaction $V$, we assume the scalar and vector fields combine in such a
way that leads to a two-body Klein-Gordon-like potential portion of $\Phi
_{w}$ of the form $2m_{w}S+S^{2}+2\varepsilon _{w}A-A^{2}+2\varepsilon
_{w}V-V^{2}$ instead of that given in Eq. (\ref{em}).} 
\begin{equation}
{\bigg(}p_{\perp }^{2}+\Phi _{w}(A(r),V(r),S(r),p_{\perp },\hat{P},w,\sigma
_{1},\sigma _{2}){\bigg)}\psi _{+}=b^{2}(w)\psi _{+}\ ,  \label{schlike}
\end{equation}%
with $b^{2}(w)$ playing the role of the eigenvalue.\footnote{%
Due to the dependence of $\Phi _{w}$ on $w,$ this is a nonlinear eigenvalue
equation. \ } This eigenvalue equation can then be solved for the
four-component effective particle spinor wave function $\psi _{+}$ related
to the 16 component spinor $\psi (x_{\perp })$ in Appendix A.

The set of equations (\ref{tbde}) and the equivalent Schr\"{o}dinger-like
equation (\ref{schlike}) possess a number of important and desirable
features. First, they reduce to the correct one-body Dirac form when one of
the two constituents becomes very massive. (The Salpeter equation does not
have this important property.) Second, the generalized three-dimensional Schr%
\"{o}dinger equation (\ref{schlike}) is quite similar to the nonrelativistic
Schr\"{o}dinger equation and it indeed goes over to the correct
nonrelativistic Schr\"{o}dinger equation in the limit of weak binding. One
can thus employ familiar techniques to obtain its solutions. Third, Eq. (\ref%
{schlike}) can be solved nonperturbatively for both QED bound states (e.g.
positronium and muonium) and QCD bound states (i.e. bound states obtained
from two-body relativistic potential models for mesons) since every term in $%
\Phi _{w}$ is nonsingular in the sense that they are less attractive than $%
-1/4r^{2}$ (no delta functions or attractive $1/r^{3}$ potentials for
example). Thus, unlike with the $1/m$ non-relativistic and semirelativistic
expansions, the covariant Dirac formalism itself introduces natural cutoff
factors that smooth out singular spin-dependent interactions, there being no
need to introduce them by hand (- see \cite{cra82},\cite{cra84},\cite{becker}%
, \cite{yoon} and Sec. 6.4) as in other approaches. Fourth, the relativistic
potentials appearing in these equations are directly related through Eq.\ (%
\ref{bsen}) to the interactions of perturbative quantum field theory, while
for QCD bound states they may be introduced semiphenomenolgically through $%
A(r)$ and $S(r)$ (and in this paper $V(r)$). Fifth, these equations have
been tested analytically \cite{exct} and numerically \cite{becker} against
the known perturbative fine and hyperfine structures of QED bound states and
related field theoretic bound states. The (nonperturbative) successes with
these QED bound states provide strong motivation for applying the
constrained Dirac formalism to meson bound states as in \cite{crater2}.
Sixth, these equations provide a covariant three-dimensional framework in
which the local potential approximation consistently fulfills the
requirements of gauge invariance in QED \cite{saz96}. Finally, the same
general structures of the Darwin, spin-orbit, spin-spin and tensor terms in $%
\Phi _{w}(A,V,S)$ of Eq. (\ref{schlike}) and (\ref{57}), responsible for the
accurate hyperfine structures of QED bound states arising naturally from the
TBDE formalism when $A=-\alpha /r$ and $S=V=0$, are used with the only
alteration in its application to the QCD bound states being that $A,V,S$ are
apportioned appropriately from the Adler-Piran potential.

We emphasize the importance of the nonperturbative QED bound state test.
(See Appendix C for a review of the application of the TBDE to QED.) Many of
the wave equations used in the standard approaches to QED bound states have
been modified to include QCD inspired potentials and then applied to
nonperturbative numerical calculations of QCD bound states without first
testing those approaches nonperturbatively in QED. By this we mean that the
accepted perturbative results of those equations (QED spectral results
correct through order $\alpha ^{4})$ have not been replicated using
numerical methods. \ Sommerer $et~al.$ \cite{iowa} have shown that the
Blankenbecler-Sugar equation and the Gross equations fail this test. This
indicates a danger in applying such three-dimensional truncations of the
Bethe-Salpeter equation: if failure occurs in their applications to QED
bound states this brings into question the spectral results of similar
nonperturbative (i.e. numerical) approaches based on the same truncations
when applied to QCD bound states. \ This would be true especially when the
only difference between the vector portions of the QED and QCD potentials
would be the replacing of the QED $-\alpha /r$ by a similar $A(r)$ from QCD.
\ 

In \cite{crater2} we presented details of the application of this formalism
to meson spectroscopy using a covariant version of the Adler-Piran static
quark potential. Note especially that the equations used there displayed a
single $\Phi (A(r),S(r),p_{\perp },\hat{P},w,\sigma _{1},\sigma _{2},)$ in
Eq.\ (\ref{schlike}). It depends on the quark masses through factors such as
those that appear in Eq. (\ref{em}). However its dependence is the same for
all quark mass ratios - hence a single structure for all the $Q\bar{Q},\ q%
\bar{Q},$ and $q\bar{q}$ mesons in a single overall fit. We found that the
fit provided by the TBDE for the entire meson spectrum (from the pion to the
excited bottomonium states) competes with the best fits to partial spectra
provided by other approaches and does so with the smallest number of
interaction functions (just $A(r)$ and $S(r)$) without additional cutoff
parameters necessary to make those approaches numerically tractable. We also
found that the pion bound state displays some characteristics of a Goldstone
boson. That is, as the quark mass tends to zero, the pion mass (unlike the $%
\rho $ and the excited $\pi $) vanishes, in contrast to almost every other
relativistic potential model. (For more discussion on this see footnote 23
below).

\subsection{Two Body Quasipotential Approaches}

Also presented in \cite{crater2} was a detailed \ comparison \ between the
meson spectroscopy results of our model and those of several other
approaches: one based on the Breit equation \cite{bry}, two on truncated
versions of the Bethe-Salpeter \ equation \cite{wisc}, \cite{iowa}, and one
on a quasipotential approach \cite{isgr}. \ We explore in this section in
more detail the differences and similarities between our approach and the
quasipotential approach. \ The quasipotential equation was first introduced
by Logunov and Tavkhelidze \cite{quasi}. \ In its homogeneous form, that
equation describes a two-particle relativistic composite system with its
c.m. momentum space form (in the notation used here $\mathbf{p}$ is the
relative momentum given in Eq. (\ref{relm})) for spinless particles given by%
\begin{equation}
(w-\sqrt{\mathbf{p}^{2}+m_{1}^{2}}-\sqrt{\mathbf{p}^{2}+m_{2}^{2}})\Psi _{w}(%
\mathbf{p)=}\int V(\mathbf{p,q,}w)\Psi _{w}(\mathbf{q)}\frac{d^{3}q}{(2\pi
)^{3}},  \label{log}
\end{equation}%
where $\Psi _{w}(\mathbf{p)}$ is the quasipotential wave function projected
onto positive-frequency states and $V(\mathbf{p,q,}w)$ is the quasipotential
calculated by means of the off-energy shell scattering amplitude (so that
the respective c.m. energies of the two particles are not given by the above
square roots but by Eq. (\ref{es})). \ The corresponding inhomogeneous
quasipotential equation is of the general form%
\begin{equation}
T(\mathbf{p,q,}w)+V(\mathbf{p,q,}w)+\int V(\mathbf{p,k,}w)G_{w}(\mathbf{k)}V(%
\mathbf{k,q,}w)=0,  \label{quasi}
\end{equation}%
a linear integral equation of the Lippmann-Schwinger type relating the
quasipotential to the off-energy shell extrapolation of the Feynman
scattering amplitude. \ The choice of this equation and the accompanying
homogeneous equation is not unique~\cite{yaes}. \ For example, the Green
function $G_{w}(\mathbf{k)}$ has only its imaginary part determined by
requiring the condition of elastic unitarity on Eq. (\ref{quasi}). \footnote{%
For hermitian potentials, that condition has the symbolic form of $T-T^{\dag
}=T^{\dag }(G-G^{\dag })T.$} \ Todorov \cite{quasi} took advantage of this
nonuniqueness to write down a local version of the corresponding homogeneous
equation of the form%
\begin{equation}
\left( \mathbf{p}^{2}-b^{2}\right) \phi (\mathbf{p)+}\frac{2\varepsilon
_{1}\varepsilon _{2}}{w}\int V_{w}(\mathbf{p,k)}\phi (\mathbf{k)}\frac{d^{3}k%
}{(2\pi )^{3}}=0,  \label{tod0}
\end{equation}%
with$~\phi (\mathbf{p})$ the wave function in momentum space. In \cite{cra84}
Crater and Van Alstine showed that the spinless version of Eq. (\ref{schlike}%
) in the case of QED ($V=S=0$) has, in the c.m. frame, the form (see Eq.
(66b) and discussion below Eq. (73b) of that paper)\ 
\begin{equation}
(\mathbf{p}^{2}-(\varepsilon _{w}-A)^{2}+m_{w}^{2}+\frac{1}{2}\mathbf{\nabla 
}^{2}\mathcal{G}+\frac{1}{4}\left( \mathbf{\nabla }\mathcal{G}\right)
^{2})\psi =0,  \label{cra84}
\end{equation}%
where%
\begin{eqnarray}
\mathcal{G} &\mathcal{=}&\ln G,  \notag \\
G &=&\frac{1}{(1-2A/w)^{1/2}}.
\end{eqnarray}%
As was pointed out in that paper, for $A=-\alpha /r,$ this reduces for weak
potentials to the minimal or gauge structure form postulated by Todorov, 
\begin{equation}
\lbrack (\mathbf{p-A)}^{2}-(\varepsilon _{w}-A^{0})^{2}+m_{w}^{2}]\psi =0.
\label{gage}
\end{equation}%
Although not noticed at the time, Eq. (\ref{cra84}) does in fact
additionally have this minimal structure not only for arbitrary strength
couplings but also for potentials not restricted to Coulomb potentials,
provided just that%
\begin{eqnarray}
A^{0} &=&A,  \notag \\
\mathbf{A} &\mathbf{=}&\mathbf{-}\frac{i}{2}\mathbf{\nabla }\mathcal{G~}%
I_{s},
\end{eqnarray}%
where $I_{s}$ is the space reflection operator satisfying%
\begin{equation}
I_{s}f(\mathbf{r)=}f(-\mathbf{r).}
\end{equation}

Aneva, Karchev, and Rizov \cite{akr} developed the weak potential version of
this two-body Klein-Gordon equation for two spin combinations: for one
spin-zero and one spin-one half particle and for a spin-one-half
particle-antiparticle pair. \ \ For the latter it has the form, 
\begin{equation}
\left( \mathbf{p}^{2}-b^{2}\right) \phi _{\lambda _{1}\lambda _{2}}(\mathbf{%
p)+}\frac{2\varepsilon _{1}\varepsilon _{2}}{w}\int \bar{u}_{\lambda
_{1}^{\prime }}(\mathbf{p)}\bar{v}_{\lambda _{2}}(-\mathbf{k)}\mathbb{V}_{w}(%
\mathbf{p,k)}v_{\lambda _{2}^{\prime }}(-\mathbf{p)}u_{\lambda _{1}}(\mathbf{%
k)}\phi _{\lambda _{1}^{\prime }\lambda _{2}^{\prime }}(\mathbf{k)}\frac{%
d^{3}k}{(2\pi )^{3}}=0.  \label{riz}
\end{equation}%
Expressing the on shell free four-component Dirac spinors in terms of
two-component Pauli spinors and assuming the local quasipotentials 
\begin{equation}
\mathbb{V}_{w}(\mathbf{p,k)=}\mathbb{V}_{w}(\mathbf{p-k)=}\mathcal{A}(%
\mathbf{p-k})\gamma _{1}^{\mu }\gamma _{2\mu }+\mathcal{V}(\mathbf{p-k}%
)\beta _{1}\beta _{2}+\mathcal{S}(\mathbf{p-k})1_{1}1_{2},  \label{qua}
\end{equation}%
Eq. (\ref{riz}) can be brought to a four-component wave equation form
superficially similar to Eq. (\ref{schlike}) in the c.m. frame. We write it
as%
\begin{equation}
{\bigg(}\mathbf{p}^{2}+\mathcal{V}_{w}(A(r),V(r),S(r),\mathbf{p},w,\mathbf{%
\sigma }_{1}\mathbf{,\sigma }_{2}){\bigg)}\phi (\mathbf{r)}=b^{2}(w)\phi (%
\mathbf{r).}  \label{vw}
\end{equation}%
However, there are distinct differences. First of all, the spin structure of 
$\Phi _{w}$ is not identical to that of $\mathcal{V}_{w}$ even if the
functions $A(r),V(r),S(r)$ are the same. \ The reason is that the spin
dependence of the vector and scalar potentials $\tilde{A}_{i}^{\mu }~$\ and $%
\ \tilde{S}_{i}$ and in particular the minimal type of context in which they
appear in Eqs. (\ref{tbde}) arise from (nonlinear) hyperbolic functions (see
Appendix A and \cite{jmath}, \cite{long}) \ of matrices such as appear in
Eq. (\ref{qua}). Without that hyperbolic structure the external potential
forms in which the minimal structures appear would be absent. \ To reproduce
the effects of those nonlinear functions in Eq. (\ref{vw}) one would have to
supplement Eq. (\ref{qua}) with types of invariants other than just scalar
and vector \cite{saz86}, \cite{saz94}, \cite{saz97} (a pseudovector
invariant for example). Secondly, the desirable minimal scalar structures as
appear in the first line on the right hand side of Eq. (\ref{57}) below
would not appear in $\mathcal{V}_{w}$ without including higher order
diagrams that would again require invariants other than just scalar and
vector. These minimal scalar structures are not only desirable, they arise
naturally and strictly from $O(1/c^{2})$ expansions of classical and quantum
field theoretic potentials \cite{fw}, \cite{saz97} and from gauge invariance
considerations (see Todorov in \cite{quasi}, and also \cite{akr}, \cite%
{crstr}, and \cite{cra87}). In a later section we discuss the recent work of 
\cite{rusger}, which uses a quasipotential equation similar to Eq. (\ref{vw}%
) in meson spectroscopy calculations.

\section{The Adler-Piran Potential for the Two Body Dirac Equations.}

In this section we use the relativistic Schr\"{o}dinger-like Eq.(\ref%
{schlike}) to construct a relativistic naive quark model by choosing the
three invariant functions $A,V$ and $S$ to incorporate the Adler-Piran
static quark potential \cite{adl}. \ This potential was originally obtained
from the QCD field theory through a nonlinear effective action model for
heavy quark statics. Adler and Piran used the renormalization group
approximation to obtain both total flux confinement and a linear static
potential at large distances. Their model uses nonlinear electrostatics with
displacement and electric fields related through a nonlinear constitutive
equation with the effective dielectric constant given by a leading log-log
model which fixes all parameters in their model apart from a mass scale $%
\Lambda .$ \ Their static potential also contains an unknown "integration
constant" $U_{0}$ in the final form of their potential (hereafter called $%
V_{AP}(r)$). \ We insert into Eq.(\ref{schlike}) invariants $A,V$ and $S$
with forms determined so that the sum $A+V+S$ appearing as the potential in
the nonrelativistic limit of our equations becomes the Adler-Piran
nonrelativistic $Q\bar{Q}$ potential (which depends on two parameters $%
\Lambda $ and $U_{0})$ plus the Coulomb interaction between the quark and
antiquark. That is, 
\begin{equation}
V_{AP}(r)+V_{coul}=\Lambda (U(\Lambda r)+U_{0})+\frac{e_{1}e_{2}}{r}=A+V+S\ .
\label{asap}
\end{equation}%
As determined by Adler and Piran, the short and long distance behaviors of $%
U(\Lambda r)$ generate known lattice and continuum results through the
explicit appearance of an effective running coupling constant in coordinate
space. That is, the Adler-Piran potential incorporates asymptotic freedom
through 
\begin{equation}
\Lambda U(\Lambda r<<1)\sim 1/(r\ln \Lambda r),
\end{equation}%
and linear confinement through 
\begin{equation}
\Lambda U(\Lambda r>>1)\sim \Lambda ^{2}r.
\end{equation}%
In addition to obtaining these leading behavior analytic forms for short and
long distances, they converted the numerically obtained values of the
potential at all distances (short, intermediate, and long distances) to
compact analytic expressions. The explicit closed form expressions \cite{adl}
for $U(\Lambda r)$ are different for each of the four regions, $\ $and are
linked continuously. \ Letting $x=\Lambda r,$%
\begin{eqnarray}
U(x) &=&-(16\pi /27)(1+a_{1}x^{a_{2}})f(w_{p})/w_{p},~~~~~0<x\leq 0.0125, 
\notag \\
w_{p} &=&1/(a_{3}x)^{2},  \notag \\
U(x) &=&K+\alpha (x/0.125)^{E},~~~~~0.0125\leq x\leq 0.125,  \notag \\
E &=&\beta +\gamma \ln (1/x)+\delta (\ln (1/x))^{2}+\varepsilon (\ln
(1.x))^{3}  \notag \\
U(x) &=&K^{\prime }+\alpha ^{\prime }\ln x+\beta ^{\prime }(\ln
x)^{2}+\gamma ^{\prime }(lnx)^{3}+\delta ^{\prime }(\ln x)^{4}+\varepsilon
^{\prime }(lnx)^{4},~~~~0.125\leq \Lambda r\leq 2,  \notag \\
U(x) &=&\Lambda (c_{1}x+c_{2}\ln x+\frac{c_{3}}{\sqrt{x}}+\frac{c_{4}}{x}%
+c_{5}),~~~~2\leq \Lambda r<\infty .
\end{eqnarray}%
The function $f$ is defined by $w_{p}=f(\ln f+\zeta \ln \ln f);~\zeta
=2(51-19n_{f}/3)/(11-2n_{f}/3)^{2}=64/81$ for $n_{f}=3~.$ The various
constants\ $~a_{1}$ to $a_{3},~K,\alpha ,\beta ,\gamma ,\delta ,\varepsilon
,K^{\prime },\alpha ^{\prime },\beta ^{\prime },\gamma ^{\prime },\delta
^{\prime },\varepsilon ^{\prime },$and $c_{1}$ to$~c_{5}$ are given by the
Adler-Piran leading log-log model \cite{adl} and are not adjustable
parameters. \ We modify these closed forms so that the connections between
different regions are continuous in second derivatives. The nonrelativistic
analysis used by Adler and Piran, however, does not determine the
relativistic transformation properties of the potential. How this potential
is apportioned between vector and scalar is therefore somewhat, although not
completely, arbitrary.

In earlier work \cite{cra88} we divided the potential in the following way
among three relativistic invariants $A,V$ and $S$ for all $x=\Lambda r$. (In
our former construction, the additional invariant $V$ was responsible for a
possible independent timelike vector interaction.)

\begin{align}
S& =\eta (\Lambda (c_{1}x+c_{2}\ln (x)+\frac{c_{3}}{\sqrt{x}}+c_{5}+U_{0}), 
\notag \\
V& =(1-\eta )\Lambda (c_{1}x+c_{2}\ln (x)+\frac{c_{3}}{\sqrt{x}}%
+c_{5}+U_{0}),  \notag \\
A& =U(x)-\Lambda (c_{1}x+c_{2}\ln (x)+\frac{c_{3}}{\sqrt{x}}+c_{5}),
\label{old}
\end{align}%
in which $\eta ={\frac{1}{2}}$. That is, we assumed that (with the exception
of the Coulomb-like term ($c_{4}/x$)) the long distance part was equally
divided between scalar and a proposed timelike vector.

In the present investigation, we compute the best fit meson spectrum for the
following apportionment of the Adler-Piran potential: 
\begin{eqnarray}
A &=&\exp (-\beta \Lambda r)[V_{AP}-\frac{c_{4}}{r}]+\frac{c_{4}}{r}+\frac{%
e_{1}e_{2}}{r},  \notag \\
\ \ V+S &=&V_{AP}+\frac{e_{1}e_{2}}{r}-A=(V_{AP}-\frac{c_{4}}{r})(1-\exp
(-\beta \Lambda r))\equiv \mathcal{U},  \label{aps}
\end{eqnarray}%
In order to covariantly incorporate the Adler-Piran potential into our
equations, we treat the short distance portion as purely
electromagnetic-like (in the sense of the transformation properties of the
potential). The attractive ($c_{4}=-0.58$) QCD-Coulomb-like portion (not to
be confused with the electrostatic $V_{coul}=e_{1}e_{2}/r$) is assigned
completely to the electromagnetic-like\ part $A$. That is, the constant
portion of the running coupling constant corresponding to the exchange
diagram is expected to be electromagnetic-like ($\sim \gamma _{1\mu }\gamma
_{2}^{\mu }$). Through the additional parameter $\beta $, the exponential
factor gradually turns off the electromagnetic-like contribution (i.e. $A)$
to the potential at long distance except for the $1/r$ portion mentioned
above, while the scalar and timelike portions (i.e. $S$ and $V$) gradually
turn on, becoming fully responsible for the linear confining and subdominant
terms at long distance. We choose not to consider an apportionment function
with a large number of parameters as the simple exponential gives a single
length scale for the turning of the potential from electromagnetic-like to
scalar and timelike. Altogether our three invariant potential functions
depend on three parameters: $\Lambda ,U_{0},$ and $\beta $. \ \ We
furthermore let a free parameter $\xi $ divide the relative portions of $%
\mathcal{U}$ as follows 
\begin{eqnarray}
S &=&\xi \mathcal{U=\xi }(V_{AP}-\frac{c_{4}}{r})(1-\exp (-\beta \Lambda r)),
\notag \\
V &=&\mathcal{U-S=(}1-\xi \mathcal{)}(V_{AP}-\frac{c_{4}}{r})(1-\exp (-\beta
\Lambda r)).  \label{usv}
\end{eqnarray}%
This differs from the division in the earlier work \cite{cra88}. \ Also, the
earlier work did not include the effects of the tensor interaction or
spin-orbit difference terms or the $u-d$ quark mass differences\footnote{%
In the present treatment, we treat the entire interaction present in our
equations, thereby keeping each of these effects. In our former treatment 
\cite{cra88} we also performed a decoupling between the upper-upper and
lower-lower components of the wave functions for spin-triplet states which
turned out to be defective but which we subsequently corrected in our
numerical test of our formalism for QED \cite{becker}. \ } (see Eq. (\ref{57}%
) below). \ In \cite{crater2} , Crater and Van Alstine chose $\xi =1$ and
thus assumed that the scalar interaction is solely responsible for the long
distance confining terms.

When inserted into the constraint equations, $V,\ S$ and $A$ become
relativistic invariant functions of the invariant separation $r=\sqrt{%
x_{\perp }^{2}}$ . The covariant structures of the constraint formalism then
automatically determine the exact forms by which the central static
potential is supplemented with accompanying relativistic spin-dependent and
recoil terms.

\section{Meson Spectroscopy from the Schr\"{o}dinger-like form of the
Two-Body Dirac Equations}

\subsection{Center of Momentum form of Covariant Pauli-Schr\"{o}dinger
Reduction of the Two-Body Dirac Equations}

In Appendix A we outline the steps needed to obtain the explicit c.m. form
of \ Eq. (\ref{schlike}). \ That form is \cite{liu}, \cite{saz94}, \cite%
{crater2}, 
\begin{align}
& \{\mathbf{p}^{2}+\Phi (\mathbf{r,}m_{1},m_{2},w,\mathbf{\sigma }_{1},%
\mathbf{\sigma }_{2})\}\psi _{+}=  \notag \\
=& \{\mathbf{p}^{2}+2m_{w}S+S^{2}+2\varepsilon _{w}A-A^{2}+2\varepsilon
_{w}V-V^{2}+\Phi _{D}  \notag \\
& +\mathbf{L\cdot (\sigma }_{1}\mathbf{+\sigma }_{2}\mathbf{)}\Phi _{SO}+%
\mathbf{\sigma }_{1}\mathbf{\cdot \hat{r}\sigma }_{2}\mathbf{\cdot \hat{r}%
L\cdot (\sigma }_{1}\mathbf{+\sigma }_{2}\mathbf{)}\Phi _{SOT}  \notag \\
& +\mathbf{\sigma }_{1}\mathbf{\cdot \sigma }_{2}\Phi _{SS}+(3\mathbf{\sigma 
}_{1}\mathbf{\cdot \hat{r}\sigma }_{2}\mathbf{\ \cdot \hat{r}-\sigma }_{1}%
\mathbf{\cdot \sigma }_{2})\Phi _{T}  \notag \\
& +\mathbf{L\cdot (\sigma }_{1}\mathbf{-\sigma }_{2}\mathbf{)}\Phi _{SOD}+i%
\mathbf{L\cdot \sigma }_{1}\mathbf{\times \sigma }_{2}\Phi _{SOX}\}\psi _{+}
\notag \\
& =b^{2}\psi _{+}.  \label{57}
\end{align}%
The detailed \ forms of the separate quasipotentials $\Phi _{i}$ are given
in Appendix A together with their forms for weak potential and in the static
limit. In Appendix B we give the radial forms of Eq. (\ref{57}). \ The
subscripts of most of the quasipotentials are self explanatory. \footnote{%
The subscript on quasipotential $\Phi _{D}$ refers to Darwin. \ It consist
of what are called Darwin terms, those that are the two-body analogue of
terms that accompany the spin-orbit term in the one-body Pauli reduction of
the ordinary one-body Dirac equation, and ones related by canonical
transformations to Darwin interactions \cite{fw,sch73}, momentum dependent
terms arising from retardation effects. The subscripts on the other
quasipotentials refer respectively to $SO$ (spin-orbit), $SOD$ (spin-orbit
difference), $SOX$ (spin-orbit cross terms), $SS$ (spin-spin), $T$ (tensor), 
$SOT$ (spin-orbit-tensor)} \ After the eigenvalue $b^{2}$ of (\ref{57}) is
obtained, the invariant mass of the composite two-body system $w$ can then
be obtained by inverting Eq.\ (\ref{bb}). It is given explicitly by 
\begin{equation}
w=\sqrt{b^{2}+m_{1}^{2}}+\sqrt{b^{2}+m_{2}^{2}}.
\end{equation}%
The structure of the linear and quadratic terms in Eq. (\ref{57}) as well as
the Darwin and spin-orbit terms, are plausible in light of the discussion
given above Eq. (\ref{em}), and in light of the static limit Dirac
structures that come about from the Pauli reduction of the Dirac equation
(see Eqs. (\ref{571}) below for the two-component Pauli reduction of the
Dirac equation). Their appearances as well as that of the remaining spin
structures are direct outcomes of the Pauli reductions of the simultaneous
TBDE Eq. (\ref{tbde}).

\subsection{Spectral Results}

Theory1 (abbreviated by Th1) has two invariant interaction functions: $A(r)$
for the short distance behavior and $S(r)$ for scalar confinement. Theory 2
(abbreviated Th2) has three invariant interaction functions including the
previous two plus $V(r)~$for$~$ timelike vector confinement. In Appendix A.1
we outline how these invariant interaction functions are related to what we
call vertex invariants \ ($\mathcal{J}(r),~\mathcal{G(}r),\mathcal{L}(r)$),
which define the covariant gamma matrix interaction structures which enter
into the hyperbolic form of the TBDE as seen in Eqs. (\ref{hyp1}, \ref{hyp2}%
, \ref{hyp3}-\ref{th2}), and the related energy and mass potentials $E_{1,2}$
and $M_{1,2}$, which characterize the external potential forms of the TBDE
as seen in Eqs. (\ref{extd}-\ref{d3}). The distinction between Th1 and Th2
includes more than the partial cancellation $S^{2}-V^{2}=\mathcal{U}%
^{2}(2\xi -1)$ between the quadratic scalar and timelike vector
interactions. \ It also includes their partial cancellations for spin-orbit
and Darwin terms (see Appendix A.4).\ In this section we present spectral
results with (Th2) and without (Th1) the added timelike invariant function $%
V(r).$

\ We display our results in Tables \ref{1}-\ref{15.5}. The first table lists
the best fit values for the quark masses and the potential parameters $%
\Lambda ,\ \Lambda U_{0},\ \beta $ for Th1 (scalar only confinement) and
Th2. The ratio that optimized the fit for Th2 is $\xi =0.704,$ (see Eq. (\ref%
{usv})). \ In the first two columns of Tables \ref{2}-\ref{9}, we list
quantum numbers and experimental rest mass values (in GeV) and experimental
errors listed parenthetically (in MeV) for 105 known mesons. \ We include
all well known and plausible candidates listed in the standard reference (%
\cite{prtl}). We omit only those mesons with substantial flavor mixing, like
the $\eta $ and $\eta ^{\prime }$ mesons . In the tables, the quantum
numbers listed are those of the $\psi _{+}$ part of the 16-component wave
function. In the third and fourth columns are the theoretical values, with
Th1 referring to the results without the timelike vector interaction and Th2
with the timelike vector interaction. In the fifth and sixth columns we give
the differences between our theoretical results and the experimental and in
the last two columns the contributions of each theoretically computed values
to the total $\chi ^{2}$ of 237 for Th1 and 173 for Th2 \footnote{%
The reader of \cite{crater2} may notice that the total $\chi ^{2}$ of the
model there of 101 (corresponding to Th1 here) was substantially lower than
the 237 that we found here. \ There are several reasons for this difference.
The main reason is that in this paper we do not include a 5\% addition to
the calculational uncertainty based on the total meson widths. \ Also, there
are 19 more mesons in the present model, many of which were difficult to
fit. \ Thirdly, the experimental errors changed. \ Fourth, many of the newer
mesons (for example the $\eta _{b}$) not only added more to the $\chi ^{2}$
from their own fit, but also indirectly to the older ones (e.g. the $%
1^{3}S_{1}$ $\Upsilon $ meson). \ There is no difference in the parameter
sets and potentials in Th1 and those used in \cite{crater2} although the
values are different.}. \ 

To generate the fits, in addition to varying the five quark masses we vary
the parameters $\Lambda ,\ U_{0},$ $\beta ,$ and $\xi $ in the apportioned
static Adler-Piran potential in $A,V,$ and $S$. Those invariants are put
into our relativistic wave equations just as we have inserted the invariant
Coulomb potential $A=-\alpha /r$ (but with $V=S=0$) to obtain the results of
QED bound states \cite{exct,becker}. Note especially that we use a single $%
\Phi (A,S)$ for Th1 and a single $\Phi (A,V,S)$ for Th2 for all quark mass
ratios. Hence in each theory we use a single structure for all the $Q\bar{Q}%
,\ q\bar{Q},$ \textit{and} $q\bar{q}$ mesons in a single overall fit. The
entire confining part of the potential transforms as a world scalar for Th1
and combined timelike and scalar for Th2. Since $\xi >0.5$ in our equations,
this structure leads in both models to linear confinement at long distances
and quadratic confinement at extremely long distances (where the quadratic
contribution $S^{2}$ outweighs the linear term $2m_{w}S$ in Th1 and $%
S^{2}-V^{2}$ outweighing the linear terms $2m_{w}S+2\varepsilon _{w}V$ in
Th2). \ At distances at which $\exp (-\beta \Lambda r)<<1,~$the
corresponding fine and hyperfine structures producing spin-orbit, Thomas,
Darwin, spin-spin, and tensor terms (the last two are relatively small in
that domain) are dominated by the confining interaction, while at short
distances ($\exp (-\beta \Lambda r)\sim 1)$ the electromagnetic-like portion
of the interaction gives the dominant contribution to the fine and hyperfine
structures. Furthermore because the signs of each of the spin-orbit and
Darwin terms in the Pauli-form of our TBDE are opposite for the scalar and
vector interactions (see Appendix A.4), the spin-orbit contributions of
those parts of the interaction produce opposite effects with degrees of
cancellation depending on the size of the quarkonia atom. \ Another point to
make is that because of the various sizes of the quarkonia atoms and the
c.m. energy dependence the behavior of $\Phi _{w}$ is sharply different for
the light mesons compared with the heavy ones. \ This may possibly account
for the ability of our formalism to obtain good fits for the light meson
hyperfine splittings while at the same time giving good overall fits to the
heavy mesons.

We obtain the meson masses given in columns three and four as the result of
a least squares fit using the known experimental errors from the Particle
Date Group (PDG) tables \cite{prtl} and an assumed calculational error of
1.0 MeV. We employ the calculational error not to represent the uncertainty
of our algorithm but more to prevent the mesons that are stable with respect
to the strong interaction from being weighted too heavily. Our $\chi ^{2}$
is per datum (105) minus parameters (8 or 9). In Table \ref{1}, the value of 
$\beta $ for Th1 implies that (in the best fit) as the quark separation
increases, our apportioned Adler-Piran potential switches from primarily
vector to scalar at about ($\beta \Lambda )^{-1}\sim $0.60 fermi. This shift
is a relativistic effect since the effective nonrelativistic limit of the
potential ($\mathcal{A}+S$) exhibits no such shift (i.e., by construction $%
\beta $ drops out). \ For Th2, this distance is substantially less, $(\beta
\Lambda )^{-1}\sim $0.22 fermi.

\ \ \ \ \ \ \ \ \ \ \ \ \ \ \ \ \ \ \ \ \ \ \ \ \ \ \ \ \ \ \ \ \ \ \ \ \ \
\ \ \ \ \ \ \ \ \ \ \ \ \ \ \ \ \ \ \ \ \ \ \ \ \ \ \ \ \ \ \ \ \ \ \ \ \ \
\ \ \ \ \ \ \ \ \ \ \ \ \ \ \ \ \ 
\begin{table}[tbp] \centering%
$%
\begin{tabular}{|llll|}
\hline
Parameter & Th1 &  & Th2 \\ \hline
$m_{b}$ & $4.917~\text{GeV}$ &  & $4.953~\text{GeV}$ \\ 
$m_{c}$ & $1.546~\text{GeV}$ &  & $1.585~\text{GeV}$ \\ 
$m_{s}$ & $0.2874~\text{GeV}$ &  & $0.3079~\text{GeV}$ \\ 
$m_{u}$ & $0.0713~\text{GeV}$ &  & $0.0985~\text{GeV}$ \\ 
$m_{d}$ & $0.0771~\text{GeV}$ &  & $0.1045~\text{GeV}$ \\ 
$\Lambda $ & $0.2213~\text{GeV}$ &  & $0.2255~\text{GeV}$ \\ 
$\Lambda U_{0}$ & $1.815$ GeV &  & $1.770$ GeV \\ 
$\beta $ & $1.502$ &  & $4.408$ \\ 
$\xi $ & $1$ &  & $0.704$ \\ \hline
\end{tabular}%
$%
\caption{Parameters for Theory 1 and 2.  The 4 potential parameters $\Lambda$, $\Lambda U_0$,
$1/(\beta\Lambda)$, and $\xi$ are respectively the QCD scale factor, the Adler-Piran integration
constant, the vector-scalar transition distance, and the $S/(S+V)$ ratio.}%
\label{1}%
\end{table}%
\ 

\ \ \ \ \ \ \ \ \ \ \ \ \ \ \ \ \ \ \ \ \ \ \ \ \ \ \ \ \ \ \ \ \ \ \ \ \ \
\ \ \ \ \ \ \ \ \ \ \ \ \ \ \ \ \ \ \ \ \ \ \ \ \ \ \ \ \ \ \ \ \ \ \ \ \ \
\ \ \ \ \ \ \ \ \ \ \ \ \ \ \ 

$\bigskip $%
\begin{table}[tbp] \centering%
\ 
\begin{tabular}{|lcccccccl|}
\hline
$u\bar{d}$ mesons~~~~~~~~~~~ & Exp. & Th1. & Th2. & Exp.-Th1. & Exp.-Th2. & $%
\chi ^{2}$-Th1. & $\chi ^{2}$-Th2. &  \\ \hline
$\pi :u\overline{d}\ 1\,{}^{1}S_{0}$ & 0.140(0.0) & 0.141 & 0.134 & -0.002 & 
0.006 & 0.0 & 0.3 &  \\ 
$\rho :u\overline{d}\ 1\,{}^{3}S_{1}$ & 0.775(0.4) & 0.790 & 0.781 & -0.015
& -0.005 & 1.9 & 0.2 &  \\ 
$b_{1}:u\overline{d}\ 1\,{}^{1}P_{1}$ & 1.230(3.2) & 1.283 & 1.243 & -0.053
& -0.014 & 2.5 & 0.2 &  \\ 
$a_{1}:u\overline{d}\ 1\,{}^{3}P_{1}$ & 1.230(40.) & 1.425 & 1.320 & -0.195
& -0.090 & 0.2 & 0.1 &  \\ 
$\pi :u\overline{d}\ 2\,{}^{1}S_{0}$ & 1.300(100) & 1.493 & 1.435 & -0.193 & 
-0.135 & 0.0 & 0.0 &  \\ 
$a_{2}:u\overline{d}\ 1\,{}^{3}P_{2}$ & 1.318(0.6) & 1.276 & 1.310 & 0.042 & 
0.008 & 12.9 & 0.5 &  \\ 
$\rho :u\overline{d}\ 2\,{}^{3}S_{1}$ & 1.465(25.) & 1.745 & 1.684 & -0.280
& -0.219 & 1.3 & 0.8 &  \\ 
$a_{0}:u\overline{d}\ 1\,{}^{3}P_{0}$ & 1.474(19.) & 1.165 & 1.024 & 0.309 & 
0.450 & 2.6 & 5.6 &  \\ 
$b_{2}:u\overline{d}\ 1\,{}^{1}D_{2}$ & 1.672(3.2) & 1.815 & 1.763 & -0.143
& -0.090 & 18.2 & 7.2 &  \\ 
$a_{3}:u\overline{d}\ 1\,{}^{3}D_{3}$ & 1.689(2.1) & 1.663 & 1.718 & 0.026 & 
-0.029 & 1.2 & 1.5 &  \\ 
$a_{1}:u\overline{d}\ 1\,{}^{3}D_{1}$ & 1.720(20.) & 1.944 & 1.847 & -0.224
& -0.127 & 1.2 & 0.4 &  \\ 
$a_{2}:u\overline{d}\ 2\,{}^{3}P_{2}$ & 1.732(16.) & 2.025 & 2.009 & -0.293
& -0.277 & 3.3 & 3.0 &  \\ 
$\pi :u\overline{d}\ 3\,{}^{1}S_{0}$ & 1.816(14.) & 2.090 & 2.037 & -0.274 & 
-0.221 & 3.8 & 2.5 &  \\ 
$b_{2}:u\overline{d}\ 2\,{}^{1}D_{2}$ & 1.895(16.) & 2.300 & 2.267 & -0.405
& -0.372 & 6.4 & 5.4 &  \\ 
$a_{4}:u\overline{d}\ 1\,{}^{3}F_{4}$ & 2.011(12.) & 1.984 & 2.057 & 0.027 & 
-0.046 & 0.1 & 0.1 &  \\ 
$b_{2}:u\overline{d}\ 3\,{}^{1}D_{2}$ & 2.090(29.) & 2.704 & 2.700 & -0.614
& -0.610 & 4.5 & 4.4 &  \\ 
$\rho :u\overline{d}\ 3\,{}^{3}S_{1}$ & 2.149(17.) & 2.281 & 2.326 & -0.132
& -0.177 & 0.6 & 1.1 &  \\ 
$a_{3}:u\overline{d}\ 2\,{}^{3}D_{3}$ & 2.250(45.) & 2.275 & 2.290 & -0.025
& -0.040 & 0.0 & 0.0 &  \\ 
$a_{5}:u\overline{d}\ 1\,{}^{3}G_{5}$ & 2.330(35.0) & 2.258 & 2.349 & 0.072
& -0.019 & 0.0 & 0.0 &  \\ 
$a_{6}:u\overline{d}\ 1\,{}^{3}H_{6}$ & 2.450(130) & 2.500 & 2.609 & -0.050
& -0.159 & 0.0 & 0.0 &  \\ \hline
\end{tabular}%
\caption{ $u\bar d$ Mesons, Theory 1 and 2 - In this table and the ones below, the
meson masses are in units of GeV, with experimental errors given parenthetically in units
of MeV.}\label{2}%
\end{table}%

\begin{table}[tbp] \centering%
\begin{tabular}{|lcccccccl|}
\hline
$s\bar{u},~s\bar{d}~$Mesons~~~~~~~~~~~~ & Exp. & Th1. & Th2. & Exp.-Th1. & 
Exp.-Th2. & $\chi ^{2}$-Th1. & $\chi ^{2}$-Th2. &  \\ \hline
$K\,{}^{-}:s\overline{u}\ 1\,{}^{1}S_{0}$ & 0.494(0.0) & 0.485 & 0.519 & 
0.008 & -0.025 & 0.7 & 6.4 &  \\ 
$K\,{}^{0}:s\overline{d}\ 1\,{}^{1}S_{0}$ & 0.498(0.0) & 0.488 & 0.520 & 
0.010 & -0.022 & 1.0 & 5.0 &  \\ 
$K^{\ast }\,{}^{-}:s\overline{u}\ 1\,{}^{3}S_{1}$ & 0.892(0.3) & 0.917 & 
0.896 & -0.025 & -0.004 & 6.0 & 0.2 &  \\ 
$K^{\ast }\,{}^{-}:s\overline{d}\ 1\,{}^{3}S_{1}$ & 0.896(0.3) & 0.919 & 
0.897 & -0.023 & -0.001 & 4.8 & 0.0 &  \\ 
$K\,{}^{-}:s\overline{u}\ 1\,{}^{1}P_{1}$ & 1.272(7.0) & 1.356 & 1.339 & 
-0.084 & -0.067 & 1.4 & 0.9 &  \\ 
$K^{\ast }\,{}^{-}:s\overline{u}\ 1\,{}^{3}P_{1}$ & 1.403(7.0) & 1.419 & 
1.359 & -0.016 & 0.044 & 0.1 & 0.4 &  \\ 
$K^{\ast }\,{}^{-}:s\overline{u}\ 2\,{}^{3}S_{1}$ & 1.414(15.) & 1.759 & 
1.706 & -0.345 & -0.292 & 5.3 & 3.8 &  \\ 
$K^{\ast }\,{}^{-}:s\overline{u}\ 1\,{}^{3}P_{0}$ & 1.425(50.) & 1.132 & 
1.079 & 0.293 & 0.346 & 0.3 & 0.5 &  \\ 
$K^{\ast }\,{}^{-}:s\overline{u}\ 1\,{}^{3}P_{2}$ & 1.426(50.) & 1.379 & 
1.404 & 0.047 & 0.022 & 6.8 & 1.4 &  \\ 
$K^{\ast }\,{}^{-}:s\overline{d}\ 1\,{}^{3}P_{2}$ & 1.432(1.3) & 1.380 & 
1.405 & 0.052 & 0.027 & 10.2 & 2.8 &  \\ 
$K\,{}^{-}:s\overline{u}\ 2\,{}^{1}S_{0}$ & 1.460(40.) & 1.523 & 1.476 & 
-0.063 & -0.016 & 0.0 & 0.0 &  \\ 
$K^{\ast }\,{}^{-}:s\overline{u}\ 1\,{}^{3}D_{1}$ & 1.717(27.) & 1.922 & 
1.837 & -0.205 & -0.120 & 0.6 & 0.2 &  \\ 
$K\,{}^{-}:s\overline{u}\ 1\,{}^{1}D_{2}$ & 1.773(8.0) & 1.835 & 1.803 & 
-0.062 & -0.030 & 0.6 & 0.1 &  \\ 
$K^{\ast }\,{}^{-}:s\overline{u}\ 1\,{}^{3}D_{3}$ & 1.776(7.0) & 1.740 & 
1.792 & 0.036 & -0.016 & 0.3 & 0.0 &  \\ 
$K^{\ast }\,{}^{-}:s\overline{u}\ 1\,{}^{3}D_{2}$ & 1.816(13.) & 1.824 & 
1.795 & -0.008 & 0.021 & 0.0 & 0.0 &  \\ 
$K\,{}^{-}:s\overline{u}\ 3\,{}^{1}S_{0}$ & 1.830(13.) & 2.115 & 2.081 & 
-0.285 & -0.251 & 0.5 & 0.4 &  \\ 
$K^{\ast }\,{}^{-}:s\overline{u}\ 2\,{}^{3}P_{2}$ & 1.973(33.) & 2.078 & 
2.060 & -0.105 & -0.087 & 0.1 & 0.1 &  \\ 
$K^{\ast }\,{}^{-}:s\overline{u}\ 1\,{}^{3}F_{4}$ & 2.045(9.0) & 2.045 & 
2.117 & 0.000 & -0.072 & 0.0 & 0.6 &  \\ 
$K^{\ast }\,{}^{-}:s\overline{u}\ 2\,{}^{3}D_{2}$ & 2.247(17.) & 2.326 & 
2.313 & -0.079 & -0.066 & 0.2 & 0.1 &  \\ 
$K^{\ast }\,{}^{-}:s\overline{u}\ 2\,{}^{3}F_{3}$ & 2.324(24.) & 2.642 & 
2.600 & -0.318 & -0.276 & 1.7 & 1.3 &  \\ 
$K^{\ast }\,{}^{-}:s\overline{u}\ 1\,{}^{3}G_{5}$ & 2.382(14.) & 2.309 & 
2.401 & 0.073 & -0.019 & 0.3 & 0.0 &  \\ 
$K^{\ast }\,{}^{-}:s\overline{u}\ 2\,{}^{3}F_{4}$ & 2.490(20.) & 2.555 & 
2.600 & -0.065 & -0.110 & 0.1 & 0.3 &  \\ \hline
\end{tabular}%
\caption{ $s\bar u$ and $s\bar d$ Mesons, Theory 1 and 2}\label{3}%
\end{table}%

\begin{table}[tbp] \centering%
\ 
\begin{tabular}{|lcccccccl|}
\hline
$s\bar{s}$ Mesons~~~~~~~~~~~~ & Exp. & Th1. & Th2. & Exp.-Th1. & Exp.-Th2. & 
$\chi ^{2}$-Th1. & $\chi ^{2}$-Th2. &  \\ \hline
$\phi :s\overline{s}\ 1\,{}^{3}S_{1}$ & 1.019(0.0) & 1.050 & 1.013 & -0.031
& 0.006 & 9.3 & 0.4 &  \\ 
$\phi :s\overline{s}\ 1\,{}^{3}P_{0}$ & 1.370(100) & 1.211 & 1.175 & 0.159 & 
0.195 & 0.0 & 0.0 &  \\ 
$\phi :s\overline{s}\ 1\,{}^{3}P_{1}$ & 1.518(5.0) & 1.480 & 1.437 & 0.038 & 
0.081 & 0.5 & 2.5 &  \\ 
$\phi :s\overline{s}\ 1\,{}^{3}P_{2}$ & 1.525(5.0) & 1.496 & 1.506 & 0.029 & 
0.019 & 0.3 & 0.1 &  \\ 
$\phi :s\overline{s}\ 2\,{}^{3}S_{1}$ & 1.680(20.) & 1.811 & 1.875 & -0.131
& -0.195 & 0.4 & 0.9 &  \\ 
$\phi :s\overline{s}\ 1\,{}^{3}D_{3}$ & 1.854(7.0) & 1.839 & 1.879 & 0.015 & 
-0.025 & 0.0 & 0.1 &  \\ 
$\phi :s\overline{s}\ 2\,{}^{3}P_{2}$ & 2.011(70) & 2.149 & 2.128 & -0.138 & 
-0.117 & 0.0 & 0.0 &  \\ 
$\phi :s\overline{s}\ 3\,{}^{3}P_{2}$ & 2.297(28.) & 2.612 & 2.603 & -0.315
& -0.306 & 1.3 & 1.2 &  \\ \hline
\end{tabular}%
\caption{$s\bar s$  Mesons, Theory 1 and 2}\label{4}%
\end{table}%

\begin{table}[tbp] \centering%
\begin{tabular}{|lcccccccl|}
\hline
$c\bar{u},~c\bar{d}~$Mesons~~~~~~~~~~~ & Exp. & Th1. & Th2. & Exp.-Th1. & 
Exp.-Th2. & $\chi ^{2}$-Th1. & $\chi ^{2}$-Th2. &  \\ \hline
$D^{0}:c\overline{u}\ 1\,{}^{1}S_{0}$ & 1.865(0.2) & 1.865 & 1.876 & 0.000 & 
-0.011 & 0.0 & 1.1 &  \\ 
$D^{+}:c\overline{d}\ 1\,{}^{1}S_{0}$ & 1.870(0.2) & 1.872 & 1.883 & -0.003
& -0.013 & 0.1 & 1.7 &  \\ 
$D^{\ast 0}:c\overline{u}\ 1\,{}^{3}S_{1}$ & 2.007(0.2) & 2.013 & 2.007 & 
-0.006 & 0.000 & 0.4 & 0.0 &  \\ 
$D^{\ast +}:c\overline{d}\ 1\,{}^{3}S_{1}$ & 2.010(0.2) & 2.019 & 2.013 & 
-0.008 & -0.002 & 0.7 & 0.1 &  \\ 
$D^{\ast 0}:c\overline{u}\ 1\,{}^{3}P_{0}$ & 2.352(50.) & 2.224 & 2.221 & 
0.128 & 0.131 & 0.1 & 0.1 &  \\ 
$D^{\ast +}:c\overline{d}\ 1\,{}^{3}P_{0}$ & 2.403(14.) & 2.232 & 2.230 & 
0.171 & 0.173 & 1.5 & 1.5 &  \\ 
$D^{+}:c\overline{d}\ 1\,{}^{3}P_{2}$ & 2.460(3.0) & 2.398 & 2.414 & 0.062 & 
0.046 & 3.9 & 2.1 &  \\ 
$D^{\ast 0}:c\overline{u}\ 1\,{}^{3}P_{2}$ & 2.461(1.6) & 2.393 & 2.409 & 
0.069 & 0.052 & 13.2 & 7.7 &  \\ \hline
\end{tabular}%
\ \caption{$c\bar u$ and $c\bar d$ Mesons, Theory 1 and 2}\label{5}%
\end{table}%

\begin{table}[tbp] \centering%
\begin{tabular}{|lcccccccl|}
\hline
$c\bar{s}$ Mesons~~~~~~~~~~~~ & Exp. & Th1. & Th2. & Exp.-Th1. & Exp.-Th2. & 
$\chi ^{2}$-Th1. & $\chi ^{2}$-Th2. &  \\ \hline
$D_{s}:c\overline{s}\ 1\,{}^{1}S_{0}$ & 1.968(0.3) & 1.972 & 1.974 & -0.004
& -0.006 & 0.1 & 0.3 &  \\ 
$D_{s}^{\ast }:c\overline{s}\ 1\,{}^{3}S_{1}$ & 2.112(0.5) & 2.138 & 2.119 & 
-0.026 & -0.007 & 5.4 & 0.4 &  \\ 
$D_{s}^{\ast }:c\overline{s}\ 1\,{}^{3}P_{0}$ & 2.318(0.6) & 2.348 & 2.340 & 
-0.031 & -0.022 & 6.9 & 3.5 &  \\ 
$D_{s}:c\overline{s}\ 1\,{}^{1}P_{1}$ & 2.535(0.3) & 2.505 & 2.499 & 0.030 & 
0.036 & 8.3 & 11.6 &  \\ 
$D_{s}^{\ast }:c\overline{s}\ 1\,{}^{3}P_{2}$ & 2.573(0.9) & 2.534 & 2.532 & 
0.039 & 0.040 & 8.4 & 8.9 &  \\ 
$D_{s}^{\ast }:c\overline{s}\ 2\,{}^{3}S_{1}$ & 2.690(7.0) & 2.714 & 2.702 & 
-0.024 & -0.012 & 0.1 & 0.0 &  \\ \hline
\end{tabular}%
\ \caption{$c\bar s$  Mesons, Theory 1 and 2}\label{6}%
\end{table}%

\begin{table}[tbp] \centering%
\ 
\begin{tabular}{|lcccccccl|}
\hline
$c\bar{c}$ Mesons~~~~~~~~~~~~ & Exp. & Th1. & Th2. & Exp.-Th1. & Exp.-Th2. & 
$\chi ^{2}$-Th1. & $\chi ^{2}$-Th2. &  \\ \hline
$\eta _{c}:c\overline{c}\ 1\,{}^{1}S_{0}$ & 2.980(1.2) & 2.965 & 2.973 & 
0.015 & 0.007 & 1.0 & 0.2 &  \\ 
$J/\psi (1S):c\overline{c}\ 1\,{}^{3}S_{1}$ & 3.097(0.0) & 3.131 & 3.128 & 
-0.034 & -0.031 & 11.4 & 9.7 &  \\ 
$\chi _{0}:c\overline{c}\ 1\,{}^{3}P_{0}$ & 3.415(0.3) & 3.395 & 3.397 & 
0.020 & 0.018 & 3.7 & 3.0 &  \\ 
$\chi _{1}:c\overline{c}\ 1\,{}^{3}P_{1}$ & 3.511(0.1) & 3.506 & 3.505 & 
0.005 & 0.006 & 0.2 & 0.4 &  \\ 
$h_{1}:c\overline{c}\ 1\,{}^{1}P_{1}$ & 3.526(0.3) & 3.522 & 3.523 & 0.004 & 
0.003 & 0.2 & 0.1 &  \\ 
$\chi _{2}:c\overline{c}\ 1\,{}^{3}P_{2}$ & 3.556(0.1)) & 3.562 & 3.557 & 
-0.006 & -0.001 & 0.4 & 0.0 &  \\ 
$\eta _{c}:c\overline{c}\ 2\,{}^{1}S_{0}$ & 3.637(4.0) & 3.606 & 3.602 & 
0.031 & 0.035 & 0.6 & 0.7 &  \\ 
$\psi (2S):c\overline{c}\ 2\,{}^{3}S_{1}$ & 3.686(0.0) & 3.688 & 3.689 & 
-0.002 & -0.002 & 0.0 & 0.1 &  \\ 
$\psi (1D):c\overline{c}\ 1\,{}^{3}D_{1}$ & 3.773(0.4) & 3.807 & 3.807 & 
-0.034 & -0.034 & 0.9 & 0.9 &  \\ 
$\chi _{2}:c\overline{c}\ 2\,{}^{3}P_{2}$ & 3.929(5.0) & 3.980 & 3.983 & 
-0.051 & -0.054 & 1.0 & 1.1 &  \\ 
$\psi (3S):c\overline{c}\ 3\,{}^{3}S_{1}$ & 4.039(10.) & 4.086 & 4.092 & 
-0.047 & -0.053 & 0.2 & 0.3 &  \\ 
$\psi (2D):c\overline{c}\ 2\,{}^{3}D_{1}$ & 4.153(3.0) & 4.164 & 4.169 & 
-0.011 & -0.016 & 0.1 & 0.3 &  \\ 
$\psi (4S):c\overline{c}\ 4\,{}^{3}S_{1}$ & 4.421(4.0) & 4.410 & 4.426 & 
0.011 & -0.005 & 0.1 & 0.0 &  \\ 
$\psi (3D):c\overline{c}\ 3\,{}^{3}D_{1}$ & 4.421(4.0) & 4.467 & 4.483 & 
-0.046 & -0.062 & 1.2 & 2.3 &  \\ 
$\psi (5S):c\overline{c}\ 5\,{}^{3}S_{1}$ & 4.800(100) & 4.690 & 4.719 & 
0.110 & 0.081 & 0.0 & 0.0 &  \\ 
$\psi (4D):c\overline{c}\ 4\,{}^{3}D_{1}$ & 4.880(100) & 4.735 & 4.764 & 
0.145 & 0.116 & 0.0 & 0.0 &  \\ 
$\psi (6S):c\overline{c}\ 6\,{}^{3}S_{1}$ & 5.180(100) & 4.940 & 4.983 & 
0.203 & 0.197 & 0.0 & 0.0 &  \\ 
$\psi (5D):c\overline{c}\ 5\,{}^{3}D_{1}$ & 5.290(100) & 4.977 & 5.020 & 
0.350 & 0.270 & 0.1 & 0.1 &  \\ \hline
\end{tabular}%
\caption{$c\bar c$  Mesons, Theory 1 and 2}\label{7}%
\end{table}%

\begin{table}[tbp] \centering%
\begin{tabular}{|lcccccccl|}
\hline
$b\bar{u},~b\bar{d}$ $b\bar{s}~$Mesons~~~~~~~~~~~~ & Exp. & Th1. & Th2. & 
Exp.-Th1. & Exp.-Th2. & $\chi ^{2}$-Th1. & $\chi ^{2}$-Th2. &  \\ \hline
$B^{-}:b\overline{u}\ 1\,{}^{1}S_{0}$ & 5.279(0.3) & 5.281 & 5.283 & -0.002
& -0.004 & 0.0 & 0.2 &  \\ 
$B^{0}:b\overline{d}\ 1\,{}^{1}S_{0}$ & 5.280(0.3) & 5.282 & 5.284 & -0.003
& -0.005 & 0.1 & 0.2 &  \\ 
$B^{\ast -}:b\overline{u}\ 1\,{}^{3}S_{1}$ & 5.325(0.5) & 5.335 & 5.333 & 
-0.010 & -0.008 & 0.8 & 0.5 &  \\ 
$B^{\ast -}:b\overline{u}\ 1\,{}^{3}P_{2}$ & 5.747(2.9) & 5.671 & 5.687 & 
0.076 & 0.059 & 6.2 & 3.8 &  \\ 
$B_{s}^{0}:b\overline{s}\ 1\,{}^{1}S_{0}$ & 5.366(0.6) & 5.373 & 5.367 & 
-0.007 & -0.001 & 0.3 & 0.0 &  \\ 
$B_{s}^{\ast 0}:b\overline{s}\ 1\,{}^{3}S_{1}$ & 5.413(1.3) & 5.441 & 5.430
& -0.029 & -0.017 & 3.0 & 1.0 &  \\ 
$B_{s}^{\ast 0}:b\overline{s}\ 1\,{}^{3}P_{1}$ & 5.829(0.7) & 5.789 & 5.792
& 0.040 & 0.037 & 10.9 & 9.4 &  \\ 
$B_{s}^{\ast 0}:b\overline{s}\ 1\,{}^{3}P_{2}$ & 5.840(0.6) & 5.805 & 5.805
& 0.035 & 0.035 & 8.9 & 9.0 &  \\ 
$B_{c}^{-}:b\overline{c}\ 1\,{}^{1}S_{0}$ & 6.276(21.) & 6.249 & 6.251 & 
0.027 & 0.025 & 0.4 & 0.4 &  \\ \hline
\end{tabular}%
\ \caption{$b\bar u$ , $b\bar d$ and $b\bar s$ Mesons, Theory 1 and 2}\label%
{8}%
\end{table}%

\begin{table}[tbp] \centering%
\ 
\begin{tabular}{|lcccccccl|}
\hline
$b\bar{b}$ Mesons~~~~~~~~~~~~ & Exp. & Th1. & Th2. & Exp.-Th1. & Exp.-Th2. & 
$\chi ^{2}$-Th1. & $\chi ^{2}$-Th2. &  \\ \hline
$\eta _{b}:b\overline{b}\ 1\,{}^{1}S_{0}$ & 9.389(4.0) & 9.337 & 9.330 & 
0.052 & 0.059 & 1.6 & 2.0 &  \\ 
$\Upsilon (1S):b\overline{b}\ 1\,{}^{3}S_{1}$ & 9.460(0.3) & 9.444 & 9.444 & 
0.016 & 0.017 & 2.4 & 2.6 &  \\ 
$\chi _{b0}:b\overline{b}\ 1\,{}^{3}P_{0}$ & 9.859(0.4) & 9.836 & 9.834 & 
0.023 & 0.026 & 4.6 & 5.6 &  \\ 
$\chi _{b1}:b\overline{b}\ 1\,{}^{3}P_{1}$ & 9.893(0.3) & 9.886 & 9.886 & 
0.007 & 0.007 & 0.4 & 0.4 &  \\ 
$\chi _{b2}:b\overline{b}\ 1\,{}^{3}P_{2}$ & 9.912(0.3) & 9.922 & 9.920 & 
-0.010 & -0.008 & 0.9 & 0.6 &  \\ 
$\Upsilon (2S):b\overline{b}\ 2\,{}^{3}S_{1}$ & 10.023(0.3) & 10.022 & 10.022
& 0.001 & 0.002 & 0.0 & 0.0 &  \\ 
$\Upsilon (D):b\overline{b}\ 2\,{}^{3}D_{2}$ & 10.161(0.6) & 10.178 & 10.179
& -0.017 & -0.018 & 2.1 & 2.3 &  \\ 
$\chi _{b0}:b\overline{b}\ 2\,{}^{3}P_{0}$ & 10.232(0.4) & 10.230 & 10.229 & 
0.002 & 0.003 & 0.1 & 0.1 &  \\ 
$\chi _{b1}:b\overline{b}\ 2\,{}^{3}P_{1}$ & 10.255(0.5) & 10.261 & 10.262 & 
-0.006 & -0.007 & 0.3 & 0.4 &  \\ 
$\chi _{b2}:b\overline{b}\ 2\,{}^{3}P_{2}$ & 10.269(0.4) & 10.284 & 10.286 & 
-0.015 & -0.017 & 1.9 & 2.5 &  \\ 
$\Upsilon (3S):b\overline{b}\ 3\,{}^{3}S_{1}$ & 10.355(0.6) & 10.366 & 10.368
& -0.011 & -0.013 & 0.8 & 1.2 &  \\ 
$\Upsilon (4S):b\overline{b}\ 4\,{}^{3}S_{1}$ & 10.579(1.2) & 10.626 & 10.633
& -0.046 & -0.053 & 8.8 & 11.7 &  \\ 
$\Upsilon (5S):b\overline{b}\ 5\,{}^{3}S_{1}$ & 10.865(8.0) & 10.844 & 10.857
& 0.021 & 0.008 & 0.1 & 0.0 &  \\ 
$\Upsilon (6S):b\overline{b}\ 6\,{}^{3}S_{1}$ & 11.019(8.0) & 11.036 & 11.055
& -0.017 & -0.036 & 0.0 & 0.2 &  \\ \hline
\end{tabular}%
\caption{$b\bar b$  Mesons, Theory 1 and 2}\label{9}%
\end{table}%

\begin{table}[tbp] \centering%
\begin{tabular}{|lccccc|}
\hline
& \multicolumn{2}{c}{$\chi ^{2}$} &  & \multicolumn{2}{c|}{\textbf{Average }$%
\chi ^{2}$} \\ \cline{2-3}\cline{5-6}
\textbf{Meson Family} & \textbf{Th1} & \textbf{Th2} & \textbf{\# Mesons} & 
\textbf{Th1} & \textbf{Th2} \\ \hline
$u\overline{d}$ & 62.1 & 30.5 & 20 & 3.1 & 1.5 \\ 
$s\overline{u}~,~s\overline{d}$ & 37.7 & 26.5 & 22 & 1.8 & 1.3 \\ 
$s\overline{s}$ & 11.0 & 4.5 & 8 & 1.4 & 0.6 \\ 
$c\overline{u}~,~c\overline{d}$ & 20.9 & 14.6 & 8 & 2.6 & 1.8 \\ 
$c\overline{s}$ & 29.4 & 26.6 & 6 & 4.9 & 4.4 \\ 
$c\overline{c}$ & 28.2 & 23.4 & 18 & 1.6 & 1.3 \\ 
$b\overline{u}~,~b\overline{s}~,~b\overline{c}$ & 31.2 & 25.9 & 9 & 3.5 & 2.9
\\ 
$b\overline{b}$ & 24.8 & 25.1 & 14 & 1.8 & 1.8 \\ 
Total & 245.3 & 177.1 & 105 & 2.4 & 1.7 \\ \hline
\end{tabular}%
\caption{$\chi^2$ by Family for Th1 and Th2}\label{10}%
\end{table}%

Table \ref{10} lists the 8 meson families, their respective $\chi ^{2}$
contributions and their averages. \ The most striking feature is that as the
quark masses increase from the lightest to the heaviest, the differences of
the respective $\chi ^{2}$ shifts from about a factor of 2 \ to almost even.
\ The heaviest mesons are also the smallest in mean radius. \ This means
that they are less likely to experience the effects of the $S^{2}$ and $%
-V^{2}$ portions of the confining interactions. \ As the mesons become
large, they experience more of the effects of these parts of the potentials.
\ The most dramatic improvement from the inclusion of the timelike vector
confining potential $V(r)$ (Th2) is with the light quark $u\bar{d}$ family.
Referring now to Tables \ref{2}-\ref{9} \ most of the improvement comes from
that of the fits to the $a_{2},b_{2}$ and $b_{1}$ mesons.\ For the $s\bar{u}%
,s\bar{d}$ family the largest improvement comes from the lowest lying $%
^{3}P_{2}$ mesons. The ground state singlet-triplet splitting changes from
an overestimation to an underestimation. \ For the $s\bar{s}$ family the
most significant improvement is in the ground state, although somewhat
off-balanced by a worse fit for the lowest lying $^{3}P_{1}$ state. \ In the
case of the $c\bar{u},c\bar{d}$ family the main improvement is from the $%
^{3}P_{2}$ mesons. \ For the $c\bar{s}$ mesons there is a slight overall
improvement for Th2 with offsetting changes for the $^{3}P_{0}$ and $%
^{1}P_{1}$ mesons. \ There is only a very slight improvement for the $c\bar{c%
}$ mesons. \ For both Th1 and Th2 the worst fit is to the $J/\psi $ meson
with a mass too large by about 30 MeV. \ It cannot be adjusted downward by
lowering the charm mass due to the fact that other mesons in this family
would be pushed further from the data. \ With the heavy-light family, a
single $b$ quark, there again is not much overall change and even less in
the $b\bar{b}$ family although another significant improvement is in the $%
^{3}P_{2}$\ $b\bar{u}$ state. \ An oddity with the $b\bar{b}$ is the sudden
increase in $\chi ^{2}$ at the $4^{3}S_{1}$ meson, the worst fit of all the
mesons in terms of the incremental $\chi ^{2}$. \ Since that meson is
closest to threshold, its mass will be most affected by it, whereas our
theoretical model does not take threshold effects into account.

A possible explanation of why most improvements come for the $^{3}P_{2}$
states is that the effect on the spin-orbit coupling due to the Thomas terms
is opposite in sign for the timelike vector and scalar mesons. Without the
balancing effect of the timelike vector confining interaction, the scalar
interaction enforces an inversion of the spin-orbit splittings of the light
mesons that are far too distorted for the $u\bar{d}$ and $s\bar{u}$
multiplets\footnote{%
Another possible source of the strange multiplet inversion, is that the
observed $^{3}P_{0}$ states of 1450 for the $u\bar{d}$ and 1430 for the $u%
\bar{s}$ systems are, in fact not zero node states, but rather one node
excited states. \ This may, in our formalism, give room to an interpretation
of the $u\bar{d}(980)$ and the $\kappa (700-900)$ mesons as possible
candidates for the zero node states. \ Our tables support that more than the
identification of \ a zero node $1474$ for the $u\bar{d}$ and $1425$ for the 
$u\bar{s}$ systems. \ However, using the parameters on our model, we would
obtain the $2^{3}P_{0}$ values of $1800$ and $1850$ for those one node
states, well above the $1474$ and $1425$ experimental values. \ So this does
not appear to be a plausible alternative for either Th1 or Th2.}. \ Also,
the long range scalar parts contribute oppositely in sign from the short
range vector part attributable to the $A(r)$ potential.

\begin{table}[tbp] \centering%
\begin{tabular}{|ccccccc|}
\hline
Family & \ \ \ \ \ \ \ \ \  & Exp. & \ \ \ \ \ \ \ \ \  & Th1. & \ \ \ \ \ \
\ \ \  & Th2. \\ \hline
$u\overline{d}$ &  & 635 &  & 649 &  & 647 \\ 
$s\overline{u}$ &  & 398 &  & 432 &  & 377 \\ 
$s\overline{d}$ &  & 398 &  & 431 &  & 377 \\ 
$c\overline{u}$ &  & 142 &  & 148 &  & 131 \\ 
$c\overline{d}$ &  & 140 &  & 147 &  & 130 \\ 
$c\overline{s}$ &  & 144 &  & 166 &  & 145 \\ 
$c\overline{c}$ &  & 117 &  & 166 &  & 155 \\ 
$b\overline{u}$ &  & 46 &  & 54 &  & 50 \\ 
$s\overline{s}$ &  & 47 &  & 68 &  & 63 \\ 
$b\overline{b}$ &  & 71 &  & 107 &  & 114 \\ \hline
\end{tabular}%
\caption{Ground State Singlet-Triplet Splittings (MeV)}\bigskip \label{11}%
\end{table}%

We now examine another important feature of our method: the goodness with
which our equations account for spin-dependent effects (both fine- and
hyperfine- splittings). Table \ref{11} shows the best fit versus
experimental ground state singlet-triplet splittings and six versus four of
the ten hyperfine splittings are improved using Th2 over Th1. \ Both give
good fits for all hyperfine ground state splittings except for the $\eta
_{c}-\psi $ system and $\eta _{b}-\Upsilon $ system which for the latter
over estimate the splittings by about 50\%!\footnote{%
We point out, however, that our model does much better in simultaneously
working with heavy and light $q\bar{q}$ hyperfine splittings than that
obtained by typical constituent (non-chiral) quark models (a major exception
discussed here in detail is the quasipotential model of \cite{rusger})} One
problem with the fit for the $c\bar{c}$ system of mesons may be due to the
fact that the $D^{\ast }\ ^{3}P_{2}$, $^{1}P_{1}$ and $D_{s}^{\ast }\
^{3}P_{2}$ fits are significantly low while the $J/\psi $ fit is
significantly high. Lowering the $c$ quark mass corrects the $J/\psi $ mass
while raising the $D^{\ast },D_{s}^{\ast }\ P$ state masses would require
raising the $c$ quark mass. Reducing one discrepancy would worsen the other,
at least in our three invariant function approach

\begin{table}[tbp] \centering%
\begin{tabular}{|ccccccc|}
\hline
Family & \ \ \ \ \ \ \ \ \  & Exp. & \ \ \ \ \ \ \ \ \  & Th1. & \ \ \ \ \ \
\ \ \  & Th2. \\ \hline
$u\overline{d}$ &  & -0.36 &  & -0.57 &  & -0.03 \\ 
$s\overline{u}$ &  & -1.05 &  & -0.14 &  & 0.16 \\ 
$s\overline{s}$ &  & 0.05 &  & 0.06 &  & 0.26 \\ 
$c\overline{c}$ &  & 0.47 &  & 0.50 &  & 0.48 \\ 
$b\overline{b}\ \ \ (1^{3}P_{2,1,0})$ &  & 0.56 &  & 0.72 &  & 0.65 \\ 
$b\overline{b}\ \ \ (2^{3}P_{2,1,0})$ &  & 0.61 &  & 0.74 &  & 0.73 \\ \hline
\end{tabular}%
\caption{Spin-Orbit Splitting R Ratios}\bigskip \label{12}%
\end{table}%

For the spin-orbit splittings Table \ref{12} gives the $R$ ratios $%
(^{3}P_{2}-^{3}P_{1})/(^{3}P_{1}-^{3}P_{0}))$. \ Both sets of fits are very
poor for the two lightest multiplets.\ \ The fact that Th1 has the same sign
for the $u\bar{d}$ as the experiment values is not an indication that it
gives reasonable results since the negative sign originates from the
numerator instead of the denominator. \ \ Of the four remaining multiplets,
Th2 gives a better fit on 3. \ It must be said, however, that none of the
better fits are very good except for the $c\bar{c}.$ From the experimental
point of view the poor $R$ value for the $u\bar{d}$ and $u\bar{s}$ may be
the uncertain status of the $^{3}P_{0}$ light quark meson bound states, or
theoretically our low $^{3}P_{0}$ theoretical meson masses. Also, the lack
of any mechanism in our model to account for the effects of decay rates on
level shifts undoubtedly has an effect. Another likely cause for the poorer
performance of Th1 as one goes from heavier to light mesons is that the
radial size of the meson grows so that the long distance interactions, in
which the scalar interaction becomes dominant, play a more important role.
This effect is blunted by Th2 as seen in the $u\bar{d}$ numbers. In Table %
\ref{2} The spin-orbit terms due to scalar interactions are opposite in sign
and tend (at long distance) to dominate the spin-orbit terms due to vector
interactions for Th1, but less so in Th2. This results in partial to full
multiplet inversions as we proceed from the $s\bar{s}$ to the $u\bar{d}$
mesons. This inversion mechanism is less for\ Th2 than for Th1 because of
the value of $\xi $.

\begin{table}[tbp] \centering%
\begin{tabular}{|ccccccc|}
\hline
Family & \ \ \ \ \ \ \ \ \  & Exp. & \ \ \ \ \ \ \ \ \  & Th1. & \ \ \ \ \ \
\ \ \  & Th2. \\ \hline
$u\overline{d}$ &  & 76 &  & 30 &  & 36 \\ 
$s\overline{u}$ &  & 146 &  & 9 &  & 14 \\ 
$c\overline{c}$ &  & -1 &  & 3 &  & -1 \\ \hline
\end{tabular}%
\caption{Splitting Between  $^1P_1$ and weighted triplet states (MeV)}%
\bigskip \label{13}%
\end{table}%

The hyperfine structure of our equations also influences the splitting
between the $^{1}P_{1}$ and the weighted sum $%
[5(^{3}P_{2})+3(^{3}P_{1})+1(^{3}P_{0})]/9$ of bound states. Table \ref{13}
indicates the agreement of the theoretical and experimental mass differences
is excellent for the $c\bar{c}$ system, too small but of the right sign for
the $u\bar{s}$ system and $u\bar{d}$ systems. The agreement, however, for
the light systems is nevertheless considerably better than that in the case
of the fine structure splitting $R$ ratios. Note that in the case of unequal
mass $P$ states, our calculations of the two values incorporate the effects
of $\mathbf{L}\cdot (\mathbf{\sigma }_{1}-\mathbf{\sigma }_{2})$ and $%
\mathbf{L}\cdot \mathbf{\sigma }_{1}\times \mathbf{\sigma }_{2}~$which mix
spin.

\begin{table}[tbp] \centering%
\begin{tabular}{|ccccccc|}
\hline
Family & \ \ \ \ \ \ \ \ \  & Exp. & \ \ \ \ \ \ \ \ \  & Th1. & \ \ \ \ \ \
\ \ \  & Th2. \\ \hline
$u\overline{d}$ &  & 255 &  & 199 &  & 163 \\ 
$s\overline{u}$ &  & 303 &  & 163 &  & 131 \\ 
$c\overline{c}\ \ \ (1^{3}D_{1}-2^{3}S_{1})$ &  & 87 &  & 119 &  & 118 \\ 
$c\overline{c}\ \ \ (2^{3}D_{1}-3^{3}S_{1})$ &  & 114 &  & 78 &  & 77 \\ 
\hline
\end{tabular}%
\caption{Mixing Due to the Tensor Term Between Orbital D and Radial S
Excitations of the Spin-Triplet Ground States (in MeV)}\bigskip \label{14}%
\end{table}%

Next consider the mixing due to the tensor term between orbital $D$ and
radial $S~$excitations of the spin-triplet ground states. This mixing occurs
most notably in the $c\bar{c},u\bar{s}$ and $u\bar{d}$ systems. Table \ref%
{14} show that Th1 is better than Th2 although both are pretty far off the
mark. For the charmonium system, the lower doublet results are high whereas
the higher doublet results are low.

\begin{table}[tbp] \centering%
\begin{tabular}{|ccccccc|}
\hline
Family &  & Exp. Difference &  & Th1. Difference &  & Th2. Difference \\ 
\hline
\multicolumn{1}{|l}{$u\overline{d}\ \ \ \pi :\ 1,2,3^{1}S_{0}$} &  & 1160,
516 &  & 1352, 597 &  & 1301, 602 \\ 
\multicolumn{1}{|l}{$u\overline{d}\ \ \ \rho :\ 1,2,3^{3}S_{1}$} &  & 690,
684 &  & 955, 536 &  & 903, 642 \\ 
\multicolumn{1}{|l}{$s\overline{u}\ \ \ K^{-}:\ 1,2,3^{1}S_{0}$} &  & 966,
370 &  & 1038, 592 &  & 957, 605 \\ 
\multicolumn{1}{|l}{$s\overline{u}\ \ \ K^{\ast -}:\ 1,2^{2}S_{1}$} &  & 522
&  & 842 &  & 810 \\ 
\multicolumn{1}{|l}{$s\overline{s}\ \ \ \phi :\ 1,2^{2}S_{1}$} &  & 661 &  & 
761 &  & 862 \\ 
\multicolumn{1}{|l}{$c\overline{c}\ \ \ \eta _{c}:\ 1,2^{1}S_{0}$} &  & 657
&  & 641 &  & 629 \\ 
\multicolumn{1}{|l}{$c\overline{c}\ \ \ \psi :\ 1,2,3^{3}S_{1}$} &  & 589,
353 &  & 557, 398 &  & 561, 403 \\ 
\multicolumn{1}{|l}{$b\overline{b}\ \ \ \Upsilon :\ 1,2,3,4,5,6^{3}S_{1}$} & 
& 563, 332, 224, 286, 154 &  & 578, 344, 260, 218, 192 &  & 578, 346, 265,
224,198 \\ \hline
\end{tabular}%
\caption{Radial Excitations (MeV)}\bigskip \label{15}%
\end{table}%

Next we consider the effects of the change from Th1 to Th 2 on the radial
excitations. \ Table \ref{15} shows that for the most part Th1 gives
slightly better results for the radially excited states, although where both
theories are furthest off (the $u\bar{d}$ states) Th2 gives better results.
\ The radially excited $u\bar{d}$ mesons have a larger mean radius than for
the heavier meson and thus the temporizing effects of the $-V^{2}$ term
tends to counteract more the increased confining potential for large $r$
from linear to quadratic due to the $S^{2}$ terms.

Finally we comment on the isospin splittings shown in table \ref{15.5}.
There are two effects we must consider here: the positive $d-u$ mass
differences of about $6$ MeV for both theories and the Coulomb interaction
between the quarks on the order of $\alpha \times 197$ MeV or less depending
on the meson sizes. The Coulomb interaction is counter to the $d-u$ mass
difference for the $s\bar{d}-s\bar{u}$ and $b\bar{d}-b\bar{u}$ splittings
while enhancing the $d-u$ mass difference for the $c\bar{d}-c\bar{u}$
splittings. \ These alternatively competing and enhancing effects are seen
in the sizes of the splittings for both theories as you read down the table
from the $s\bar{d}-s\bar{u}$ \ through the $c\bar{d}-c\bar{u}$ to the $b\bar{%
d}-b\bar{u}$ splitting. For the $K-K^{\ast }$ family the values for the
isospin splittings are $3~$and $0$ MeV for Th1 and 2 vs the experimental
value 4 MeV for the singlet ground states while for the triplet the isospin
splittings are $2~$and $1$ MeV vs the experimental value 4 MeV. The
experimental splitting grows for the orbital excitation ($K_{2}^{\ast }$) to
6 MeV. The probable reason for the increase is that at the larger distances,
the influence of the Coulomb differences becomes small so that only the $d-u$
mass difference influences the result. Our theories do not show a similar
increase for the orbital excitations. In the case of the $D^{+}-D^{0}$
splitting our mass differences for Th1 and Th2 are 7 and 7 MeV respectively
versus the experimental mass difference of just 5 MeV. Here we see the
opposite overall effect between the combined effects of the $d-u$ mass
difference and the slightly increased electromagnetic binding present in the
case of the $D^{0}$ and the slightly decreased binding in the case of the $%
D^{+}$. Whereas in the kaon system the results are too small, for the $D$
the results are too large . This can be partially understood since the
Coulomb and $d-u$ mass differences work in concert with the Coulomb
potential for these doublets. These effects work in the same way for the
spin-triplet splitting resulting in the theoretical values of 6 and 6 MeV
for the two theories compared with the experimental value 3 MeV. For the $%
^{3}P_{2}$ isodoublet we obtain -5 and -5 MeV versus about 1 for the
experimental value again showing the expected opposite trend from that of
the kaon system. \ The experimental splitting between the $^{3}P_{0}$
isodoublet of 51 MeV appears incomprehensibly large. Our two values are 8
and 9 MeV. \ \ The isospin splittings that we obtain for the spin singlet $B$
meson system are 2 and 1 MeV for Th1 and Th2 versus 1 MeV. Here the
competing effects cancel as in the kaon system only more so since the mesons
are smaller and thus the Coulomb parts play a stronger role than for the
kaon. \ 

\begin{table}[tbp] \centering%
\begin{tabular}{|lcccccc|}
\hline
Family & \ \ \ \ \ \ \ \ \  & Exp & \ \ \ \ \ \ \ \ \  & Theory 1 & \ \ \ \
\ \ \ \ \  & Theory 2 \\ \hline
$s\bar{d}-s\overline{u}:\ 1^{1}S_{0}$ &  & 4 &  & 3 &  & 0 \\ 
$s\overline{d}-s\overline{u}:\ 1^{3}S_{1}$ &  & 4 &  & 2 &  & 1 \\ 
$s\overline{d}-s\overline{u}:\ 1^{3}P_{2}$ &  & 6 &  & 1 &  & 1 \\ \hline
$c\overline{d}-c\overline{u}:\ 1^{1}S_{0}\ $ &  & 5 &  & 7 &  & 7 \\ 
$c\overline{d}-c\overline{u}:\ 1^{3}S_{1}\ $ &  & 3 &  & 6 &  & 6 \\ 
$c\overline{d}-c\overline{u}:\ 1^{3}P_{0}\ $ &  & 51 &  & 8 &  & 9 \\ 
$c\overline{d}-c\overline{u}:\ 1^{3}P_{2}\ $ &  & 1 &  & -5 &  & -5 \\ \hline
$b\bar{d}-b\overline{u}:\ 1^{1}S_{0}$ &  & 1 &  & 2 &  & 1 \\ 
&  &  &  &  &  &  \\ 
$d-u\ $ mass &  &  &  & 5.8 &  & 6.0 \\ \hline
\end{tabular}%
\ \caption{Isospin Splitting (MeV)}\label{15.5}%
\end{table}%

\section{The Effective Relativistic Schr\"{o}dinger Equation with Flavor
Mixing for Spin-Zero Isoscalar Mesons.}

Consider the general eigenvalue equation (\ref{57}) for an isoscalar meson,
one with quark structure $q\bar{q}$. \ As seen in Appendix A the mass
dependence appearing in $\Phi _{w}$ directly or indirectly through $%
m_{w},\varepsilon _{w},\varepsilon _{1},\varepsilon _{2},$ is of four types: 
$m_{1}m_{2},$ $m_{1}^{2}\ $and $m_{2}^{2}$, $m_{1}^{2}+m_{2}^{2}$ and $%
m_{1}^{2}-m_{2}^{2}.$ The actual isoscalar mesons consist of mixtures of
three equal mass quark-antiquark pairs. We write the three separate equal
mass versions of (\ref{57}), using Eq. (\ref{bb}), together in shorthand as
\ 
\begin{equation}
\lbrack \mathbf{p}^{2}+\Phi _{w}(\mathbf{r,}m_{1}=m_{2},w,\mathbf{\sigma }%
_{1},\mathbf{\sigma }_{2})]%
\begin{bmatrix}
\psi _{u\bar{u}} \\ 
\psi _{d\bar{d}} \\ 
\psi _{s\bar{s}}%
\end{bmatrix}%
\equiv \lbrack \mathbf{p}^{2}+\Phi _{w}(\mathbf{r,}\mathbb{M})]%
\begin{bmatrix}
\psi _{u\bar{u}} \\ 
\psi _{d\bar{d}} \\ 
\psi _{s\bar{s}}%
\end{bmatrix}%
=\frac{1}{4}(w^{2}-4\mathbb{M}^{2})%
\begin{bmatrix}
\psi _{u\bar{u}} \\ 
\psi _{d\bar{d}} \\ 
\psi _{s\bar{s}}%
\end{bmatrix}%
.  \label{mx}
\end{equation}%
in which%
\begin{equation}
\mathbb{M=}%
\begin{bmatrix}
m_{u} & 0 & 0 \\ 
0 & m_{d} & 0 \\ 
0 & 0 & m_{s}%
\end{bmatrix}%
.
\end{equation}%
Eq. (\ref{mx}) does not include mixing between the pairs. Motivated by ideas
presented by Brayshaw \footnote{%
Brayshaw considered the modification of the meson mass by \ $w\mathbf{1}%
\rightarrow w\mathbf{1}+|G\rangle Z\langle G|$ with $Z$ a fixed parameter
and $|G\rangle $ in the $ns$ subspace.\ This is equivalent to the mixing
matrix 
\begin{equation*}
\mathbb{T=}Z%
\begin{bmatrix}
\langle n\bar{n}|G\rangle \langle G|n\bar{n}\rangle & \langle n\bar{n}%
|G\rangle \langle G|s\bar{s}\rangle \\ 
\langle s\bar{s}|G\rangle \langle G|n\bar{n}\rangle & \langle s\bar{s}%
|G\rangle \langle G|s\bar{s}\rangle%
\end{bmatrix}%
\equiv 
\begin{bmatrix}
t_{11} & t_{12} \\ 
t_{21} & t_{22}%
\end{bmatrix}%
,
\end{equation*}%
with the property that (for real matrix elements)%
\begin{equation*}
t_{11}t_{22}=t_{12}t_{21}=\left\vert t_{12}\right\vert ^{2}=t_{12}^{2},
\end{equation*}%
just as with the choice of Eq.(\ref{53}).} \cite{bry}, we model the effects
of $q_{i}\bar{q}_{i}\rightarrow q_{j}\bar{q}_{j}$ via two gluon annihilation
and creation as an effective scalar potential by postulating a symmetric
matrix $\mathbb{M}$ that is not diagonal, 
\begin{equation}
\begin{bmatrix}
m_{u} & 0 & 0 \\ 
0 & m_{d} & 0 \\ 
0 & 0 & m_{s}%
\end{bmatrix}%
\rightarrow \mathbb{M=}%
\begin{bmatrix}
m_{u} & 0 & 0 \\ 
0 & m_{d} & 0 \\ 
0 & 0 & m_{s}%
\end{bmatrix}%
+%
\begin{bmatrix}
\delta m_{u} & \sqrt{\delta m_{u}\delta m_{d}} & \sqrt{\delta m_{u}\delta
m_{s}} \\ 
\sqrt{\delta m_{u}\delta m_{d}} & \delta m_{d} & \sqrt{\delta m_{d}\delta
m_{s}} \\ 
\sqrt{\delta m_{u}\delta m_{s}} & \sqrt{\delta m_{d}\delta m_{s}} & \delta
m_{s}%
\end{bmatrix}%
.  \label{53}
\end{equation}%
Suppose that an orthogonal matrix $\mathbb{R}$ diagonalizes $\mathbb{M~}$%
\footnote{%
We ignore here the coupling that would result from this diagonalization that
would be brought on by the Coulomb interactions between the equal mass $q%
\bar{q}$ pairs.} 
\begin{equation}
\mathbb{RMR}^{-1}=\mathbb{M}_{D}.
\end{equation}%
Then Eq. (\ref{mx}) becomes 
\begin{equation*}
\lbrack \mathbf{p}^{2}+\Phi _{w}(\mathbf{r,}\mathbb{M}_{D})]\psi =\frac{1}{4}%
(w^{2}-4\mathbb{M}_{D}^{2})\psi .
\end{equation*}%
This gives us, in essence, three new effective families of equal
quark-anti-quark mesons, like ones that contain $b,c,s,u,d$ except that
mixtures are involved. In this paper we see if this idea is successful for
the ground state pseudoscalar isoscalar family of mesons alone. With the
three parameters, one obtains three different effective quark masses, one
for each isoscalar family. \ The three $\delta m_{i}$ are adjusted to give
the best fit to the correct $\pi ^{0},\ \eta ,\ \eta ^{\prime }$ masses. 
\footnote{%
These fits are simultaneous with the fits of the earlier 105 mesons, with
the same quark masses and potential parameters used in the generation of \
Tables (\ref{2}-\ref{9}). \ Oddly, a precise fit to the $\pi ^{0}$ was not
possible in either theory, in spite of the three extra parameters $\delta
m_{u},\delta m_{d},$ and $\delta m_{s}$ available.} \ Table \ref{16} gives
the values of $\delta m_{u},\ \delta m_{d},\ \delta m_{s}$ together with the
three effective quark masses, the eigenvalues of $\mathbb{M}$ which we call $%
m_{q(\pi ^{0})},\ m_{q(\eta )},\ m_{q(\eta ^{\prime })}$.

\ \ \ \ \ \ \ \ \ \ \ \ \ \ \ 
\begin{table}[tbp] \centering%
$%
\begin{tabular}{|lccc|}
\hline
Parameter & Th1 & \ \ \ \ \  & Th2 \\ \hline
$\delta m_{u}$ & $0.1004$ &  & $0.1070$ \\ 
$\delta m_{d}$ & $0.1378$ &  & $0.1055$ \\ 
$\delta m_{s}$ & $0.0468$ &  & $0.0578$ \\ 
$m_{q(\pi ^{0})}$ & $0.0737$ &  & $0.1015$ \\ 
$m_{q(\eta )}$ & $0.2175$ &  & $0.2261$ \\ 
$m_{q(\eta ^{\prime })}$ & $0.4297$ &  & $0.4536$ \\ \hline
\end{tabular}%
$\caption{Mixing Parameters for Th 1 and 2 (GeV)}\bigskip \label{16}%
\end{table}%

Paralleling the earlier Tables \ref{2}-\ref{9}, Table \ref{17} gives the
best fit values for the $\pi ^{0},~\eta ,$ and $\eta ^{\prime }$ mesons. \
The predicted quark content becomes a further test of our model.

\begin{table}[tbp] \centering%
\begin{tabular}{|lcccccccl|}
\hline
Mesons~~~~~~~~~~~~ & Exp. & Th1. & Th2. & Exp.-Th1. & Exp.-Th2. & $\chi ^{2}$%
-Th1. & $\chi ^{2}$-Th2. &  \\ \hline
$\pi ^{0}:1\,{}^{1}S_{0}$ & 0.135(0.0) & 0.139 & 0.134 & -0.004 & 0.001 & 0.2
& 0.0 &  \\ 
$\eta :\ 1\,{}^{1}S_{0}$ & 0.548(0.0) & 0.548 & 0.548 & 0.000 & 0.000 & 0.0
& 0.0 &  \\ 
$\eta ^{\prime }:\ 1\,{}^{1}S_{0}$ & 0.958(0.2) & 0.958 & 0.958 & 0.000 & 
0.000 & 0.0 & 0.0 &  \\ \hline
\end{tabular}%
\ \caption{$c\bar s$  Mesons,Theory 1 and 2 (GeV)}\label{17}%
\end{table}%

The corresponding eigenvectors are quite close to the mixtures%
\begin{eqnarray}
|\pi ^{0}\rangle &=&\frac{1}{\sqrt{2}}\left[ 
\begin{array}{c}
1 \\ 
-1 \\ 
0%
\end{array}%
\right] \equiv |\pi _{3}\rangle ,  \notag \\
~|\eta \rangle &=&\frac{\cos \theta }{\sqrt{6}}\left[ 
\begin{array}{c}
1 \\ 
1 \\ 
-2%
\end{array}%
\right] -\frac{\sin \theta }{\sqrt{3}}\left[ 
\begin{array}{c}
1 \\ 
1 \\ 
1%
\end{array}%
\right] \equiv \cos \theta ~|\eta _{8}\rangle -\sin \theta ~|\eta
_{1}\rangle ,  \notag \\
~|\eta ^{\prime }\rangle &=&\frac{\sin \theta }{\sqrt{6}}\left[ 
\begin{array}{c}
1 \\ 
1 \\ 
-2%
\end{array}%
\right] +\frac{\cos \theta }{\sqrt{3}}\left[ 
\begin{array}{c}
1 \\ 
1 \\ 
1%
\end{array}%
\right] \equiv \sin \theta |\eta _{8}\rangle +\cos \theta ~|\eta _{1}\rangle
,
\end{eqnarray}%
With the three eigenvectors in matrix form we find%
\begin{equation}
\begin{bmatrix}
|\pi ^{0} & |\eta \rangle & |\eta ^{\prime }\rangle%
\end{bmatrix}%
=%
\begin{bmatrix}
0.770 & 0.470 & 0.431 \\ 
-0.638 & 0.574 & 0.514 \\ 
-0.006 & -0.671 & 0.742%
\end{bmatrix}%
,
\end{equation}%
corresponding to $\theta =-12.6$ degrees for Th1 and%
\begin{equation}
\begin{bmatrix}
|\pi ^{0} & |\eta \rangle & |\eta ^{\prime }\rangle%
\end{bmatrix}%
=%
\begin{bmatrix}
0.717 & 0.542 & 0.438 \\ 
-0.697 & 0.565 & 0.442 \\ 
-0.008 & -0.622 & 0.783%
\end{bmatrix}%
,
\end{equation}%
corresponding to $\theta =-16.3$ degrees for Th2. \ The Th2 value is
consistent with chiral perturbation theory results corresponding to using
the formula \cite{prtl}. 
\begin{equation}
\tan ^{2}\theta _{\lbrack quad]}=\frac{4m_{K}^{2}-m_{\pi }^{2}-3m_{\eta }^{2}%
}{m_{\pi }^{2}+3m_{\eta ^{\prime }}^{2}-4m_{K}^{2}}.  \label{mix}
\end{equation}%
With the meson masses listed in Tables \ref{2} and \ref{3} \ for Th2 we
obtain $\theta _{\lbrack quad]}=-17.2$ degrees, reasonably close to our Th2
result of $-16.3$ degrees. \ On the other hand, using the values listed
there for Th1 gives $\theta _{\lbrack quad]}=-7.3~$degrees which is
significantly different from $-12.6$ degrees. Using the experimental masses
gives $\theta _{\lbrack quad]}=-11.5$ degrees. Radiative vector meson decays
give an angle between $-10$ and $-20$ degree while fits to tensor decay
widths give $-17~$degrees. \ \ Note that even though the chiral symmetry and
its breaking are not built into our model as in for example it is in \cite%
{est}, it is of interest that the use of Eq. (\ref{mix}) does produce a
consistent result by use of our theoretically computed masses\footnote{%
There are at least three other additional hints of a possible emergent
chiral symmetry and its breaking in the TBDE model . \ a) \ the relatively
small quark $u,d,$masses $\sim 60-100$ MeV compared with $\sim 300$ MeV with
most other quark models b) Our $\pi $ mass decreases toward zero as$%
~m_{q}\rightarrow 0$, c) the matrix element of the divergence of the axial
vector current being proportional to the quark mass\cite{saz86},\cite{cra88}}%
. \ 

\section{The Quasipotential Equation of Ebert, Faustov and Galkin}

\subsection{The Model \ and Comparisons with TBDE}

The model which we critically examine here gives excellent fits to the meson
spectrum as well as numerous meson decay rates. The quasipotential approach
of Ebert, Faustov and Galkin (EFG) \cite{rusger}\ is a local one very
similar to that of Todorov \cite{quasi} and Aneva, Karchev, and Rizov \cite%
{akr}, discussed in Sec. 2 (see Eq. (\ref{riz})). \ Insofar as our
discussion given in that section is concerned, the main difference between
the TBDE and EFG approach is the replacement of the timelike vector
confining interaction in Eq. (\ref{qua}) with a different confining vector
interaction: 
\begin{eqnarray}
\mathcal{V}(\mathbf{p-k})\beta _{1}\beta _{2} &\rightarrow &\mathcal{V}(%
\mathbf{p-k})\Gamma _{1\mu }\Gamma _{2}^{\mu },  \notag \\
\Gamma _{i\mu } &=&\gamma _{i\mu }-\frac{i\kappa (p-k)_{\nu }\sigma _{i\mu
\nu }}{2m_{i}},~i=1,2.
\end{eqnarray}%
They include as do we a scalar confining potential. In coordinate space
their choice is%
\begin{eqnarray}
V(r) &=&(1-\varepsilon )(Ar+B),  \notag \\
S(r) &=&\varepsilon (Ar+B).
\end{eqnarray}%
For their electromagnetic-like vector interaction they use the Coulomb gauge
(instead of the Feynman gauge used in the TBDE).%
\begin{equation*}
\frac{4}{3}\alpha _{s}D_{\mu \nu }(\mathbf{p-k)}\gamma _{1}^{\mu }\gamma
_{2}^{\nu }.
\end{equation*}%
Its momentum space form is%
\begin{equation}
D^{00}(\mathbf{p-k)}\mathbf{=-}\frac{4\pi }{\left( \mathbf{p-k}\right) ^{2}}%
,~D^{ij}=-\frac{4\pi }{\left( p-k\right) ^{2}}(\delta ^{ij}-\frac{%
(p-k)^{i}(p-k)^{j}}{\left( \mathbf{p-k}\right) ^{2}}).
\end{equation}%
\ The addition of the Pauli term with their value $\kappa =-1$ has the
effect of cancelling the lowest order spin-dependent contributions in each
factor of $\Gamma _{i\mu }$ when sandwiched between on energy shell spinors.
\ In the nonrelativistic limit their scalar and vector confining
interactions combine to $(Ar+B)$ with $A=0.18$ GeV$^{2}$, B=$-0.3$GeV$.$ \
In essence their approach embodies a modified version of the Cornell
potential into the local quasipotential approach.\ \ Their choice of $\kappa 
$ was fixed by an analysis of the fine structure splitting of heavy
quarkonia $^{3}P_{J}$ states \cite{rusger1} and their choice of $\varepsilon
=-1$ is determined from considerations of charmonium radiative decay. \ 

\bigskip 
\begin{table}[tbp] \centering%
\begin{tabular}{|lllllllll|}
\hline
Old Mesons~~~~~~~~~~~~ & Exp. & TBDE & EFG & Exp.- TBDE & Exp. - EFG. & $%
\chi ^{2}$(TBDE) & $\chi ^{2}$(EFG). &  \\ \hline
$\pi :u\overline{d}\ 1\,{}^{1}S_{0}$ & 0.140 & 0.134 & 0.154 & 0.005 & -0.014
& 0.4 & 2.5 &  \\ 
$\rho :u\overline{d}\ 1\,{}^{3}S_{1}$ & 0.775 & 0.779 & 0.776 & -0.003 & 
-0.001 & 0.1 & 0.0 &  \\ 
$b_{1}:u\overline{d}\ 1\,{}^{1}P_{1}$ & 1.230 & 1.237 & 1.258 & -0.007 & 
-0.028 & 0.1 & 0.9 &  \\ 
$a_{1}:u\overline{d}\ 1\,{}^{3}P_{1}$ & 1.230 & 1.311 & 1.254 & -0.081 & 
-0.024 & 0.0 & 0.0 &  \\ 
$\pi :u\overline{d}\ 2\,{}^{1}S_{0}$ & 1.300 & 1.426 & 1.292 & -0.126 & 0.008
& 0.0 & 0.0 &  \\ 
$a_{2}:u\overline{d}\ 1\,{}^{3}P_{2}$ & 1.318 & 1.303 & 1.317 & 0.015 & 0.001
& 2.0 & 0.0 &  \\ 
$\rho :u\overline{d}\ 2\,{}^{3}S_{1}$ & 1.465 & 1.674 & 1.486 & -0.209 & 
-0.021 & 0.8 & 0.0 &  \\ 
$a_{0}:u\overline{d}\ 1\,{}^{3}P_{0}$ & 1.474 & 1.015 & 1.176 & 0.459 & 0.298
& 7.0 & 3.0 &  \\ 
$b_{2}:u\overline{d}\ 1\,{}^{1}D_{2}$ & 1.672 & 1.752 & 1.643 & -0.080 & 
0.029 & 6.8 & 0.9 &  \\ 
$a_{3}:u\overline{d}\ 1\,{}^{3}D_{3}$ & 1.689 & 1.706 & 1.714 & -0.017 & 
-0.025 & 0.7 & 1.4 &  \\ 
$a_{1}:u\overline{d}\ 1\,{}^{3}D_{1}$ & 1.720 & 1.836 & 1.742 & -0.116 & 
-0.022 & 0.4 & 0.0 &  \\ 
$a_{2}:u\overline{d}\ 2\,{}^{3}P_{2}$ & 1.732 & 1.997 & 1.779 & -0.265 & 
-0.047 & 3.3 & 0.1 &  \\ 
$\pi :u\overline{d}\ 3\,{}^{1}S_{0}$ & 1.816 & 2.022 & 1.788 & -0.206 & 0.028
& 2.6 & 0.0 &  \\ 
$b_{2}:u\overline{d}\ 2\,{}^{1}D_{2}$ & 1.895 & 2.252 & 1.960 & -0.357 & 
-0.065 & 6.0 & 0.2 &  \\ 
$a_{4}:u\overline{d}\ 1\,{}^{3}F_{4}$ & 2.011 & 2.042 & 2.018 & -0.031 & 
-0.007 & 0.1 & 0.0 &  \\ 
$b_{2}:u\overline{d}\ 3\,{}^{1}D_{2}$ & 2.090 & 2.682 & 2.216 & -0.592 & 
-0.126 & 5.0 & 0.2 &  \\ 
$\rho :u\overline{d}\ 3\,{}^{3}S_{1}$ & 2.149 & 2.309 & 1.921 & -0.160 & 
0.228 & 1.1 & 2.2 &  \\ 
$a_{6}:u\overline{d}\ 1\,{}^{3}H_{6}$ & 2.450 & 2.590 & 2.475 & -0.140 & 
-0.025 & 0.0 & 0.0 &  \\ \hline
$K\,{}^{-}:s\overline{u}\ 1\,{}^{1}S_{0}$ & 0.494 & 0.528 & 0.482 & -0.034 & 
0.012 & 14.2 & 1.6 &  \\ 
$K^{\ast }\,{}^{-}:s\overline{u}\ 1\,{}^{3}S_{1}$ & 0.892 & 0.898 & 0.897 & 
-0.007 & -0.005 & 0.5 & 0.3 &  \\ 
$K\,{}^{-}:s\overline{u}\ 1\,{}^{1}P_{1}$ & 1.272 & 1.336 & 1.294 & -0.064 & 
-0.022 & 1.0 & 0.1 &  \\ 
$K^{\ast }\,{}^{-}:s\overline{u}\ 1\,{}^{3}P_{1}$ & 1.403 & 1.354 & 1.412 & 
0.049 & -0.009 & 0.6 & 0.0 &  \\ 
$K^{\ast }\,{}^{-}:s\overline{u}\ 2\,{}^{3}S_{1}$ & 1.414 & 1.698 & 1.675 & 
-0.284 & -0.261 & 4.3 & 3.6 &  \\ 
$K^{\ast }\,{}^{-}:s\overline{u}\ 1\,{}^{3}P_{0}$ & 1.425 & 1.075 & 1.362 & 
0.350 & 0.063 & 0.6 & 0.0 &  \\ 
$K^{\ast }\,{}^{-}:s\overline{u}\ 1\,{}^{3}P_{2}$ & 1.426 & 1.401 & 1.424 & 
0.025 & 0.002 & 0.0 & 0.0 &  \\ 
$K\,{}^{-}:s\overline{u}\ 2\,{}^{1}S_{0}$ & 1.460 & 1.414 & 1.538 & 0.046 & 
-0.078 & 0.0 & 0.0 &  \\ 
$K^{\ast }\,{}^{-}:s\overline{u}\ 1\,{}^{3}D_{1}$ & 1.717 & 1.828 & 1.699 & 
-0.111 & 0.018 & 0.2 & 0.0 &  \\ 
$K\,{}^{-}:s\overline{u}\ 1\,{}^{1}D_{2}$ & 1.773 & 1.795 & 1.709 & -0.022 & 
0.064 & 0.1 & 0.8 &  \\ 
$K^{\ast }\,{}^{-}:s\overline{u}\ 1\,{}^{3}D_{3}$ & 1.776 & 1.784 & 1.789 & 
-0.008 & -0.013 & 0.0 & 0.0 &  \\ 
$K^{\ast }\,{}^{-}:s\overline{u}\ 1\,{}^{3}D_{2}$ & 1.816 & 1.787 & 1.824 & 
0.029 & -0.008 & 0.1 & 0.0 &  \\ 
$K\,{}^{-}:s\overline{u}\ 3\,{}^{1}S_{0}$ & 1.830 & 2.069 & 2.065 & -0.239 & 
-0.235 & 4.1 & 3.9 &  \\ 
$K^{\ast }\,{}^{-}:s\overline{u}\ 2\,{}^{3}P_{2}$ & 1.973 & 2.050 & 1.896 & 
-0.077 & 0.077 & 0.1 & 0.1 &  \\ 
$K^{\ast }\,{}^{-}:s\overline{u}\ 1\,{}^{3}F_{4}$ & 2.045 & 2.106 & 2.096 & 
-0.061 & -0.051 & 0.5 & 0.4 &  \\ 
$K^{\ast }\,{}^{-}:s\overline{u}\ 2\,{}^{3}D_{2}$ & 2.247 & 2.301 & 2.163 & 
-0.054 & 0.084 & 0.1 & 0.3 &  \\ 
$K^{\ast }\,{}^{-}:s\overline{u}\ 2\,{}^{3}F_{3}$ & 2.324 & 2.585 & 2.348 & 
-0.261 & -0.024 & 1.4 & 0.0 &  \\ 
$K^{\ast }\,{}^{-}:s\overline{u}\ 1\,{}^{3}G_{5}$ & 2.382 & 2.387 & 2.356 & 
-0.005 & 0.026 & 0.0 & 0.0 &  \\ 
$K^{\ast }\,{}^{-}:s\overline{u}\ 2\,{}^{3}F_{4}$ & 2.490 & 2.585 & 2.436 & 
-0.095 & 0.054 & 0.3 & 0.1 &  \\ \hline
$\phi :s\overline{s}\ 1\,{}^{3}S_{1}$ & 1.019 & 1.017 & 1.038 & 0.002 & 
-0.019 & 0.1 & 4.1 &  \\ 
$\phi :s\overline{s}\ 1\,{}^{3}P_{0}$ & 1.370 & 1.175 & 1.420 & 0.195 & 
-0.050 & 0.0 & 0.0 &  \\ 
$\phi :s\overline{s}\ 1\,{}^{3}P_{1}$ & 1.518 & 1.436 & 1.464 & 0.082 & 0.054
& 3.1 & 1.4 &  \\ 
$\phi :s\overline{s}\ 1\,{}^{3}P_{2}$ & 1.525 & 1.505 & 1.529 & 0.020 & 
-0.004 & 0.2 & 0.0 &  \\ 
$\phi :s\overline{s}\ 2\,{}^{3}S_{1}$ & 1.680 & 1.868 & 1.698 & -0.188 & 
-0.018 & 1.1 & 0.0 &  \\ 
$\phi :s\overline{s}\ 1\,{}^{3}D_{3}$ & 1.854 & 1.874 & 1.950 & -0.020 & 
-0.096 & 0.1 & 2.2 &  \\ 
$\phi :s\overline{s}\ 2\,{}^{3}P_{2}$ & 2.011 & 2.120 & 2.030 & -0.109 & 
-0.019 & 0.0 & 0.0 &  \\ 
$\phi :s\overline{s}\ 3\,{}^{3}P_{2}$ & 2.297 & 2.590 & 2.412 & -0.293 & 
-0.115 & 1.3 & 0.2 &  \\ \hline
\end{tabular}%
\ 
\caption{Comparison of Th2($A,S,V$) Fits with Ebert et al \cite{rusger},
\cite{rusger1}-\cite{rusger5}}\label{18}%
\end{table}%

\bigskip 
\begin{table}[tbp] \centering%
\begin{tabular}{|lllllllll|}
\hline
New Mesons~~~~~~~~~~~~ & Exp. & TBDE. & EFG & Exp. - TBDE & Exp.-EFG & $\chi
^{2}$-TBDE & $\chi ^{2}$-EFG &  \\ 
$D^{+}:c\overline{d}\ 1\,{}^{1}S_{0}$ & 1.870 & 1.881 & 1.871 & -0.012 & 
-0.001 & 1.5 & 0.0 &  \\ 
$D^{\ast +}:c\overline{d}\ 1\,{}^{3}S_{1}$ & 2.010 & 2.010 & 2.010 & 0.000 & 
0.000 & 0.0 & 0.0 &  \\ 
$D^{\ast +}:c\overline{d}\ 1\,{}^{3}P_{0}$ & 2.403 & 2.224 & 2.406 & 0.179 & 
-0.003 & 2.0 & 0.0 &  \\ 
$D^{\ast 0}:c\overline{u}\ 1\,{}^{3}P_{2}$ & 2.460 & 2.408 & 2.460 & 0.052 & 
0.000 & 3.3 & 0.0 &  \\ \hline
$D_{s}:c\overline{s}\ 1\,{}^{1}S_{0}$ & 1.968 & 1.976 & 1.969 & -0.007 & 
-0.001 & 0.5 & 0.0 &  \\ 
$D_{s}^{\ast }:c\overline{s}\ 1\,{}^{3}S_{1}$ & 2.112 & 2.120 & 2.111 & 
-0.008 & 0.001 & 0.7 & 0.0 &  \\ 
$D_{s}^{\ast }:c\overline{s}\ 1\,{}^{3}P_{0}$ & 2.318 & 2.338 & 2.509 & 
-0.020 & -0.191 & 3.7 & 320 &  \\ 
$D_{s}:c\overline{s}\ 1\,{}^{1}P_{1}$ & 2.535 & 2.498 & 2.536 & 0.038 & 
-0.001 & 15.4 & 0.0 &  \\ 
$D_{s}^{\ast }:c\overline{s}\ 1\,{}^{3}P_{2}$ & 2.573 & 2.531 & 2.571 & 0.042
& 0.002 & 11.8 & 0.0 &  \\ 
$D_{s}^{\ast }:c\overline{s}\ 2\,{}^{3}S_{1}$ & 2.690 & 2.698 & 2.731 & 
-0.008 & -0.041 & 0.0 & 0.4 &  \\ \hline
$\eta _{c}:c\overline{c}\ 1\,{}^{1}S_{0}$ & 2.980 & 2.972 & 2.978 & 0.008 & 
0.002 & 0.3 & 0.0 &  \\ 
$J/\psi (1S):c\overline{c}\ 1\,{}^{3}S_{1}$ & 3.097 & 3.126 & 3.097 & -0.029
& 0.000 & 10.3 & 0.0 &  \\ 
$\chi _{0}:c\overline{c}\ 1\,{}^{3}P_{0}$ & 3.415 & 3.393 & 3.423 & 0.022 & 
-0.008 & 5.2 & 0.7 &  \\ 
$\chi _{1}:c\overline{c}\ 1\,{}^{3}P_{1}$ & 3.511 & 3.500 & 3.509 & 0.010 & 
0.002 & 1.3 & 0.0 &  \\ 
$h_{1}:c\overline{c}\ 1\,{}^{1}P_{1}$ & 3.526 & 3.519 & 3.525 & 0.007 & 0.001
& 0.6 & 0.0 &  \\ 
$\chi _{2}:c\overline{c}\ 1\,{}^{3}P_{2}$ & 3.556 & 3.553 & 3.556 & 0.004 & 
0.000 & 0.1 & 0.0 &  \\ 
$\eta _{c}:c\overline{c}\ 2\,{}^{1}S_{0}$ & 3.637 & 3.597 & 3.663 & 0.040 & 
-0.026 & 1.1 & 0.5 &  \\ 
$\psi (2S):c\overline{c}\ 2\,{}^{3}S_{1}$ & 3.686 & 3.683 & 3.684 & 0.004 & 
0.002 & 0.1 & 0.1 &  \\ 
$\psi (1D):c\overline{c}\ 1\,{}^{3}D_{1}$ & 3.773 & 3.801 & 3.795 & -0.028 & 
-0.022 & 8.4 & 5.2 &  \\ 
$\chi _{2}:c\overline{c}\ 2\,{}^{3}P_{2}$ & 3.929 & 3.975 & 3.972 & -0.046 & 
-0.043 & 1.0 & 0.9 &  \\ 
$\psi (3S):c\overline{c}\ 3\,{}^{3}S_{1}$ & 4.039 & 4.083 & 4.088 & -0.044 & 
-0.049 & 0.2 & 0.3 &  \\ 
$\psi (2D):c\overline{c}\ 2\,{}^{3}D_{1}$ & 4.153 & 4.160 & 4.194 & -0.007 & 
-0.041 & 0.1 & 2.0 &  \\ \hline
$B^{-}:b\overline{u}\ 1\,{}^{1}S_{0}$ & 5.279 & 5.285 & 5.280 & -0.006 & 
-0.001 & 0.4 & 0.0 &  \\ 
$B^{0}:b\overline{d}\ 1\,{}^{1}S_{0}$ & 0.000 & 0.000 & 0.000 & 0.000 & 0.000
& 0.0 & 0.0 &  \\ 
$B^{\ast -}:b\overline{u}\ 1\,{}^{3}S_{1}$ & 5.325 & 5.334 & 5.326 & -0.009
& -0.001 & 0.8 & 0.0 &  \\ 
$B^{\ast -}:b\overline{u}\ 1\,{}^{3}P_{2}$ & 5.747 & 5.687 & 5.741 & 0.060 & 
0.006 & 4.7 & 0.0 &  \\ 
$B_{s}^{0}:b\overline{s}\ 1\,{}^{1}S_{0}$ & 5.366 & 5.370 & 5.372 & -0.004 & 
-0.006 & 0.1 & 0.3 &  \\ 
$B_{s}^{\ast 0}:b\overline{s}\ 1\,{}^{3}S_{1}$ & 5.413 & 5.432 & 5.414 & 
-0.019 & -0.001 & 1.7 & 0.0 &  \\ 
$B_{s}^{\ast 0}:b\overline{s}\ 1\,{}^{3}P_{1}$ & 5.829 & 5.793 & 5.831 & 
0.037 & -0.002 & 11.0 & 0.0 &  \\ 
$B_{s}^{\ast 0}:b\overline{s}\ 1\,{}^{3}P_{2}$ & 5.840 & 5.805 & 5.842 & 
0.034 & -0.002 & 10.5 & 0.0 &  \\ \hline
$\eta _{b}:b\overline{b}\ 1\,{}^{1}S_{0}$ & 9.389 & 9.334 & 9.400 & 0.055 & 
-0.011 & 2.1 & 0.1 &  \\ 
$\Upsilon (1S):b\overline{b}\ 1\,{}^{3}S_{1}$ & 9.460 & 9.447 & 9.460 & 0.014
& 0.000 & 2.1 & 0.0 &  \\ 
$\chi _{b0}:b\overline{b}\ 1\,{}^{3}P_{0}$ & 9.859 & 9.835 & 9.864 & 0.024 & 
-0.005 & 6.1 & 0.2 &  \\ 
$\chi _{b1}:b\overline{b}\ 1\,{}^{3}P_{1}$ & 9.893 & 9.887 & 9.892 & 0.006 & 
0.001 & 0.4 & 0.0 &  \\ 
$\chi _{b2}:b\overline{b}\ 1\,{}^{3}P_{2}$ & 9.912 & 9.921 & 9.912 & -0.009
& 0.000 & 0.8 & 0.0 &  \\ 
$\Upsilon (2S):b\overline{b}\ 2\,{}^{3}S_{1}$ & 10.023 & 10.021 & 10.020 & 
0.002 & 0.003 & 0.0 & 0.1 &  \\ 
$\Upsilon (1D):b\overline{b}\ 1\,{}^{3}D_{2}$ & 10.161 & 10.178 & 10.157 & 
-0.017 & 0.004 & 2.4 & 0.1 &  \\ 
$\chi _{b0}:b\overline{b}\ 2\,{}^{3}P_{0}$ & 10.232 & 10.228 & 10.232 & 0.005
& 0.001 & 0.2 & 0.0 &  \\ 
$\chi _{b1}:b\overline{b}\ 2\,{}^{3}P_{1}$ & 10.255 & 10.261 & 10.253 & 
-0.005 & 0.002 & 0.3 & 0.1 &  \\ 
$\chi _{b2}:b\overline{b}\ 2\,{}^{3}P_{2}$ & 10.269 & 10.284 & 10.267 & 
-0.015 & 0.002 & 2.4 & 0.0 &  \\ 
$\Upsilon (3S):b\overline{b}\ 3\,{}^{3}S_{1}$ & 10.355 & 10.366 & 10.355 & 
-0.011 & 0.000 & 1.0 & 0.0 &  \\ 
$\Upsilon (4S):b\overline{b}\ 4\,{}^{3}S_{1}$ & 10.579 & 10.628 & 10.604 & 
-0.049 & -0.025 & 11.7 & 3.0 &  \\ \hline
\end{tabular}%
\caption{Comparison of Th2($A,S,V$) Fits with Ebert et al \cite{rusger},
\cite{rusger1}-\cite{rusger5}}\label{19}%
\end{table}%

We compare in Tables \ref{18} and \ref{19} their results to our Th 2 by
listing the deviations from the experimental results and respective $\chi
^{2}$. \ Of the 86 common mesons fit to the respective models, the collected
results of quasipotential approach of \cite{rusger}, \cite{rusger1}-\cite%
{rusger5} (EFG) are more accurate in 69 of the fits (61 when comparing $\chi
^{2}$)\footnote{%
Due to their large descrepancy on the (controversial) $^{3}P_{0}$ $%
D_{s}(2370)$ meson, their $\chi ^{2}$ \ of 389 is actually larger than our
255 (our fit here is altered to include less mesons and $m_{u}=m_{d}$ \ so
there are slight differences from the previous tables). If one eliminates
that meson from the fit our $\chi ^{2}$ reduces to 251 while their $\chi ^{2}
$ reduces to 69. \ It should be pointed out that their fit did not appear to
be a least $\chi ^{2}$ one like our fit. \ It also was not an overall fits
like ours.}. This includes the difficult singlet-triplet splittings for the
ground and excited states of charmonium and the ground state of bottomonium,
as well as good fits to most of the radial and orbital\ excitations of the
ground states of the light mesons. Particularly interesting examples for the 
$\pi -\rho $ family are the$~^{3}P_{1}-^{1}P_{1}$ splitting and the radial
excitation of the singlet and triplet $S-$states, and for the $K$-$K^{\ast }$
families the $~^{3}P_{1}-^{1}P_{1}$ splitting and the $^{3}P_{0}$ $\ $\
mass. These three areas of their spectrum are noteworthy improvements over
the TBDE approach. It is hard, however, to give an even theoretical
comparison between the two approaches for a number of different reasons. It
is worthwhile, however, to point out the differences in the two approaches,
summarized in Table \ref{20}. \ First of all in our approach we give an
overall fit to the entire spectrum. \ The approach of EFG to the spectrum is
spread over several papers and it is not clear that a uniform
parametrization would yield the same results as given in the tables
(summarized here) from their separate papers. \ They do use the same values
for the constants $A,\ B,\ \kappa ,$ and $\varepsilon $ as well as the quark
masses in the various papers. \ However, the $\Lambda $ parameter they use
in the parametrization of the short distance QCD-Coulomb part of the
potential is different \footnote{%
In their ealier papers where the fits to the heavier mesons are given they
use $\alpha _{s}=4\pi /(\beta _{0}\ln (\mu ^{2}/\Lambda ^{2}))$ where $\mu $
is the reduced mass and in the recent papers where the fits to lighter
mesons are given they use $\alpha _{s}=4\pi /(\beta _{0}\ln ((\mu
^{2}+M_{B}^{2})/\Lambda ^{2})).$ In the former papers they use $\Lambda =169$
or $178$ MeV and in the recent ones $\Lambda =413$ MeV. It is not clear how
using just one form for all the mesons would affect the overall fit. \ }. \
Another difference is that the static QCD potential in the Adler-Piran model
displays explicitly the asymptotic freedom behavior by its radial dependence 
\footnote{%
In \cite{yoon} the Adler-Piran potential was replaced by a form that
displays asymptotic freedom in the QCD coupling via $\left( 8\pi /27\right)
/\ln (K+B/(\Lambda r)^{2})$. \ Although the model used there does not give
as good a fit to the meson spectrum as the Adler-Piran model it does display
asymptototic freedom in a simpler form.} whereas in the Coulomb potential
used by EFG, the asymptotic freedom behavior is displayed indirectly in
their $\alpha _{s}$ by the quark mass dependence as seen in footnote 20. \
On the other hand, for their heavy quark bound states a radial modification
displaying short distance QCD asymptotic freedom corrections was used but
was not for the light quark bound states. \ Both the quasipotential approach
of EFG and the TBDE used here, extending earlier work of Crater and Van
Alstine, \ have three invariant interaction functions. \ \ Both use
electromagnetic-like four-vector interactions; in the Feynman gauge for the
TBDE and the Coulomb gauge for the quasipotential approach. \ In the EFG
quasipotential approach the third interaction is a Pauli-modified vector
interaction whereas in our approach it is timelike vector interaction. \ The
potentials used in each of the three parts are of course different in the
two approaches. \ The spin dependence, although similar for the most part
have distinctly different origins. \ In the 16 component TBDE the kinematics
and dynamics and spinors are tied together in one wave equation (see. e.g.
Eq. (\ref{schlike})). \ Its off shell dependence is fixed by the wave
equation and the spinors are all interacting. \ In the quasipotential
approach the potential is constructed in part from the actions of free
particle spinors. \ This leaves substantial leeway in how the off-shell
behavior is fixed.

\begin{table}[tbp] \centering%
\ 
\begin{tabular}{|lll|}
\hline
Properties & TBDE & Quasipotential Approach of EFG \\ \hline
Invariant Interactions & 3-$A($EM-like vector)$,\ S$(scalar), & 3-$A($%
EM-like vector)$,\ S$(scalar), \\ 
& $V$(timelike vector) & $V$(Pauli-modified vector) \\ 
&  &  \\ 
QCD Coupling & Coordinate Space Dependent & Quark Mass Dependent \\ 
&  &  \\ 
Meson Fits & Overall Spectrum, Same & Spectrum in Parts with Different \\ 
& Parametrizations & Parametrizations \\ 
&  &  \\ 
Spin- Dependence & Fixed by the TBDE given $A,\ S,\ V$ & Fixed by the
Quasipotential and $A,\ S,\ V$ \\ 
&  &  \\ 
Kinematics & Exact & Exact \\ 
&  &  \\ 
Singular potentials & Avoided by Dirac Equation Formalism & Avoided by Adhoc
Substitutions \\ 
&  &  \\ 
Numerical evaluations & Yes & Not for Heavy Mesons \\ 
&  &  \\ 
Chiral Symmetry & Zero Pion Mass for $m_{q}\rightarrow 0$ & Not Tested \\ 
&  &  \\ 
QED Spectral Tests & Both Perturbative and Nonperturbative & Perturbative
Only \\ \hline
&  &  \\ \hline
Static Limit ($m_{2}\rightarrow \infty )$ & Reduces to One-Body Dirac Eq. & 
Does Not Reduce to One-Body Dirac Eq. \\ \hline
\end{tabular}%
\caption{Comparison of TBDE with EFG}\label{20}%
\end{table}%

Both approaches have exact relativistic kinematics (from the use of $%
b^{2}(w) $) and do not use either $v/c$ or $1/m_{q}$ expansions. Both
approaches lead to non-linear eigenvalue equations and give good values of
the pion and kaon masses. Both approaches avoid singular effective
potentials that would otherwise prevent nonperturbative spectral
calculations. \ In the quasipotential approach of EFG, those singularities
are avoided in an adhoc though plausible fashion (see Eq. (12) in \cite%
{rusger}). In the case of the TBDE, natural smoothing mechanisms appear in
the Dirac formalism allowing one to avoid these adhoc assumptions \cite%
{cra82}, \cite{becker} \footnote{%
In \cite{yoon}, Crater, Yoon, and Wong described some unusual singularity
structures of the effective potentials and wave functions that show up in
Eq. (\ref{schlike}) for singlet and triplet states, in both QED and QCD. In
these cases the TBDE lead to effective potentials and wave functions that
are nevertheless not singular. The most noteworthy case was for the coupled $%
{{}^{3}}S_{1}$-${{}^{3}}D_{1}~$ triplet system, when the tensor coupling is
properly taken into account. There it was shown that including the tensor
coupling is essential in order that the effective potentials and wave
functions are well behaved at short distances, with the $S$-state and $D$%
-state wave functions becoming simply proportional to each other at short
distance (see Appendix C).}. In addition, in the approach of the TBDE,
strictly nonperturbative (i.e. numerical techniques) are used for the
spectral evaluations. This does not appear to be the case for the work of
EFG, particularly for the heavy mesons, where use of perturbation theory is
required because of singular potentials. It may very well be that their
adhoc substitutions used in the later paper \cite{rusger} will render the
use of perturbation theory unnecessary for those mesons. But in that case
clear tests must ensure that\ not only do the adhoc substitutions give the
same results as the perturbative treatments, but that this hold in the
sensitive testing grounds of QED \cite{iowa}, \cite{becker} ground states
and those of related field theories . \ Both approaches display chiral
symmetry breaking through the appearance of quark masses. \ It has been
demonstrated in the TBDE, however, that the pion mass vanishes in the limit
in which the quark mass vanishes \cite{crater2}. That is not demonstrated in
the EFG formalism nor for any other potential model formalism that we know
of for the mesons (In an exception, Sazdjian has demonstrated this using
pseudoscalar interactions for the TBDE \cite{saz86}).\footnote{%
\ H. Sazdjian has considered chiral symmetry and its breaking in the context
of a closely related version of the TBDE. \ This was later discussed by
Crater and Van Alstine, \cite{cra88}. \ In particular, it is found that the
matrix element of the divergence of the axial vector current is proportional
to the quark mass. \ This demsonstrates that the quark masses in the TBDE
play the same role as the quark masses in QFT. \ In addition, Sazdjian shows
that the pion decay constant in the context of the TBDE does not vanish in
the limit of $m_{q}\rightarrow 0$. \ Sazdjian shows analytically in the
context of a pseudoscalar confining potential, the existence of a massless
pseudoscalar meson when $m_{q}\rightarrow 0$. \ These are two of the main
effects of the spontaneous breakdown of chiral symmetry. \ In our earlier
work with Van Alstine~\cite{cra88}, we showed numerically for the case of
scalar confinining intereractions that the calculated pion mass tends to
zero as $m_{q}\rightarrow 0$. \ In a later work \ (\cite{crater2}) we
showed, \ however, that the behavior is not of the square-root relation ($%
m_{\pi }\sim \sqrt{m_{q}}/F_{\pi }$). \ The same behavior appears to hold
with the present calculations.} \ 

The final point we want to make about the differences is that the wave
equation arising from the TBDE has been tested in perturbative QED and
related field theories. Todorov, and others \cite{quasi}, \cite{akr} and 
\cite{tod75} showed using perturbative methods how their local version of
the quasipotential equation displays the accepted QED spectral results
through order $\alpha ^{4}$ for two oppositely charged particles with
arbitrary mass ratios. \ The work by Crater and Van Alstine \cite{exct} and
others \cite{becker} go beyond this and show that not only do the TBDE
display the correct fine and hyperfine spectral results when treated
perturbatively, but those same results can be recovered when the equations
are treated nonperturbatively. In \cite{mfa}, \ a local quasipotential
equation closely related to that used by EFG in the meson spectrum was shown
to also reproduce perturbatively the spectral results through order $\alpha
^{4}$ of QED for two oppositely charged particles with arbitrary mass
ratios. \ However, the important nonperturbative tests as done in \cite%
{becker} of the bound state formalism for QED has not been carried out with
the local quasipotential equation of EFG.

\ \ It is our contention that any relativistic potential model that includes
four-vector interactions should, when the vector interaction is replaced by
its QED counterpart, and confining potentials are set $=0$, reproduce the
the standard QED spectral results. \ There are two models for which we carry
out this test. We limit our test to the singlet positronium ground state.
The fundamental question we ask is do these two approaches, \cite{isgr} and 
\cite{rusger}, which are used quite successfully for meson spectroscopy give
the correct spectral results when restricted to QED. \ The first one we
examine is that of Ebert, Faustov and Galkin \cite{rusger}.

\subsection{Positronium Ground State Spectral Test of the Quasipotential
Equation of Ebert, Faustov and Galkin.}

The effective Schr\"{o}dinger equation of this approach restricted to an
equal mass bound system for vector interactions is given in Eqs. (1-4),
(13-22) of \cite{rusger}. For that restriction we have, in the notation of
the present paper,%
\begin{eqnarray}
\left( \frac{\mathbf{p}^{2}}{2\mu _{R}}+V(r)\right) \Psi _{w} &=&\frac{%
b^{2}(w)}{2\mu _{R}}\Psi _{w},  \notag \\
\mu _{R} &=&\frac{\varepsilon _{1}\varepsilon _{2}}{w}=\frac{w}{4},  \notag
\\
b^{2}(w) &=&\frac{1}{4}(w^{2}-4m^{2}),  \notag \\
V(r) &=&A(r)[1+(\frac{w-2m}{w})^{2}]+(\frac{2}{w(w+2m)}+\frac{8}{3w^{2}}%
\mathbf{S}_{1}\cdot \mathbf{S}_{2})\nabla ^{2}A,  \notag \\
A &=&-\frac{\alpha }{r}.  \label{eff}
\end{eqnarray}%
The unperturbed effective Hamiltonian is%
\begin{equation}
H_{0}=\frac{\mathbf{p}^{2}}{2\mu _{R}}-\frac{\alpha }{r},
\end{equation}%
and the perturbation is for singlet states%
\begin{equation}
H_{1}=-(\frac{w-2m}{w})^{2}\frac{\alpha }{r}-(\frac{8\pi \alpha }{w(w+2m)}-%
\frac{8\pi \alpha }{w^{2}})\delta ^{3}(\mathbf{r).}  \label{prt}
\end{equation}%
Comparing with the nonrelativistic hydrogenic Schr\"{o}dinger equation 
\begin{equation}
\left( \frac{\mathbf{p}^{2}}{2\mu }-\frac{\alpha }{r}\right) \psi =\mathcal{E%
}\psi _{w},  \label{s1}
\end{equation}%
with ground state energy%
\begin{equation}
\mathcal{E=-}\frac{\mu \alpha ^{2}}{2},  \label{s2}
\end{equation}%
we see that for the eigenvalue equation (\ref{eff}), the total c.m.
invariant energy from $H_{0}$ is determined by analogy with Eqs. (\ref{s1}, %
\ref{s2}) from%
\begin{equation}
\frac{b^{2}(w)}{2\mu _{R}}=-\frac{\mu _{R}\alpha ^{2}}{2}.  \label{ev}
\end{equation}%
Thus with 
\begin{equation}
w^{2}-4m^{2}=-4\mu _{R}^{2}\alpha ^{2}=-\frac{w^{2}}{4}\alpha ^{2}
\end{equation}%
we find that 
\begin{equation}
w=2m-\frac{m\alpha ^{2}}{4}+\frac{3m\alpha ^{4}}{64}.
\end{equation}%
Substituting this into Eq. (\ref{prt}) at the appropriate order gives 
\begin{eqnarray}
H_{1} &=&-(\frac{m\alpha ^{2}}{8m})^{2}\frac{\alpha }{r}+\frac{\pi \alpha }{%
m^{2}}\delta ^{3}(\mathbf{r)}  \notag \\
&\rightarrow &\frac{\pi \alpha }{m^{2}}\delta ^{3}(\mathbf{r)},
\end{eqnarray}%
with their spin-spin terms partially canceling their spin independent
contact (Darwin) term while the first term is of higher order. The ground
state unperturbed wave function is%
\begin{eqnarray}
\Psi _{w} &=&\frac{\exp (-r/a_{eff})}{\left( \pi a_{eff}^{3}\right) ^{1/2}},
\notag \\
a_{eff} &=&\frac{1}{\mu _{R}\alpha }\rightarrow \frac{2}{m\alpha }.
\end{eqnarray}%
The expectation value is%
\begin{equation}
\langle H_{1}\rangle =\frac{\alpha }{m^{2}\left( \frac{2}{m\alpha }\right)
^{3}}=\frac{m\alpha ^{4}}{8},
\end{equation}%
and so Eq. (\ref{ev}), including $\langle H_{1}\rangle $ then becomes, to
the appropriate order%
\begin{eqnarray}
\frac{b^{2}(w)}{2\mu _{R}} &=&-\frac{\mu _{R}\alpha ^{2}}{2}+\frac{m\alpha
^{4}}{8},  \notag \\
w^{2}-4m^{2} &=&-\frac{w^{2}}{4}\alpha ^{2}+\frac{m^{2}\alpha ^{4}}{2},
\end{eqnarray}%
so that 
\begin{equation}
w=2m-\frac{m\alpha ^{2}}{4}+\frac{3m\alpha ^{4}}{64}+\frac{m\alpha ^{4}}{8}%
=2m-\frac{m\alpha ^{2}}{4}+\frac{11m\alpha ^{4}}{64}.  \label{evq}
\end{equation}

This result is in disagreement with the accepted fine structure result of 
\begin{equation}
w=2m-\frac{m\alpha ^{2}}{4}-\frac{21m\alpha ^{4}}{64}.  \label{exct}
\end{equation}%
In the Two-Body Dirac equation this spectrum results from an exact solution
of the Schr\"{o}dinger-like form \cite{exct}, \cite{becker}, \cite{crstr} 
\footnote{%
This equation and its gauge structure can also be seen to result from the
equal mass singlet equation version of (\ref{57}) for $S=V=0,$ $A=-\alpha /r$
or its radial version given by Eq. (\ref{ss}). (It is noted that under these
conditions, $\Phi _{D}=3\Phi _{SS},$and $\Phi _{SOD}=\Phi _{SOX}=0$)} 
\begin{equation}
(\mathbf{p}^{2}+2\varepsilon _{w}A-A^{2})\psi =b^{2}\psi ,
\end{equation}%
which yields 
\begin{equation}
w=m\sqrt{2+2/\sqrt{1+\frac{\alpha ^{2}}{\left( 1+\sqrt{\frac{1}{4}-\alpha
^{2}}-\frac{1}{2}\right) ^{2}}}}=2m-\frac{m\alpha ^{2}}{4}-\frac{21m\alpha
^{4}}{64}+..,  \label{ect}
\end{equation}%
obtained by steps similar to those outlined in Eq.(\ref{evq}).

The local quasipotential approach of \cite{rusger} does not include the term 
$-A^{2}=$ $-\alpha ^{2}/r^{2}$ which gauge invariance considerations would
demand. \ It is of interest that if their approach includes this terms then
the added contribution from this potential is%
\begin{equation}
\frac{1}{2\mu }\langle -\frac{\alpha ^{2}}{r^{2}}\rangle =\frac{1}{m}\langle
-\frac{\alpha ^{2}}{r^{2}}\rangle =-\frac{m\alpha ^{4}}{2}.
\end{equation}%
This, together with the fact that combining this with the earlier results
gives the correct added $O(\alpha ^{4})$ correction,%
\begin{equation}
\frac{11m\alpha ^{4}}{64}-\frac{m\alpha ^{4}}{2}=-\frac{21m\alpha ^{4}}{64},
\label{df}
\end{equation}%
points strongly to the lack of this term as being the cause of the incorrect
QED spectral prediction in this approach. \ 

We should emphasize that the closely related formalism of \cite{mfa} \textit{%
does} produce the correct $-\frac{21m\alpha ^{4}}{64}$ relativistic
correction. \ The difference between the formalism of EFG and that of \cite%
{mfa} is that the former does not include two loop and iterated Born diagrams%
\footnote{%
The authors are grateful to Professor R. N. Faustov for pointing out to us
the results of \cite{mfa} and for their reason that the bound state equation
used in \cite{rusger} did not include the $-A^{2}$ term.} contained in the
latter.\ Those combined diagrams to lowest order in $\alpha $ do produce the 
$-A^{2}=$ $-\alpha ^{2}/r^{2}$ $\ $term which would account for the spectral
difference seen in Eq. (\ref{df}) (see also \cite{tod75}). \ Since that $%
-A^{2}$ term is not included in the EFG meson spectral formalism it is
likely that possibly important relativistic corrections for their meson
spectrum will be missing. In a private communication, Faustov stated that
the reason they did not include the contributions of two and more gluon
exchange diagrams within QCD in calculations of the meson spectra, is that
the effects of these diagrams would be contained in the confining , long
range potential, the origin of which is not known and which is thus added
phenomenologically. However the $-A^{2}$ contribution from those two-gluon
exchange diagrams due to the Coulomb-like potential $A$ is short range and
therefore would not by itself contribute to the confining potential. In
other words we claim that since its effects are short range, it should be
considered apart from the phenomenologically added confining interaction. \ 

\subsection{Positronium Test of the Approach of Godfrey and Isgur.}

Although Crater and Van Alstine carried out an earlier comparison \cite%
{crater2} with this approach \cite{isgr}, in light of the problem with the
above quasipotential approach it is instructive to include a parallel
perturbative treatment on their different quasipotential equation. \ Their
relativistic Schr\"{o}dinger equation (see their Eqs. (1-4)) relevant for
the case considered here has the Hamiltonian which include the spin-spin
term in addition to the modified Coulomb term (see their Eq. (A15) \footnote{%
In their appendix A Godfrey and Isgur modify the Coulomb and contact
spin-spin term used here with smearing functions and extra non-local parts
in order to account for the off mass shell effects not present in the on
shell scattering amplitudes from which they extract their potentials. We do
not include the effects of the gaussian smoothing factors in the
determination of the modification of the Coulomb term from their Eq. (A15).}%
). \ Their equation was of the quasipotential type given in Eq. (\ref{log})
extended to include spin. \ Here we consider its semirelativistic expansion.%
\begin{eqnarray}
H &=&2\sqrt{\mathbf{p}^{2}+m^{2}}-\frac{\alpha }{r}+\frac{2}{3m^{2}}\mathbf{S%
}_{1}\cdot \mathbf{S}_{2}\nabla ^{2}A-\frac{\alpha }{2m^{2}}\{\mathbf{p}^{2},%
\frac{1}{r}\}  \notag \\
&\rightarrow &2m+H_{0}+H_{1},  \notag \\
H_{0} &=&\frac{\mathbf{p}^{2}}{m}-\frac{\alpha }{r},  \notag \\
H_{1} &=&-\frac{\left( \mathbf{p}^{2}\right) ^{2}}{4m^{3}}+\frac{2\pi \alpha 
}{m^{2}}\delta ^{3}(\mathbf{r)}-\frac{\alpha }{2m^{2}}\{\mathbf{p}^{2},\frac{%
1}{r}\}.
\end{eqnarray}%
The ground state unperturbed wave function is%
\begin{eqnarray}
\Psi &=&\frac{\exp (-r/a)}{\left( \pi a^{3}\right) ^{1/2}},  \notag \\
a &=&\frac{2}{m\alpha }.
\end{eqnarray}%
We find that%
\begin{eqnarray}
\langle H_{1}\rangle &=&-\frac{1}{4m}\langle \Psi \frac{\mathbf{p}^{2}}{m}%
\frac{\mathbf{p}^{2}}{m}\Psi \rangle +\langle \Psi \frac{2\pi \alpha }{m^{2}}%
\delta ^{3}(\mathbf{r)}\Psi \rangle -\frac{\alpha }{2m^{2}}\langle \Psi \{%
\mathbf{p}^{2},\frac{1}{r}\}\Psi \rangle \\
&=&-\frac{1}{4m}\langle \Psi (\mathcal{E+}\frac{\alpha }{r})^{2}\Psi \rangle
+\frac{m\alpha ^{4}}{4}-\frac{\alpha }{2m^{2}}\langle \Psi \{\mathbf{p}^{2},%
\frac{1}{r}\}\Psi \rangle ,
\end{eqnarray}%
and with $\mathcal{E}\mathcal{=-}\frac{m\alpha ^{2}}{4}$ we have 
\begin{eqnarray}
-\frac{1}{4m}\mathcal{E}^{2} &=&-\frac{m\alpha ^{4}}{64},  \notag \\
-\frac{\mathcal{E}}{2m}\langle \Psi \frac{\alpha }{r}\Psi \rangle &=&\frac{%
\alpha ^{2}}{8}(-2\mathcal{E)=}\frac{m\alpha ^{4}}{16},  \notag \\
-\frac{1}{4m}\langle \Psi \frac{\alpha ^{2}}{r^{2}}\Psi \rangle &=&-\frac{1}{%
4m}\frac{\alpha ^{2}}{\pi a^{3}}4\pi \int_{0}^{\infty }dr\exp (-2r/a)=-\frac{%
m\alpha ^{4}}{8},
\end{eqnarray}%
and using%
\begin{eqnarray}
-\frac{\alpha }{2m^{2}}\langle \Psi \{\mathbf{p}^{2},\frac{1}{r}\}\Psi
\rangle &=&-\frac{2(2\mu )\alpha }{2m^{2}}\langle \Psi \frac{1}{r}(\mathcal{%
E+}\frac{\alpha }{r})\Psi \rangle  \notag \\
&=&-\frac{\mathcal{E}}{m}\langle \Psi \frac{\alpha }{r}\Psi \rangle -\frac{1%
}{m}\langle \Psi \frac{\alpha ^{2}}{r^{2}}\Psi \rangle =\frac{m\alpha ^{4}}{8%
}-\frac{m\alpha ^{4}}{2}
\end{eqnarray}%
and find that%
\begin{equation}
\langle H_{1}\rangle =\frac{m\alpha ^{4}}{4}(-\frac{1}{16}+\frac{1}{4}-\frac{%
1}{2}+1+\frac{1}{2}-2),
\end{equation}%
and so 
\begin{equation}
w=2m+\langle H_{0}\rangle +\langle H_{1}\rangle =2m-\frac{m\alpha ^{2}}{4}-%
\frac{13m\alpha ^{4}}{64}.
\end{equation}%
This also does not agree with the accepted result of Eq. (\ref{exct}). \
Since the addition of the $-\alpha ^{2}/r^{2}$ term would drive this to the
other side of the accepted value it is not clear where the error is in this
approach.

\subsection{Comparison of Two Approaches to Dirac Equation in the Static
Limit}

The dynamics of the heavy-light $q\bar{Q}$ bound states, particularly the $u%
\bar{b}$ and $d\bar{b},$ should be well approximated by the ordinary
one-body Dirac equation. \ In the limit when say $m_{2}\rightarrow \infty $
details outlined in Appendix A.5 show that our TBDE reduce to the single
particle Dirac equation for a spin-one-half particle in an external scalar
and vector potential,%
\begin{equation}
(\mathbf{\gamma }\cdot \mathbf{p-}\beta (\varepsilon -A)+m+S)\psi =0,
\label{obde}
\end{equation}%
in which $\varepsilon $ is the total energy of the single particle of mass $%
m $. In this same limit Eq. (\ref{57}) (see Appendix A5 ) becomes%
\begin{eqnarray}
&&(\mathbf{p}^{2}+2mS+S^{2}+2\varepsilon A-A^{2}+\frac{1}{2}\frac{\nabla
^{2}A-\nabla ^{2}S}{m+S+\varepsilon -A}+\frac{3}{4}\left( \frac{S^{\prime
}-A^{\prime }}{m+S+\varepsilon -A}\right) ^{2}+\frac{A^{\prime }-S^{\prime }%
}{m+S+\varepsilon -A}\frac{\mathbf{L\cdot \sigma }_{1}}{r})\psi _{+}  \notag
\\
&=&(\varepsilon ^{2}-m^{2})\psi _{+},  \label{571}
\end{eqnarray}%
which agrees with the Pauli reduction of the Dirac equation (\ref{obde}) for
a single particle in an external scalar and vector potential when the first
order momentum terms are scaled away (see for example \cite{yoon}). \ The
wave function $\psi _{+}$ is the upper two-component spinor. \ From the
point of view of the single particle Dirac equation the quadratic $S^{2}$
and $-A^{2}$ terms above are not put in by hand but arise naturally from the
Pauli reduction. \ 

We compare Eq. (\ref{571}) with the corresponding equations from the two
quasipotential approaches, including scalar and vector interactions\footnote{%
In \cite{faustdrc} the static limit Dirac equation was recovered from a
two-body quasipotential equation by techniques with some similarity to the
Gross equation\cite{gross1}. \ In that equation the relative energy is
constrained by restricting one of the spin-one half particles to its
positive energy mass shell. \ This differs from the TBDE which treats both
particles off shell but yet constraining the relative energy covariantly
through $P\cdot p\psi =0.$ The EFG equation also has this constraint, but
unlike the equation derived in \cite{faustdrc}, the Gross equation and the
TBDE, it does not have the Dirac equation as a static limit as seen by a
comparison of Eq. (\ref{rgst}) with Eq. (\ref{571}).}. \ Referring to Eqs.
(1-4), (13-22) of \cite{rusger} we have the $m_{2}\rightarrow \infty $ limit
of%
\begin{equation}
(\mathbf{p}^{2}+2\varepsilon (A+S)+\frac{1}{2}\frac{\nabla ^{2}A}{%
m+\varepsilon }+\frac{A^{\prime }-S^{\prime }}{m+\varepsilon }\frac{\mathbf{%
L\cdot \sigma }_{1}}{r})\psi =(\varepsilon ^{2}-m^{2})\psi ,  \label{rgst}
\end{equation}%
in which we have used $\mu _{R}\rightarrow \varepsilon .~$It is evident that
for the vector interaction alone ($S=0$), this equation will not yield a
spectrum perturbatively equivalent to the Dirac spectrum for $A=-\alpha /r$
and for the same reason as with positronium, that is, the lacking of the $%
-A^{2}$ term. This would result in incorrect fine structure for
hydrogen-like atoms (see discussions and footnotes below Eq. (\ref{df}) for
the reason for this omission). \ Also, there are three parts for the scalar
potential interaction that differ from the Dirac equation: \ the appearance
of $2\varepsilon S$ instead of the expected $2mS$, the lacking of the $S^{2}$
term, and the absence of any scalar Darwin term. \ The appearance of $%
\varepsilon $ instead of $m$ can be traced to the use of a common reduced
mass for multiplying both vector and scalar interactions. \ Beyond that is
the absence of the potential energy terms in the denominators of the
spin-orbit and Darwin terms. \ The authors correct this by hand, but their
correction does not match the forms in the Dirac equation. \ On the other
hand those potential energy terms in the denominators of the Darwin and
spin-orbit terms of Eq. (\ref{571}) provide a natural smoothing mechanism
that eliminates such singular potentials as delta functions and $1/r^{3}$
potentials. \ For example take the case of $S=0,~A=-\alpha /r$. \ The
Laplacian term would produce $4\pi \delta ^{3}(\mathbf{r)}$. \ However, the $%
A$ term in the denominator would then be \ evaluated at the origin and
completely cancel the effects of the delta function. \ Its perturbative
effects are reproduced by the adjacent $3/4$ term. Similarly the $1/r^{3}$
behavior of the spin-orbit weak potential form in which the $A$ in the
denominator is ignored is modified to very near the origin to a less
singular $1/r^{2}$ potential by the effect of the $A$ term in the
denominator as well as the $3/4$ term. \ Similar smoothing mechanics
naturally built in to the Pauli structure of the Dirac equation occur in the
Pauli reduction of the TBDE ( see \cite{cra82},\cite{cra84},\cite{becker}, 
\cite{yoon}).

Referring to A-15, 16 of \cite{isgr} we have the $m_{2}\rightarrow \infty $
limit of%
\begin{equation}
(\sqrt{\mathbf{p}^{2}+m^{2}}+A+S+\frac{A^{\prime }-S^{\prime }}{4m^{2}}\frac{%
\mathbf{L\cdot \sigma }_{1}}{r})\psi =w\psi .
\end{equation}%
This Hamiltonian form is missing Darwin terms for not only scalar
interactions, but also for vector interactions as well. \ Those are as
important for spectral studies as the spin-orbit terms and their lack is a
serious defect in both these equations. The lack of the vector Darwin term
would result in incorrect fine structure for $L=0$ hydrogen levels.

\section{Conclusion and Future Directions}

\ The application of Dirac's constraint dynamics applied to the relativistic
two-body problem leads quite naturally to the Two Body Dirac equations of
constraint dynamics when both particles have spin-one-half. This paper
follows many earlier ones analyzing the structures and applications of those
equations. \ It has several sets of aims and results.\ First we showed that
when the interaction structure used in these equations is extended from two
invariant functions (generated by what we have called $A(r)$ and $S(r)$) to
three (not only the two that generate an electromagnetic-like interaction\
and a confining world scalar interaction but also the $V(r)$ that generates
a confining timelike vector interaction), that the fit to the meson spectrum
is improved substantially. However, there is still a considerable amount of
improvement that is needed, primarily in the radial and orbital excitations
of the singlet and triplet ground states. \ Work in progress seeks to extend
invariant functions to ones that generate covariant pseudoscalar,
pseudovector, and tensor interactions. Second, this paper also included 19
mesons not included in earlier work \cite{crater2} where only two invariant
functions were used. \ Among those 19 were the isoscalar $\eta $ and $\eta
^{\prime }$ mesons. Here we developed an approach motivated by some work of
Brayshaw \cite{bry} \ which introduces a constant symmetric but nondiagonal
mass matrix that couples isoscalar $q\bar{q}$ channels. The three parameters
introduced are adjusted (when possible) to fit the $\pi ^{0},~\eta $ and $%
\eta ^{\prime }$ meson masses and then used to predict accurately, at least
in Th2, the $SU(3)$ pseudoscalar mixing angle . \ Missing is an attempt to
connect this mass matrix to the compatibility condition between the
constraints $\mathcal{S}_{1}$ and $\mathcal{S}_{2}$.

A glance at Ref. \cite{yaes} \ shows that there are no shortages of attempts
to stake claims of which relativistic two-body truncation of the
Bethe-Salpeter equation is most successful. \ One of the purposes of \cite%
{crater2} was to clarify some guidelines and important benchmarks that such
equations should have when applied to any relativistic two-body problem be
it for QED bound states, QCD bound states, or nucleon-nucleon scattering. \
This clarification continues in this paper with a fairly detailed analysis
of the local quasipotential approach of Ebert, Faustov, and Galikin. \ This
approach was chosen among numerous others for two reasons: a) \ the close
connection between the minimal dynamical structure of the constraint
approach and the early work done on the local quasipotential approach by
Todorov and his coauthors and b) the extensive phenomenological studies by
the local quasipotential approach of EFG in meson spectral and decay studies.

The two quasipotential methods that we discussed have three weaknesses. \
Neither method uses a wave equation to uncover their respective
spin-dependent corrections. \ Rather, they use an on shell version of the
scattering amplitude and the quasipotential equation for the potential. \
However, as discussed in both papers, they each must make assumptions that
allow them to include expected off shell effects. \ In contrast, the TBDE of
constraint dynamics include automatically by their mathematically consistent
construction, off mass shell effects. \ The second weakness, is that both
approaches, when applied to QED, do not produce the correct hyperfine
structure. This, in our opinion, is a serious but easily correctable
drawback in both approaches. \ Let us be precise here about our concerns. In
both of these two quasipotential approaches\ to the relativistic QCD
potential model, if one turns off the confining interaction and replaces the
nonconfining vector potential by the Coulomb potential their resultant QED
spectrum will be incorrect whether computed perturbatively or numerically. \
This calls into question their QCD spectral results since the resultant
relativistic corrections which the omitted term $-A^{2}$ would have
contributed is of the same order as the spin-dependent corrections
(dependent on $A^{\prime }$ and $\nabla ^{2}A$) which, of course, are not
omitted. Even if that correction is made, the wave equation should be shown
to have (just as can be shown with the one-body Dirac equation) the same
spectral results whether treated perturbatively or nonperturbatively.\ \ In
addition, their modeling of the scalar interactions does not conform with
the approach of the classical and quantum field theories used by Crater, Van
Alstine, Yang, Sazdjian and Jolluli which supports our choice of vector
interaction structures\footnote{%
In particular, their approach does not include the quadratic structure $%
2m_{w}S+S^{2}$ implied by the those authors' approaches to scalar field
theories.}. \ The third weakness is the lack of agreement in the $%
m_{2}\rightarrow \infty $ limit with the Pauli-form of the one-body Dirac
equation. \ In particular, the lack of scalar Darwin terms in both
approaches and vector Darwin terms in the approach of \cite{isgr} is a
serious weakness.

The constraint approach has been tested against both classical and quantum
field theories for both scalar and vector interactions. \ In our
construction of the vector potential \cite{cra84}, \cite{cra87}, \cite%
{becker} three primary guideposts were used beyond that of a minimal
structure. \ The first is the use of just the barest (lowest order) input
from field theory, the nonrelativistic Coulomb potential in QED. \ 

The second is the gauge-like minimal coupling structure Eq. (\ref{gage}) of
the potential which Todorov postulated and was later confirmed in three
independent ways: \ a) Rizov, Todorov, and Aneva \cite{tod75} demonstrated
how the gauge structure (particularly the form$~(\varepsilon
_{w}-A)^{2}-m_{w}^{2}$ ) arises in perturbation theory at higher order than
the Born approximation \footnote{%
In particular, the quasipotential equation Eq. (\ref{quasi}) for the
potential to be used in Eq. (\ref{tod0}) must be iterated to second order ($%
V^{(2)}=T_{1}GT_{1}-T_{2}$). \ It is remarkable that this this gauge like
structure postulate Eq. (\ref{gage}) anticipates the systematic inclusion of
higher order terms by the quasipotential formalism.}, b) by a comparison of
the Fokker-Tetrode classical field theory $O(1/c^{2})$ expansion for the
Hamiltonian with the general quasipotential structure of $\mathbf{p}%
^{2}+\Phi (\mathbf{r,}w)=(\varepsilon _{w}^{2}-m_{w}^{2})~$\cite{fw} 
\footnote{%
In this, not only does the gauge like minimal structure $(\varepsilon
_{w}-A)^{2}$ appear to be a natural outgrowth of classical $O(1/c^{2})$
expansion to at least order $1/c^{4},$ but also a minimal scalar interaction
structure appears so that combined they yield $(\varepsilon
_{w}-A)^{2}-(m_{w}+S)^{2}$ , again, at least through order $1/c^{4}$.}, and
c) Jollouli and Sazdjian \cite{saz97} who found a similar structure from
nonperturbative quantum field theoretic arguments both for scalar and vector
interactions. \ These three arguments demonstrate that the Born
approximation structures (particularly the first term of (\ref{qua}) used to
model the QCD potentials in \cite{isgr}) cannot possibly yield that gauge
structure and thus cannot yield the correct positronium spectrum when
applied to QED. \ The higher order structures in \cite{saz97} and \cite%
{tod75} as well as the nonlinear hyperbolic structures in Eqs. (\ref{hyp1}, %
\ref{hyp2}, \ref{hyp3}, \ref{one}-\ref{three}) argue that the Born structure
of the first term in (\ref{qua}) are insufficient and must be supplemented
by other invariant couplings. \ The papers by Sazdjian and collaborators 
\cite{saz86}, \cite{saz94}, \cite{saz97} demonstrate this explicitly,
showing that pseudovector coupling is essential if done perturbatively in
order to get the Dirac equation into an external field form in which the
minimal structure can be demonstrated. \ The work in \cite{jmath}, \cite%
{long} as well as by Sazdjian and collaborators show that the vector
coupling (see Eq. (\ref{vecc}) ) alone will, when placed in the nonlinear
context of the hyperbolic parametrization given in Eqs. (\ref{hyp1},\ref%
{hyp2}), yield that external field form in which the minimal structure can
be demonstrated. \ 

The third guidepost is the use of the relativistic reduced mass $m_{w}$ and
energy $\varepsilon _{w}$ (see Eqs. (\ref{mw}) and (\ref{ew})). \ \ In
addition to the discussions given in \cite{tod}, their appearance in the
forms $2m_{w}S+2\varepsilon _{w}A$ as part of the minimal structure is also
dictated by the same \ field theory mechanisms discussed in item b) and c)
above. \ \ Note that even though in \ Eq. (\ref{tod0}) there is the
appearance of another relativistic reduced mass ($\varepsilon
_{1}\varepsilon _{2}/w$), in working out the quasipotential using the
spinors as in Eq. (\ref{riz}) that reduced mass is replaced by $m_{w}$ in
the scalar and $\varepsilon _{w}$ in the vector case. \ This replacement
does not appear in the work of \cite{rusger}.

So, while the works of Ebert, Faustov, and Galkin, and Godfrey and Isgur are
quite impressive in terms of spectral and decay agreements, it would be of
value for adherents of those approaches to consider the criticisms presented
in this paper, related to the ability of those equations to reproduce the
static limit Dirac equations structures (which would include the $-A^{2}$
term) and general short range structure related to the $-A^{2}$ gauge term.
\ These criticisms are not about their choice of QCD inspired potentials,
but rather about how their relativistic wave equations translate the physics
of those potentials into spectral results. \ These thus concern the impact
on their QCD meson spectral results those two approaches would have based on
their field theory connections to QED bound states .

Finally a few words about future direction lines of research related to this
paper. The approach given in our paper views the meson as a two-body bound
state in a first quantized formalism. In place of the nonrelativistic Schr%
\"{o}dinger equation for two interacting particles, we use the TBDE of
constraint dynamics. \ There are systematic corrections that should follow
the completion of this first step which would end with the inclusion of
covariant pseudoscalar, pseudovector, and tensor interactions, in addition
to the scalar and vector interactions we have included in this paper. \
First is a second quantized version the TBDE similar that what has been
accomplished for nonrelativistic second quantized formalisms by \ the
Cornell group \cite{cornell}, Tornqvist \cite{trv}, and more recently by
Barnes and Swanson \cite{bsw}. \ The latter includes a microscopic theory 
\cite{bsa} of the $^{3}P_{0}$ model which includes pair production. \ The
aim, as in their recent paper, would be to gain a measure of the effects of
two-body meson decays on the observed rest mass of the decaying meson. \
Beyond that would be many body formalisms which view the meson more
generally as a linear combination of $q\bar{q}+q\bar{q}g+...~$

The primary weakness in our results of this paper are: 1) the radial and
orbital excitations of the old meson spectroscopy; 2) the less than good
hyperfine splittings of the $c\bar{c}$ and $b\bar{b}$ families compared with
the good splitting results we obtained with all the other families,
including the lightest and most highly relativistic $u\bar{d}~$; 3) \ the
failure to account for the light ($^{3}P_{0})$ scalar mesons; 4) Failure to
reproduce the square root Gell-Mann-Oakes-Renner relation. \ It is too early
to say if the completion of the first quantization program will rectify any
of these problems. We point out, however, that should the weakness of our
model, relating to the radial excitations of the light $q\bar{q}$ bound
states be substantially improved by including pseudoscalar, pseudovector,
and tensor interactions, this would offer the opportunity of allowing the
hypothesis discussed in footnote 13 to be actively considered. \ Whether
including these other interactions lead to substantial improvements or not,
it will be essential to follow in parallel with our relativistic formalism,
the second quantized nonrelativistic formalisms developed earlier.

\appendix\setcounter{equation}{0} \ \renewcommand{\theequation}{%
\Alph{section}.\arabic{equation}}

\section{Relativistic Schr\"{o}dinger Equation Details}

\subsection{\protect\bigskip\ Connections Between the TBDE and Eq. (  
\protect\ref{57}) and Forms for $\tilde{A}_{i}^{\protect\mu },\tilde{S}_{i}$
in Terms of the Invariants $A(r),V(r),$ and $S(r)$}

Here we present an outline of some details of Eq. (\ref{tbde}) and its
Pauli-Schr\"{o}dinger reduction given in full elsewhere (see \cite%
{cra87,jmath,long,liu}). Each of the two Dirac equations in (\ref{tbde}) has
a form similar to a single particle Dirac equation in an external\
four-vector and scalar potential but here acting on a 16 component wave
function $\Psi $ which is the product of an external part (being a plane
wave eigenstate of $P)~$multiplying the internal wave function $\psi $ 
\begin{equation}
\psi =%
\begin{bmatrix}
\psi _{1} \\ 
\psi _{2} \\ 
\psi _{3} \\ 
\psi _{4}%
\end{bmatrix}%
.
\end{equation}%
The four $\psi _{i}$ are each four-component spinor wave functions. To
obtain the actual general spin-dependent forms of those $\tilde{A}_{i}^{\mu
},\tilde{S}_{i}$ potentials which were required by the compatibility
condition $[\mathcal{S}_{1},\mathcal{S}_{2}]\psi =0$ was a most perplexing
problem, involving the discovery of underlying supersymmetries in the case
of scalar and timelike vector interactions\cite{cra82}, \cite{cra87}. \
Extending those external potential forms to more general covariant
interactions necessitated an entirely different approach leading to what is
called the hyperbolic form of the TBDE. \ The most general hyperbolic form
for compatible TBDE is 
\begin{eqnarray}
\mathcal{S}_{1}\psi  &=&(\cosh (\Delta )\mathbf{S}_{1}+\sinh (\Delta )%
\mathbf{S}_{2})\psi =0\mathrm{,}  \notag \\
\mathcal{S}_{2}\psi  &=&(\cosh (\Delta )\mathbf{S}_{2}+\sinh (\Delta )%
\mathbf{S}_{1})\psi =0,  \label{hyp1}
\end{eqnarray}%
where $\Delta $ represents any invariant interaction singly or in
combination. \ It has a matrix structure in addition to coordinate
dependence. Depending on that matrix structure we have either vector, scalar
or more general tensor interactions \cite{jmath}. The operators $\mathbf{S}%
_{1}$ and $\mathbf{S}_{2}$ are auxiliary constraints satisfying 
\begin{eqnarray}
\mathbf{S}_{1}\psi  &\equiv &(\mathcal{S}_{10}\cosh (\Delta )+\mathcal{S}%
_{20}\sinh (\Delta )~)\psi =0,  \notag \\
\mathbf{S}_{2}\psi  &\equiv &(\mathcal{S}_{20}\cosh (\Delta )+\mathcal{S}%
_{10}\sinh (\Delta )~)\psi =0,  \label{hyp2}
\end{eqnarray}%
in which the\ $\mathcal{S}_{i0}$ are the free Dirac operators 
\begin{equation}
\mathcal{S}_{i0}=\frac{i}{\sqrt{2}}\gamma _{5i}(\gamma _{i}\cdot
p_{i}+m_{i}).  \label{es0}
\end{equation}%
This, in turn leads to the two compatibility conditions \cite%
{cww,jmath,saz86} 
\begin{equation}
\lbrack \mathcal{S}_{1},\mathcal{S}_{2}]\psi =0,
\end{equation}%
and 
\begin{equation}
\lbrack \mathbf{S}_{1},\mathbf{S}_{2}]\psi =0,
\end{equation}%
provided that $\ \Delta (x)=\Delta (x_{\perp }).$ These compatibility
conditions do not restrict the gamma matrix structure of $\Delta $. \ That
matrix structure is determined by the type of vertex-vertex structure we
wish to incorporate in the interaction. \ \ The three types of invariant
interactions $\Delta $ that we use in this paper are%
\begin{eqnarray}
\Delta _{\mathcal{L}}(x_{\perp }) &=&-1_{1}1_{2}\frac{\mathcal{L}(x_{\perp })%
}{2}\mathcal{O}_{1},\ \mathcal{O}_{1}=-\gamma _{51}\gamma _{52},~~~\text{%
scalar}\mathrm{,}  \notag \\
\Delta _{\mathcal{J}}(x_{\perp }) &=&\beta _{1}\beta _{2}\frac{\mathcal{J}%
(x_{\perp })}{2}\mathcal{O}_{1},~~~\text{timelike\ vector}\mathrm{,}  \notag
\\
\Delta _{\mathcal{G}}(x_{\perp }) &=&\gamma _{1\perp }\cdot \gamma _{2\perp }%
\frac{\mathcal{G}(x_{\perp })}{2}\mathcal{O}_{1},~~\text{spacelike\ vector}{,%
}  \label{hyp3}
\end{eqnarray}%
where%
\begin{eqnarray}
\gamma _{i\perp }^{\mu } &=&(\eta ^{\mu \nu }+\hat{P}^{\mu }\hat{P}^{\nu
})\gamma _{\nu i},  \notag \\
\gamma _{5i} &=&\gamma _{i}^{0}\gamma _{i}^{1}\gamma _{i}^{2}\gamma _{i}^{3},
\notag \\
\beta _{i} &=&-\gamma _{i}\cdot \hat{P},~\ i=1,2.  \label{beta}
\end{eqnarray}%
$~$For For general independent scalar, timelike vector, and spacelike vector
interactions we have%
\begin{equation}
\Delta (x_{\perp })=\Delta _{\mathcal{L}}+\Delta _{\mathcal{J}}+\Delta _{%
\mathcal{G}}.  \label{Delta}
\end{equation}%
The special case of an electromagnetic-like interaction (in the Feynman
gauge) corresponds to $\mathcal{J}=-\mathcal{G}$ or 
\begin{eqnarray}
\Delta _{\mathcal{J}}+\Delta _{\mathcal{G}} &\equiv &\Delta _{\mathcal{EM}%
}=(-\gamma _{1}\cdot \hat{P}\gamma _{2}\cdot \hat{P}+\gamma _{1\perp }\cdot
\gamma _{2\perp })\frac{\mathcal{G}(x_{\perp })}{2}\mathcal{O}_{1}  \notag \\
&=&\gamma _{1}\cdot \gamma _{2}\frac{\mathcal{G}(x_{\perp })}{2}\mathcal{O}%
_{1}.  \label{vecc}
\end{eqnarray}%
Our Th1 corresponds to a scalar and electromagnetic interaction, 
\begin{equation}
\Delta (x_{\perp })=\Delta _{\mathcal{L}}+\Delta _{\mathcal{EM}}.
\label{A10}
\end{equation}%
Our Th2 corresponds to a modification of the timelike portion of $\Delta _{%
\mathcal{EM}}$ to 
\begin{eqnarray}
\Delta (x_{\perp }) &=&\Delta _{\mathcal{L}}+\Delta _{\mathcal{J}}+\Delta _{%
\mathcal{G}}=(-1_{1}1_{2}\mathcal{L}(x_{\perp })+\beta _{1}\beta _{2}%
\mathcal{J}(x_{\perp })+\gamma _{1\perp }\cdot \gamma _{2\perp }\mathcal{G}%
(x_{\perp }))\frac{\mathcal{O}_{1}}{2},  \notag \\
\mathcal{J}\mathcal{\neq -G} &&.  \label{th2}
\end{eqnarray}%
\ This leads to\footnote{%
In short, \ one inserts Eq. (\ref{hyp2}) into (\ref{hyp1}) and brings the
free Dirac operator (\ref{es0}) to the right of the matrix hyperbolic
functions. \ Using commutators and $\cosh ^{2}\Delta -\sinh ^{2}\Delta =1$
one arrives at Eq. (\ref{extd}). \ The structure of these equations are very
much the same as that of a Dirac equation for each of the two particles,
with $M_{i}$ and $E_{i}$ playing the roles that $m+S$ and $\varepsilon -A$
do in the single particle Dirac equation (\ref{obde}). Over and above the
usual kinetic part, the spin-dependent modifications involving $G\mathcal{P}%
_{i}$ and the last set of derivative terms are two-body recoil effects
essential for the compatibility (consistency) of the two equations} \cite%
{jmath,long} $\left( \partial _{\mu }=\partial /\partial x^{\mu }\right) $%
\begin{align}
\mathcal{S}_{1}\psi & =\big(-G\beta _{1}\Sigma _{1}\cdot \mathcal{P}%
_{2}+E_{1}\beta _{1}\gamma _{51}+M_{1}\gamma _{51}-G\frac{i}{2}\Sigma
_{2}\cdot \partial (\mathcal{L}\beta _{2}\mathcal{-J}\beta _{1})\gamma
_{51}\gamma _{52}\big)\psi =0,  \notag \\
\mathcal{S}_{2}\psi & =\big(G\beta _{2}\Sigma _{2}\cdot \mathcal{P}%
_{1}+E_{2}\beta _{2}\gamma _{52}+M_{2}\gamma _{52}+G\frac{i}{2}\Sigma
_{1}\cdot \partial (\mathcal{L}\beta _{1}\mathcal{-J}\beta _{2})\gamma
_{51}\gamma _{52}\big)\psi =0,  \label{extd}
\end{align}%
\ \ with 
\begin{eqnarray}
G &=&\exp \mathcal{G},  \notag \\
\mathcal{P}_{i} &\equiv &p_{\perp }-\frac{i}{2}\Sigma _{i}\cdot \partial 
\mathcal{G}\Sigma _{i}.  \label{d2}
\end{eqnarray}%
The connections between what we call the vertex invariants $\mathcal{L},%
\mathcal{J},\mathcal{G}$ and the mass and energy potentials $M_{i},E_{i}$
are found to be 
\begin{eqnarray}
M_{1} &=&m_{1}\ \cosh \mathcal{L}\ +m_{2}\sinh \mathcal{L},  \notag \\
M_{2} &=&m_{2}\ \cosh \mathcal{L}\ +m_{1}\ \sinh \mathcal{L},  \notag \\
E_{1} &=&\varepsilon _{1}\ \cosh \mathcal{J}\ +\varepsilon _{2}\sinh 
\mathcal{J},  \notag \\
E_{2} &=&\varepsilon _{2}\ \cosh \mathcal{J}+\varepsilon _{1}\sinh \mathcal{J%
}.  \label{d2b}
\end{eqnarray}%
Eq. (\ref{extd}) depends on the standard Pauli-Dirac representation of gamma
matrices in block forms (see Eq.\ (2.28) in \cite{crater2} for their
explicit forms) and where\footnote{%
\ Just as $x^{\mu }$ is a four vector, so are $\gamma ^{\mu }$ (in the sense
of Dirac) and $P^{\mu }.$ \ Thus, the matrix structures of the time-like and
space-like interactions in Eq. (\ref{hyp3}) are $\gamma _{1}^{0}\gamma
_{2}^{0}$ and $\mathbf{\gamma }_{1}\cdot \mathbf{\gamma }_{2}$ only in the
c.m. system due to the fact that from Eq. (\ref{beta}), $\beta _{i}=\gamma
_{i}^{0}$ only in the c.m. frame. \ Likewise, $\Sigma _{i}^{\mu }=(0,\mathbf{%
\Sigma )}$ only in the c.m. frame just as is $x_{\perp }^{\mu }=(0,\mathbf{r)%
}$ in that frame only.} 
\begin{equation}
\Sigma _{i}=\gamma _{5i}\beta _{i}\gamma _{\perp i}.  \label{d3}
\end{equation}%
Comparing Eq. (\ref{extd}) with Eq. (\ref{tbde}) we find that the
spin-dependent vector interactions of Eq. (\ref{tbde}) are \cite%
{cra87,becker}%
\begin{align}
\tilde{A}_{1}^{\mu }& =\big((\varepsilon _{1}-E_{1})-i\frac{G}{2}\left(
\gamma _{2}\cdot \partial \mathcal{J}\right) \gamma _{2}\cdot \hat{P}\big )%
\hat{P}^{\mu }+(1-G)p_{\perp }^{\mu }-\frac{i}{2}\partial G\cdot \gamma
_{2}\gamma _{2\perp }^{\mu },  \notag \\
A_{2}^{\mu }& =\big((\varepsilon _{2}-E_{2})+i\frac{G}{2}\left( \gamma
_{1}\cdot \partial \mathcal{J}\right) )\gamma _{1}\cdot \hat{P}\big )\hat{P}%
^{\mu }-(1-G)p_{\perp }^{\mu }+\frac{i}{2}\partial G\cdot \gamma _{1}\gamma
_{1\perp }^{\mu }.  \label{vecp}
\end{align}%
Note that the first portion of the vector potentials is timelike (parallel
to $\hat{P}^{\mu })$ while the next portion is spacelike (perpendicular to $%
\hat{P}^{\mu })$. The spin-dependent scalar potentials $\tilde{S}_{i}$ are 
\begin{align}
\tilde{S}_{1}& =M_{1}-m_{1}-\frac{i}{2}G\gamma _{2}\cdot \partial \mathcal{L}%
,  \notag \\
\tilde{S}_{2}& =M_{2}-m_{2}+\frac{i}{2}G\gamma _{1}\cdot {\partial }\mathcal{%
L}{.}  \label{scalp}
\end{align}%
Eq. (\ref{vecp}) simplifies to%
\begin{align}
\tilde{A}_{1}^{\mu }& =\big((\varepsilon _{1}-E_{1})\big )\hat{P}^{\mu
}+(1-G)p_{\perp }^{\mu }-\frac{i}{2}\partial G\cdot \gamma _{2}\gamma
_{2}^{\mu },  \notag \\
A_{2}^{\mu }& =\big((\varepsilon _{2}-E_{2})\big )\hat{P}^{\mu
}-(1-G)p_{\perp }^{\mu }+\frac{i}{2}\partial G\cdot \gamma _{1}\gamma
_{1}^{\mu },
\end{align}%
for electromagnetic-like interactions. \ 

We have chosen a parametrization for $\mathcal{L},~\mathcal{J},$ and $%
\mathcal{G}$ that takes advantage of the Todorov effective external
potential forms and at the same time will display the correct static limit
form for the Pauli reduction (see Eq. (\ref{571})). \ The choice for these
parametrizations is fixeded due to the fact that for classical \cite{fw} or
quantum field theories \cite{saz97} for separate scalar and vector
interactions the spin independent part of the quasipotential $\Phi _{w}~$
involves the difference of squares of the invariant mass and energy
potentials ($M_{i}$ and $E_{i}$ respectively)%
\begin{equation}
M_{i}^{2}=m_{i}^{2}+2m_{w}S+S^{2};\ E_{i}^{2}=\varepsilon
_{i}^{2}-2\varepsilon _{w}V+V^{2},  \label{kg1}
\end{equation}%
so that \ 
\begin{equation}
M_{i}^{2}-E_{i}^{2}=2m_{w}S+S^{2}+2\varepsilon _{w}V-V^{2}-b^{2}(w).
\label{kg}
\end{equation}

Strictly speaking, the forms in Eq.\ (\ref{kg1}) and Eq. (\ref{kg}) are for
scalar and timelike vector interactions. Eqs. (\ref{tbde}) and (\ref{extd})
involve combined scalar, electromagnetic-like, and separate timelike vector
interactions. Without the separate timelike interactions this amounts to
working in the Feynman gauge with the simplest relation between space- and
timelike parts, (see Eqs. (\ref{vecc}), (\ref{A10}), and \cite{cra88,crater2}%
). \ In the general case the\ mass and energy potentials in place of Eq.\ (%
\ref{kg1}) are respectively 
\begin{eqnarray}
M_{i}^{2} &=&m_{i}^{2}+\exp (2\mathcal{G)(}2m_{w}S\mathcal{+}S^{2}),
\label{one} \\
E_{i}^{2} &=&\exp (2\mathcal{G(A))(}\left( \varepsilon
_{i}-A)^{2}-2\varepsilon _{w}V+V^{2}\right) ,  \label{two}
\end{eqnarray}%
so that from Eq. (\ref{d2b}), 
\begin{eqnarray}
\exp (\mathcal{L}) &=&\exp (\mathcal{L}(S,A))=\frac{\sqrt{m_{1}^{2}+\exp (2%
\mathcal{G)(}2m_{w}S\mathcal{+}S^{2})}+\sqrt{m_{2}^{2}+\exp (2\mathcal{G)(}%
2m_{w}S\mathcal{+}S^{2})}}{m_{1}+m_{2}},\   \label{1.5} \\
\exp (\mathcal{J}) &=&\exp (\mathcal{J}(V,A))=\exp (\mathcal{G)}\frac{\sqrt{%
(\varepsilon _{1}-A)^{2}-2\varepsilon _{w}V+V^{2}}+\sqrt{(\varepsilon
_{2}-A)^{2}-2\varepsilon _{w}V+V^{2}}}{\varepsilon _{1}+\varepsilon _{2}}, 
\notag
\end{eqnarray}%
with 
\begin{equation}
\exp (2\mathcal{G(}A\mathcal{))=}\frac{1}{(1-2A/w)}\equiv G^{2}.
\label{three}
\end{equation}

Below we present, the consequent connections to the invariant interaction
functions $\ A,V,$and $S$.

\bigskip a) \ In the case of electromagnetic interactions $(V=0$) with
scalar confinement (Th1), we have 
\begin{eqnarray}
\mathcal{J} &=&-\mathcal{G=}\frac{1}{2}\log (1-2A/w)=\log \frac{E_{1}+E_{2}}{%
w},  \notag \\
E_{i}^{2} &=&\exp (2\mathcal{G)(\varepsilon }_{i}-A)^{2},  \notag \\
M_{i}^{2} &=&m_{i}^{2}+\exp (2\mathcal{G)}\left( 2m_{w}S+S^{2}\right) ,
\end{eqnarray}%
and the spin-independent ($SI)$ minimal coupling comes from%
\begin{eqnarray}
\exp (2\mathcal{G)}p^{2}+M_{i}^{2}-E_{i}^{2} &\rightarrow &p^{2}+\exp (-2%
\mathcal{G)(}M_{i}^{2}-E_{i}^{2})  \notag \\
&=&p^{2}+\Phi _{SI}-b^{2}
\end{eqnarray}%
and appears as%
\begin{equation}
\Phi _{SI}-b^{2}=2m_{w}S+S^{2}+m_{i}^{2}(1-2A/w)-(\varepsilon
_{i}-A)^{2}=2m_{w}S+S^{2}+2\varepsilon _{w}A-A^{2}-b^{2}.
\end{equation}

\QTP{Body Math}
b) In the case of pure timelike vector interactions and scalar interactions
(with no electromagnetic-like interactions this model is not appropriate for
meson spectroscopy) we have%
\begin{eqnarray}
\mathcal{J} &=&\mathcal{J}_{0}\equiv \log \frac{E_{10}+E_{20}}{w},  \notag \\
E_{i0}^{2} &\equiv &\varepsilon _{i}^{2}-2\varepsilon _{w}V+V^{2},  \notag \\
\mathcal{G} &\mathcal{=}&0,
\end{eqnarray}%
and the spin-independent minimal coupling appears as%
\begin{equation}
\Phi _{SI}=2m_{w}S+S^{2}+2\varepsilon _{w}V-V^{2}.
\end{equation}

\QTP{Body Math}
c) When we include independent timelike and electromagnetic-like
simultaneously together with scalar interactions (Th2) then we have%
\begin{eqnarray}
-\mathcal{G} &\mathcal{=}&\frac{1}{2}\log (1-2A/w),  \notag \\
\mathcal{J} &=&\log \frac{E_{1}+E_{2}}{w},  \notag \\
E_{i}^{2} &=&\exp (2\mathcal{G)(}\left( \varepsilon _{i}-A)^{2}-2\varepsilon
_{w}V+V^{2}\right) =\exp (2\mathcal{G)(}E_{i0}^{2}-2\varepsilon _{i}A+A^{2})
\end{eqnarray}%
and the spin-independent minimal coupling appears like%
\begin{equation}
\Phi _{SI}=2m_{w}S+S^{2}+2\varepsilon _{w}A-A^{2}+2\varepsilon _{w}V-V^{2}.
\end{equation}

\subsection{\qquad Details on Eq. (\protect\ref{57})}

The Klein-Gordon like potential energy terms appearing at the beginning of
the Pauli form (\ref{57}) arise from%
\begin{equation*}
M_{i}^{2}-E_{i}^{2}=\exp (2\mathcal{G)[}2m_{w}S+S^{2}+2\varepsilon
_{w}A-A^{2}+2\varepsilon _{w}V-V^{2}-b^{2}(w)].
\end{equation*}%
To obtain the symbolic Pauli form of Eq.\ (\ref{schlike}) and the subsequent
detailed form in Eq. (\ref{57}) involves steps similar to those used in the
Pauli reduction of the single particle Dirac equation \cite{yoon} but with
the combinations $\phi _{\pm }=\psi _{1}\pm \psi _{4}$ and $\chi _{\pm
}=\psi _{2}\pm \psi _{3}$ instead of the individual single particle wave
function. This reduces the Pauli forms to 4 uncoupled 4 component
relativistic Schr\"{o}dinger equations \cite{saz94,cra94,long,crater2,liu}.
\ We work in the c.m. frame in which $\hat{P}=(1,\mathbf{0)}$ and $\hat{r}%
=(0,\mathbf{\hat{r}).}$ The final four-component wave functions $\psi _{\pm
},\eta _{\pm }$ \ that appear in Eq. (\ref{57}) are defined by \cite{liu} 
\begin{align}
\phi _{\pm }& =\exp (\mathcal{F}+\mathcal{K}\boldsymbol{\sigma }_{1}\mathbf{%
\cdot \hat{r}}\boldsymbol{\sigma }_{2}\mathbf{\cdot \hat{r}})\psi _{\pm
}=(\exp \mathcal{F})(\cosh \mathcal{K}+\sinh \mathcal{K}\boldsymbol{\sigma }%
_{1}\mathbf{\cdot \hat{r}}\boldsymbol{\sigma }_{2}\mathbf{\cdot \hat{r}}%
)\psi _{\pm },  \notag \\
\chi _{\pm }& =\exp (\mathcal{F}+\mathcal{K}\boldsymbol{\sigma }_{1}\mathbf{%
\cdot \hat{r}}\boldsymbol{\sigma }_{2}\mathbf{\cdot \hat{r}})\eta _{\pm
}=(\exp \mathcal{F})(\cosh \mathcal{K}+\sinh \mathcal{K}\boldsymbol{\sigma }%
_{1}\mathbf{\cdot \hat{r}}\boldsymbol{\sigma }_{2}\mathbf{\cdot \hat{r}}%
)\eta _{\pm },  \label{fk}
\end{align}%
in which 
\begin{align}
\mathcal{F}& =\frac{1}{2}\log \frac{\mathcal{D}}{\varepsilon
_{2}m_{1}+\varepsilon _{1}m_{2}}-\mathcal{G},  \notag \\
\mathcal{D}& \mathcal{=}E_{2}M_{1}+E_{1}M_{2},  \notag \\
\mathcal{K}& =\frac{(\mathcal{L}-\mathcal{J})}{2}.  \label{kf}
\end{align}%
In analogy to what occurs in the decoupled form of the Schr\"{o}dinger
equation for the individual single particle wave function, this substitution
has the convenient property that in the resultant bound state equation, the
coefficients of the first order relative momentum terms vanish.

\ Using the results in \cite{liu} and \cite{yoon} we obtain for the general
case of unequal masses the relativistic Schr\"{o}dinger equation (\ref{57})
that is a detailed c.m.\ form of Eq.\ (\ref{schlike}). In that equation we
have introduced the abbreviations

\begin{align}
\Phi _{D}& =-\frac{2(\mathcal{F}^{\prime }+1/r)(\cosh 2\mathcal{K}-1)}{r}+%
\mathcal{F}^{\prime 2}+\mathcal{K}^{\prime 2}+\frac{2\mathcal{K}^{\prime
}\sinh 2\mathcal{K}}{r}-\mathbf{\nabla }^{2}\mathcal{F}+m(r),  \notag \\
\Phi _{SO}& =-\frac{\mathcal{F}^{\prime }}{r}-\frac{(\mathcal{F}^{\prime
}+1/r)(\cosh 2\mathcal{K}-1)}{r}+\frac{\mathcal{K}^{\prime }\sinh 2\mathcal{K%
}}{r},  \notag \\
\Phi _{SOD}& =(l^{\prime }\cosh 2\mathcal{K}-q^{\prime }\sinh 2\mathcal{K}),
\notag \\
\Phi _{SOX}& =(q^{\prime }\cosh 2\mathcal{K}-l^{\prime }\sinh 2\mathcal{K}),
\notag \\
\Phi _{SS}& =k(r)+\frac{2\mathcal{K}^{\prime }\sinh 2\mathcal{K}}{3r}-\frac{%
2(\mathcal{F}^{\prime }+1/r)(\cosh 2\mathcal{K}-1)}{3r}+\frac{2\mathcal{F}%
^{\prime }\mathcal{K}^{\prime }}{3}-\frac{\mathbf{\nabla }^{2}\mathcal{K}}{3}%
,  \notag \\
\Phi _{T}& =\frac{1}{3}[n(r)+\frac{(3\mathcal{F}^{\prime }-\mathcal{K}%
^{\prime }+3/r)\sinh 2\mathcal{K}}{r}+\frac{(\mathcal{F}^{\prime }-3\mathcal{%
K}^{\prime }+1/r)(\cosh 2\mathcal{K}-1)}{r}+2\mathcal{F}^{\prime }\mathcal{K}%
^{\prime }-\mathbf{\nabla }^{2}\mathcal{K}],  \notag \\
\Phi _{SOT}& =-\mathcal{K}^{\prime }\frac{\cosh 2\mathcal{K}-1}{r}-\frac{%
\mathcal{K}^{\prime }}{r}+\frac{(\mathcal{F}^{\prime }+1/r)\sinh 2\mathcal{K}%
}{r},  \label{54}
\end{align}%
where%
\begin{align}
k(r)& =\frac{1}{3}\nabla ^{2}(\mathcal{K}+\mathcal{G)}-\frac{2\mathcal{F}%
^{\prime }(\mathcal{G}^{\prime }+\mathcal{K}^{\prime })}{3}\mathcal{-}\frac{1%
}{2}\mathcal{G}^{\prime 2},  \notag \\
n(r)& =\frac{1}{3}[\nabla ^{2}\mathcal{K}-\frac{1}{2}\nabla ^{2}\mathcal{G}+%
\frac{3(\mathcal{G}^{\prime }-2\mathcal{K}^{\prime })}{2r}+\mathcal{F}%
^{\prime }(\mathcal{G}^{\prime }-2\mathcal{K}^{\prime })],  \notag \\
m(r)& =-\frac{1}{2}\nabla ^{2}\mathcal{G+}\frac{3}{4}\mathcal{G}^{\prime 2}+%
\mathcal{G}^{\prime }\mathcal{F}^{\prime }-\mathcal{K}^{\prime 2},
\label{knm}
\end{align}%
and

\begin{align}
l^{\prime }(r)& =-\frac{1}{2r}\frac{E_{2}M_{2}-E_{1}M_{1}}{%
E_{2}M_{1}+E_{1}M_{2}}(\mathcal{L}^{\prime }+\mathcal{J}^{\prime }),  \notag
\\
q^{\prime }(r)& =\frac{1}{2r}\frac{E_{2}M_{1}-E_{1}M_{2}}{%
E_{2}M_{1}+E_{1}M_{2}}(\mathcal{L}^{\prime }+\mathcal{J}^{\prime }).
\label{A32}
\end{align}%
(The prime symbol stands for $d/dr,$ and the explicit forms of the
derivatives are given in Eq. (\ref{der})$\,$below). For $L=J$ states, the
hyperbolic terms cancel and the spin-orbit difference terms in general
produce spin mixing except for equal masses or $J=0$. For ease of use we
have listed in Appendix A3 the explicit forms that appear in the above $\Phi 
$s in Eqs. (\ref{54}) - (\ref{knm}) in terms of the general invariant
potentials $A(r),V(r),$ and $S(r).~$\ The radial components of Eq. (\ref{57}%
) are given in Appendix B.

\subsection{Explicit Expressions for terms in the Relativistic Schr\"{o}%
dinger Equation (\protect\ref{57}) from $A(r),V(r)$ and $S(r)$}

Given the functions $A(r)$ , $V(r),$ and $S(r)$ for the interaction, users
of the relativistic Schr\"{o}dinger equation (\ref{57}) will find it
convenient to have an explicit expression in an order that would be useful
for programing the terms in the associated equation (\ref{54})-(\ref{A32}).
We use the definitions above given in Eqs.\ (\ref{one} )-(\ref{three}), and (%
\ref{kf}). In order that the terms in Eq.\ (\ref{54}) be reduced to
expressions involving just $A(r),V(r),~S(r)$ and their derivatives, we list
the following formulae 
\begin{eqnarray}
\mathcal{F}^{\prime } &=&\frac{(\mathcal{L}^{\prime }+\mathcal{J}^{\prime
})(E_{2}M_{2}+E_{1}M_{1})}{2(E_{2}M_{1}+E_{1}M_{2})}-\mathcal{G}^{\prime }, 
\notag \\
\mathcal{G}^{\prime } &=&\frac{A^{\prime }}{w-2A},  \notag \\
\mathcal{L}^{\prime } &=&\frac{M_{1}^{\prime }}{M_{2}}=\frac{M_{2}^{\prime }%
}{M_{1}}=\frac{w}{M_{1}M_{2}}\left( \frac{S^{\prime }(m_{w}+S)}{w-2A}+\frac{%
(2m_{w}S+S^{2})A^{\prime }}{(w-2A)^{2}}\right) ,  \notag \\
\mathcal{J}^{\prime } &=&\frac{E_{1}^{\prime }}{E_{2}}=\frac{E_{2}^{\prime }%
}{E_{1}}=-\frac{(\mathcal{G}^{\prime }[(\varepsilon _{1}-A)(\varepsilon
_{2}-A)+2\varepsilon _{w}V-V^{2}]+(\varepsilon _{w}-V)V^{\prime })}{%
E_{1}E_{2}(w-2A)/w},  \notag \\
.\mathcal{K}^{\prime } &=&\frac{(\mathcal{L}^{\prime }-\mathcal{J}^{\prime })%
}{2}.  \label{der}
\end{eqnarray}%
Also needed are 
\begin{eqnarray}
\cosh 2\mathcal{K} &=&\frac{1}{2}\left( \frac{(\varepsilon _{1}+\varepsilon
_{2})(M_{1}+M_{2})}{(m_{1}+m_{2})(E_{1}+E_{2})}+\frac{%
(m_{1}+m_{2})(E_{1}+E_{2})}{(\varepsilon _{1}+\varepsilon _{2})(M_{1}+M_{2})}%
\right) ,  \notag \\
\sinh 2\mathcal{K} &=&\frac{1}{2}\left( \frac{(\varepsilon _{1}+\varepsilon
_{2})(M_{1}+M_{2})}{(m_{1}+m_{2})(E_{1}+E_{2})}-\frac{%
(m_{1}+m_{2})(E_{1}+E_{2})}{(\varepsilon _{1}+\varepsilon _{2})(M_{1}+M_{2})}%
\right) ,
\end{eqnarray}%
and 
\begin{eqnarray}
\mathbf{\nabla }^{2}\mathcal{F} &=&\frac{(\mathbf{\nabla }^{2}\mathcal{L}+%
\mathbf{\nabla }^{2}\mathcal{J})(E_{2}M_{2}+E_{1}M_{1})}{%
2(E_{2}M_{1}+E_{1}M_{2})}-(\mathcal{L}^{\prime }+\mathcal{J}^{\prime })^{2}%
\frac{(m_{1}^{2}-m_{2}^{2})^{2}}{2\left( E_{2}M_{1}+E_{1}M_{2}\right) ^{2}}-%
\mathbf{\nabla }^{2}\mathcal{G},  \notag \\
\mathbf{\nabla }^{2}\mathcal{L} &=&\frac{-\mathcal{L}^{\prime
2}(M_{1}^{2}+M_{2}^{2})}{M_{1}M_{2}}  \notag \\
&&+\frac{w}{M_{1}M_{2}}\left( \frac{\mathbf{\nabla }^{2}S(m_{w}+S)+S^{\prime
2}}{w-2A}+\frac{4S^{\prime }(m_{w}+S)A^{\prime }+(2m_{w}S+S^{2})\mathbf{%
\nabla }^{2}A}{(w-2A)^{2}}+\frac{4(2m_{w}S+S^{2})A^{\prime 2}}{(w-2A)^{3}}%
\right) ,  \notag \\
\mathbf{\nabla }^{2}\mathcal{J} &\mathcal{=}&\mathcal{-[(}\frac{%
E_{1}^{2}+E_{2}^{2}}{E_{1}E_{2}}\mathcal{)J}^{\prime }-2\mathcal{G}^{\prime
}]\mathcal{J}^{\prime }-\frac{\exp (2\mathcal{G)}}{E_{1}E_{2}}\{\nabla ^{2}%
\mathcal{G}[(\varepsilon _{1}-A)(\varepsilon _{2}-A)+2\varepsilon
_{w}V-V^{2}]+(\varepsilon _{w}-V)\mathbf{\nabla }^{2}V  \notag \\
&&-\mathcal{G}^{\prime 2}(w-2A)^{2}-V^{\prime 2}+2V^{\prime }\mathcal{G}%
^{\prime }(\varepsilon _{w}-V)\}  \notag \\
\mathbf{\nabla }^{2}\mathcal{G} &=&\frac{\mathbf{\nabla }^{2}A}{w-2A}+2%
\mathcal{G}^{\prime 2}.
\end{eqnarray}%
The expressions for $k(r),m(r),$ and $n(r)$ that appear in Eqs.\ (\ref{54}))
are given in Eqs.\ (\ref{knm}). They can be evaluated using the above
expressions plus 
\begin{equation}
\mathbf{\nabla }^{2}\mathcal{K}=\frac{\mathbf{\nabla }^{2}\mathcal{L}-%
\mathbf{\nabla }^{2}\mathcal{J}}{2}.  \label{blw}
\end{equation}%
The only remaining parts of Eq.\ (\ref{54}) that need expressing are those
for $l^{\prime }$ and $q^{\prime }~$given \ in Eq. (\ref{A32}) Using Eq.\ (%
\ref{kf}) they can be obtained in terms of the above formulae.\vspace*{0.3cm}

\subsection{Weak Potential Limits of Quasipotentials}

The weak potential forms of the quasipotentials are needed for working out
perturbative spectral corrections. \ For weak potentials, we take $%
\varepsilon _{i}=m_{i}$ wherever they appear in the potentials and assume $%
|A|,|V|,|S|<<m_{i}$. \ Thus%
\begin{eqnarray}
\mathcal{L} &\mathcal{\rightarrow }&\frac{m_{w}S}{m_{1}m_{2}}\rightarrow 
\frac{S}{m_{1}+m_{2}},  \notag \\
\mathcal{J} &\mathcal{\rightarrow }&\mathcal{-}\frac{A}{w}-\frac{\varepsilon
_{w}V}{\varepsilon _{1}\varepsilon _{2}}\rightarrow -\frac{A+V}{m_{1}+m_{2}},
\notag \\
\mathcal{G} &\mathcal{\rightarrow }&\frac{A}{w}\rightarrow \frac{A}{%
m_{1}+m_{2}}
\end{eqnarray}%
Among other results, this will allow us to see how the scalar interaction
contributes oppositely to the spin-orbit and Darwin terms from both vector
interactions to lowest order. In that same limit, the dominant portions of $%
\Phi ^{\prime }s$ in Eq. (\ref{57}) are%
\begin{align}
\Phi _{D}& \rightarrow -\nabla ^{2}\mathcal{F\rightarrow }-\frac{1}{4r\left(
m_{1}+m_{2}\right) }(\frac{\nabla ^{2}(\mathcal{S}-A-V)(m_{2}^{2}+m_{1}^{2})%
}{m_{2}m_{1}}-\nabla ^{2}A),  \notag \\
\Phi _{SO}& \rightarrow -\frac{\mathcal{F}^{\prime }}{r}\rightarrow -\frac{1%
}{4r\left( m_{1}+m_{2}\right) }(\frac{(\mathcal{S}^{\prime }-A^{\prime
}-V^{\prime })(m_{2}^{2}+m_{1}^{2})}{m_{2}m_{1}}-A^{\prime })  \notag \\
\Phi _{SOD}& \rightarrow l^{\prime }\rightarrow -\frac{1}{4r\left(
m_{1}+m_{2}\right) }\frac{(\mathcal{S}^{\prime }-A^{\prime }-V^{\prime
})(m_{2}^{2}-m_{1}^{2})}{m_{1}m_{2}}  \notag \\
\Phi _{SOX}& \rightarrow q^{\prime }\rightarrow 0  \notag \\
\Phi _{SS}& \rightarrow \frac{1}{3}\nabla ^{2}\mathcal{G\rightarrow }\frac{1%
}{6\left( m_{1}+m_{2}\right) }\nabla ^{2}A,  \notag \\
\Phi _{T}& \rightarrow -\mathbf{\nabla }^{2}\frac{(S+A+V)}{9(m_{1}+m_{2})}, 
\notag \\
\Phi _{SOT}& \rightarrow -\frac{\mathcal{K}^{\prime }}{r}\rightarrow -\frac{%
(S^{\prime }+A^{\prime }+V^{\prime })}{2r(m_{1}+m_{2})}.  \label{wp1}
\end{align}%
Note how in the Darwin and spin-orbit terms the $\mathcal{S}^{\prime
}-V^{\prime }$ dependence shows how the scalar and timelike confining
effects tend to cancel for $\xi <1$. As anticipated, only the Darwin and
spin-orbit terms survive in the static limit when one of the two particles
becomes very massive. \ In that limit (say $m_{2}\rightarrow \infty $), the
two spin-orbit terms $\Phi _{SO}$ and $\Phi _{SOD}~$combine to%
\begin{equation}
-\mathbf{L\cdot (\sigma }_{1}+\mathbf{\sigma }_{2})\frac{\mathcal{S}^{\prime
}-A^{\prime }-V^{\prime }}{4r}-\mathbf{L\cdot (\sigma }_{1}-\mathbf{\sigma }%
_{2})\frac{\mathcal{S}^{\prime }-A^{\prime }-V^{\prime }}{4r}=-\mathbf{%
L\cdot \sigma }_{1}\frac{\mathcal{S}^{\prime }-A^{\prime }-V^{\prime }}{2r}.
\label{wp2}
\end{equation}

\subsection{Single Particle Limit ($m_{2}\rightarrow \infty $) of the TBDE}

In addition to using $p_{2}\rightarrow (m_{2},\mathbf{0})$, $%
p_{1}=(\varepsilon _{1},\mathbf{p}$)$,$ the single particle limit of the
TBDE is obtained by substituting $m_{2}\rightarrow \infty $ in the
expressions for the various potentials listed in Appendices A1, A2, and A3$.$
$\ $To determine that limit, we use that the total c.m. energy $%
w=\varepsilon _{2}+\varepsilon _{1}\rightarrow m_{2}+\varepsilon _{1}$. \ In
that case using the expressions for $m_{w}$ and $\varepsilon _{w}$ in Eqs. (%
\ref{mw}) and (\ref{ew}) we find%
\begin{eqnarray}
m_{w} &\rightarrow &m_{1}\equiv m,  \notag \\
\varepsilon _{w} &\rightarrow &\varepsilon _{1}\equiv \varepsilon ,  \notag
\\
G &=&(1-2A/(m_{2}+\varepsilon ))^{-1/2}\rightarrow 1,  \label{a45}
\end{eqnarray}%
and thus for the quantities related to the mass potentials, we have%
\begin{eqnarray}
M_{1} &\rightarrow &\sqrt{\left( m+S\right) ^{2}}=m+S,  \notag \\
M_{2}-m_{2} &\rightarrow &\sqrt{m_{2}^{2}+2mS+S^{2}}-m_{2}\rightarrow 0, 
\notag \\
\exp (\mathcal{L)} &\mathcal{=}&\exp \mathcal{(}\frac{M_{1}+M_{2}}{%
m_{1}+m_{2}})\rightarrow 1,  \notag \\
\tilde{S}_{1} &\rightarrow &S,  \notag \\
\tilde{S}_{2} &\rightarrow &0.
\end{eqnarray}%
For the energy potentials we have, since $\hat{P}\rightarrow (1,\mathbf{0})$
that%
\begin{eqnarray}
E_{1} &\rightarrow &\sqrt{\varepsilon ^{2}-2\varepsilon (A+V)+A^{2}+V^{2}}%
\equiv \varepsilon -U  \notag \\
\varepsilon _{2}-E_{2} &\rightarrow &\varepsilon _{2}-\sqrt{\varepsilon
_{2}^{2}-2\varepsilon _{2}A-2\varepsilon V+A^{2}+V^{2}}  \notag \\
&\rightarrow &\varepsilon _{2}-\varepsilon _{2}\sqrt{1-2A/\varepsilon _{2}}%
\rightarrow A,  \notag \\
\exp (\mathcal{J)} &\mathcal{=}&\exp \mathcal{(}\frac{E_{1}+E_{2}}{%
\varepsilon _{1}+\varepsilon _{2}})\rightarrow 1,  \notag \\
\tilde{A}_{1}^{\mu } &\rightarrow &U(1,\mathbf{0}),  \notag \\
\tilde{A}_{2}^{\mu } &\rightarrow &A(1,\mathbf{0}).  \label{a47}
\end{eqnarray}%
\ The TBDE (\ref{tbde}) thus reduce to%
\begin{eqnarray}
(\mathbf{\gamma }_{1}\cdot \mathbf{p-}\beta _{1}(\varepsilon -U)+m+S)\psi
&=&0,  \notag \\
(-\beta _{2}(m_{2}-A)+m_{2})\psi &\rightarrow &-m_{2}(\beta _{2}-1)\psi =0,
\end{eqnarray}%
effectively to a single ordinary one-body Dirac equation (\ref{obde}) for a
particle in combined external scalar and time only component vector
potential. \ Note that $U\neq A+V.$ \footnote{%
Strictly speaking, in the static limit, the square roots forms obtained from
Eq. (\ref{one}) imply $M_{1}\rightarrow \left\vert m+S\right\vert >0.~$For
for large distances, $A\rightarrow 0,$ and so Eq. (\ref{two}) implies $%
E_{1}\rightarrow |\varepsilon -V|$ and for short distances $V\rightarrow 0$, 
$E_{1}\rightarrow |\varepsilon -A|.$ These could possibly be in opposition
to the forms with indefinite sign of $m+S$ and $\varepsilon -V$ or $%
\varepsilon $ $-A$ \ that would appear in the one body Dirac equation. \ In
our applications to meson spectroscopy $S$ is always positive so we need not
worry about the use of the square root form of $M_{i}$ in the computation of 
$M_{i}$. \ Since $V$ is positive and increasing with distance, it is
possible that at large enough distances $\varepsilon -V<0.$ In that case use
of the positive root for the square root would not agree with the sign of $%
\varepsilon -V$. \ We found that for the $^{1}S_{0}$ $b\bar{u}$ and $b\bar{d}
$ \ mesons that does occur near the very end of the integration cutoff
distance of about $1.92$ Fermis. That is so close to the end, however, that
use of the positive square root is unlikely to have any effect on the
spectral results (for the $^{3}P_{2}$ $b\bar{u}$ and $b\bar{d}$ mesons $%
\varepsilon -V$ remained positive throughout the integration region.) Future
work may address the theoretical problem of how to make the approach to the
static limit of the square root forms smooth and exact.} \ Doing the usual
Pauli-reduction yields for the upper component wave function the Schr\"{o}%
dinger-like form which is the same as the $m_{2}\rightarrow \infty $ limit
of Eq. (\ref{57}). \ That limit can be readily seen from Eq. (\ref{kf}) and
Eqs. (\ref{a45}-\ref{a47}) which shows 
\begin{eqnarray}
\mathcal{F} &\rightarrow &\frac{1}{2}\log \frac{m+S+\varepsilon -U}{%
m+\varepsilon },  \notag \\
\mathcal{F}^{\prime } &=&\frac{1}{2}\frac{S^{\prime }-U^{\prime }}{%
m+S+\varepsilon -A},  \notag \\
\nabla ^{2}\mathcal{F} &\mathcal{=}&\frac{1}{2}\frac{\nabla ^{2}S-\nabla
^{2}U}{m+S+\varepsilon -U}-\frac{1}{2}\left( \frac{S^{\prime }-U^{\prime }}{%
m+S+\varepsilon -U}\right) ^{2},  \notag \\
\mathcal{L}^{\prime },\mathcal{J}^{\prime },\mathcal{G}^{\prime },\mathcal{K}%
^{\prime } &\rightarrow &0,  \notag \\
\Phi _{D} &\rightarrow &\mathcal{F}^{\prime 2}-\mathbf{\nabla }^{2}\mathcal{F%
},  \notag \\
\Phi _{SO} &\rightarrow &-\frac{\mathcal{F}^{\prime }}{r},  \notag \\
\Phi _{SOD} &\rightarrow &l^{\prime }=-\frac{1}{2r}\frac{%
E_{2}M_{2}-E_{1}M_{1}}{E_{2}M_{1}+E_{1}M_{2}}(\mathcal{L}^{\prime }+\mathcal{%
J}^{\prime })  \notag \\
&\rightarrow &-\frac{1}{2r}\frac{E_{2}M_{2}}{E_{2}M_{1}+E_{1}M_{2}}(\mathcal{%
L}^{\prime }+\mathcal{J}^{\prime })\rightarrow -\frac{\mathcal{F}^{\prime }}{%
r}.
\end{eqnarray}%
From the last line we obtain%
\begin{equation}
\Phi _{SO}\mathbf{L\cdot (\sigma }_{1}+\mathbf{\sigma }_{2})+\Phi _{SOD}%
\mathbf{L\cdot (\sigma }_{1}-\mathbf{\sigma }_{2})=-\frac{2\mathcal{F}%
^{\prime }}{r}\mathbf{L\cdot \sigma }_{1}=-\frac{S^{\prime }-U^{\prime }}{%
m+S+\varepsilon -U}\mathbf{L\cdot \sigma }_{1}.
\end{equation}%
and hence Eq. (\ref{57}) reduces to Eq. (\ref{571}). \ In Section 6.4 we
display our static limit equations for Th1. \ 

\section{Radial Equations}

The following are radial eigenvalue equations, \cite{liu}, \cite{yoon},
corresponding to Eq.\ (\ref{57}) . For a general singlet $^{1}J_{J}$ wave
function $v_{LSJ}=v_{J0J}\equiv v_{0}$ coupled to a general triplet $%
^{3}J_{J}$ wave function $v_{J1J}\equiv v_{1}$, the wave equation

\begin{align}
& \{-\frac{d^{2}}{dr^{2}}+\frac{J(J+1)}{r^{2}}+2m_{w}S+S^{2}+2\varepsilon
_{w}A-A^{2}+2\varepsilon _{w}V-V^{2}+\Phi _{D}\mathbf{-}3\Phi _{SS}\}v_{0} 
\notag \\
& +2\sqrt{J(J+1)}(\Phi _{SOD}-\Phi _{SOX})v_{1}  \notag \\
& =b^{2}v_{0},  \label{ss}
\end{align}%
is coupled to

\begin{align}
& \{-\frac{d^{2}}{dr^{2}}+\frac{J(J+1)}{r^{2}}+2m_{w}S+S^{2}+2\varepsilon
_{w}A-A^{2}+2\varepsilon _{w}V-V^{2}+\Phi _{D}  \notag \\
& -2\Phi _{SO}+\Phi _{SS}+2\Phi _{T}-2\Phi _{SOT}\}v_{1}+2\sqrt{J(J+1)}(\Phi
_{SOD}+\Phi _{SOX})v_{0}  \notag \\
& =b^{2}v_{1}.  \label{pp}
\end{align}%
For a general $S=1,$ $J=L+1$ \ wave function $v_{J-11J}\equiv v_{+}~$coupled
to a general $S=1,$ $J=L-1~$wave function $v_{J+11J}\equiv v_{-}$ \ the
equation

\begin{align}
& \{-\frac{d^{2}}{dr^{2}}+\frac{J(J-1)}{r^{2}}+2m_{w}S+S^{2}+2\varepsilon
_{w}A-A^{2}+2\varepsilon _{w}V-V^{2}+\Phi _{D}  \notag \\
& +2(J-1)\Phi _{SO}+\Phi _{SS}+\frac{2(J-1)}{2J+1}(\Phi _{SOT}-\Phi
_{T})\}v_{+}  \notag \\
& +\frac{2\sqrt{J(J+1)}}{2J+1}\{3\Phi _{T}-2(J+2)\Phi _{SOT}\}v_{-}  \notag
\\
& =b^{2}v_{+},  \label{pl}
\end{align}%
$\allowbreak \allowbreak \allowbreak $is coupled to 
\begin{align}
& \{-\frac{d^{2}}{dr^{2}}+\frac{(J+1)(J+2)}{r^{2}}+2m_{w}S+S^{2}+2%
\varepsilon _{w}A-A^{2}+2\varepsilon _{w}V-V^{2}+\Phi _{D}  \notag \\
& -2(J+2)\Phi _{SO}+\Phi _{SS}+\frac{2(J+2)}{2J+1}(\Phi _{SOT}-\Phi
_{T})\}v_{-}  \notag \\
& +\frac{2\sqrt{J(J+1)}}{2J+1}\{3\Phi _{T}+2(J-1)\Phi _{SOT}\}v_{+}  \notag
\\
& =b^{2}v_{-}.  \label{mi}
\end{align}%
For the uncoupled $^{3}P_{0}$ states the single equation is%
\begin{align}
& \{-\frac{d^{2}}{dr^{2}}+\frac{2}{r^{2}}+2m_{w}S+S^{2}+2\varepsilon
_{w}A-A^{2}+2\varepsilon _{w}V-V^{2}+\Phi _{D}  \notag \\
& -4\Phi _{SO}+\Phi _{SS}+4(\Phi _{SOT}-\Phi _{T})\}v_{-}=b^{2}v_{-}.
\end{align}

\section{Two Body Dirac Equations for QED}

The Schr\"{o}dinger-like form Eq. (\ref{57}) of the TBDE given in Eq. (\ref%
{tbde}) can be used for QCD bound state (meson spectroscopy) and for QED
bound states (positronium, muonium, and hydrogen-like systems). \ In our
meson spectroscopy work presented in this paper we use the three invariant
functions $S(r),~A(r),$ and $V(r)$ specified in Sec. 3. \ Once these are
specified then so are the three vertex invariants $\mathcal{L(}r),~\mathcal{%
G(}r),$ and $\mathcal{J(}r).$ They, in turn, fix the quasipotentials given
in Eq. (\ref{54}) and below to (\ref{blw}) that appear in Eq. (\ref{57}) and
its radial forms of Appendix B. \ To make the transition from QCD meson
bound states to QED bound states we simply take $S(r)=V(r)=0$ and $%
A(r)=-\alpha /r$ (we remind the reader that only in the c.m. system is the
invariant $r=\sqrt{x_{\perp }^{2}}$ equal to $\left\vert \mathbf{r}%
\right\vert ).$ \ Our QED spectral results then follow from solving
nonperturbatively (i.e. numerically or analytically) the radial eigenvalue
equations of Appendix B. \ For example, since $\Phi _{D}=3\Phi _{SS}$ for $%
S=V=0,$\ the equal mass spin singlet equation (\ref{ss}) collapses to

\begin{equation}
\{-\frac{d^{2}}{dr^{2}}+\frac{J(J+1)}{r^{2}}-\frac{2\varepsilon _{w}\alpha }{%
r}-\frac{\alpha ^{2}}{r^{2}}\}v_{0}=b^{2}v_{0},
\end{equation}%
which has the analytic spectral solution given in Eq. (\ref{ect}) for $J=0$.
\ As long as $J(J+1)-\alpha ^{2}$ $>-1/4,$ since the above equation take the
limiting form of%
\begin{equation}
\{-\frac{d^{2}}{dr^{2}}+\frac{J(J+1)}{r^{2}}-\frac{\alpha ^{2}}{r^{2}}%
\}v_{0}=0,
\end{equation}%
it is clear that the effective potential is nonsingular, thus allowing a
well defined eigenvalue solution\footnote{%
In particular, the short distance radial behavior is $v_{J0J}\rightarrow r^{%
\sqrt{J(J+1)-\alpha ^{2}}\text{.}}$}. \ The paper \cite{yoon} demonstrated
that the corresponding short distance behavior of the potentials in the
other radial equations of Appendix B for the other QED bound states also
allow well defined eigenvalue solutions\footnote{%
The short distance radial dependencies of the wave functions that arise from
Eqs. (\ref{pl}) and (\ref{mi}) are the well behaved forms $%
v_{J-11J}(r)=r^{(1/2+\sqrt{J(J+1)-\alpha ^{2}})}$ and $v_{J+11J}(r)=\frac{J}{%
\sqrt{J(J+1)}}v_{J-11J}(r)~$\cite{yoon}. \ Such behaviors as represented
here and in the previous footnote arise because the effective potentials
that appear in our bound state equations are \ well defined (no delta
function potentials for example). \ We refer the interested reader to Eq.
(42-43) of section V of \cite{yoon} for the intriging details of the
cancellation of otherwise singular potentials due to the presence of the
tensor coupling terms.}. \ Beyond that, numerical solutions of these
eigenvalue equations yield spectra (not limited just to the singlet \ ground
states) in agreement with standard perturbative $O(\alpha ^{4})$ results 
\cite{becker}. \ In electron volts the numerical binding energy for the
singlet ground state of positronium from Eq. (\ref{ss}) is -6.8033256279
versus the perturbative result of Eq. (\ref{exct}) or $m(-\alpha
^{2}/4-21\alpha ^{4}/64)=$ -6.8033256719. \ The difference in units of $%
m\alpha ^{4}/2$ is -6.08E-05 which is on the order of $\alpha ^{2}$, so
that, as expected from (\ref{exct}) the difference is on the order of $%
m\alpha ^{6}.~$The corresponding numerical binding energy in electron volts
for the triplet ground state from the coupled equations Eqs. (\ref{pl}) and (%
\ref{mi}) is -6.8028426132 versus the perturbative result \cite{becker} of $%
m(-\alpha ^{2}/4+\alpha ^{4}/192)=~$-6.8028426636. These results do not
include the effects of the annihilation diagram. \ The difference in units
of $m\alpha ^{4}/2$ is 6.97E-05 which is on the order of $\alpha ^{2}$, so
that the difference is also on the order of $m\alpha ^{6}$. These results
(from a very extensive list of numerically computed spectral results in \cite%
{becker}) taken together, represent crucial tests of this formalism, ones
that have not been demonstrated in any other relativistic bound state
formalism. \ In fact, the authors of \cite{iowa} have found a particular
quasipotential formalism that does give such agreement, but only for the
ground state. \ They also demonstrate that several well know two-body
relativistic bound state formalisms (including the Blankenbecler-Sugar
formalism and the formalism of Gross) fail this test. Let us be explicit
about the implications of either the failure of this test or the lack of
performing this test. \ When one proposes a new bound state formalism such
as Dirac did in 1928, \ it was essential that at the very least it reproduce
the nonrelativistic hydrogenic spectrum. \ Beyond that it provides two other
remarkable results. \ First of all, by way of an order $1/c^{2}$ expansion,
it gave an order $\alpha ^{4}$ perturbative correction to the
nonrelativistic spectral results. \ Remarkably, an exact solution of the
same equation later by Darwin gave spectral results which when truncated to
order $\alpha ^{4}$ agreed precisely with the perturbative treatment, and,
at the time, with experimental fine structure measurements. \ Breit, in the
development of his two-body equation gave us an equation with interactions
beyond the Coulomb potential that ultimately reproduced, when treated
perturbatively, spectral results for two-body systems that agreed through
terms of order $\alpha ^{4}$ with experiment. \ Unlike the one-body Dirac
equation which has an exact spectral solution which agrees, at order $\alpha
^{4}$, with its perturbative solution, the Breit equation has no exact or
for that matter numerical solution which agrees, at order $\alpha ^{4}$,
with its perturbative solution. The same could be said for most all other
two-body relativistic treatments proposed since then, with the two
exceptions (\cite{becker} and \cite{iowa}) noted above. \ One's two-body
formalism having an agreement of its nonperturbative treatment with its
perturbative treatment of the spectra in a well established field theory
such as QED, in our opinion, should be regarded as a necessary hurdle to
pass \ before going on to apply these formalisms to potential models for
meson spectroscopy. \ 

\bigskip

\begin{acknowledgement}
The authors wish to thank Dr. C. Y. Wong, Prof. J. H. Yoon, and Joshua
Whitney for helpful suggestions and discussions.
\end{acknowledgement}

\end{document}